%% file: ms.tex
\documentclass[apjl]{emulateapj}

\def\spose#1{\hbox to 0pt{#1\hss}}
\def\simlt{\mathrel{\spose{\lower 3pt\hbox{$\mathchar"218$}}
     \raise 2.0pt\hbox{$\mathchar"13C$}}}
\def\simgt{\mathrel{\spose{\lower 3pt\hbox{$\mathchar"218$}}
     \raise 2.0pt\hbox{$\mathchar"13E$}}}
\providecommand\scription[2]{\scriptsize#1$\;${\scriptsize\uppercase\expandafter{\romannumeral #2}}\relax}%
\providecommand\tabion[2]{#1$\;${\tiny\uppercase\expandafter{\romannumeral #2}}\relax}%

\providecommand{\bnt}{\ensuremath{b_{\mathrm nt}}}
\providecommand{\btherm}{\ensuremath{b_{\mathrm therm}}}
\providecommand{\MH}{\ensuremath{\mbox{[M/H]}}}
\providecommand{\logT}{\ensuremath{\mbox{log}\,T}}

\providecommand{\logU}{\ensuremath{\mbox{log}\,U}}

\providecommand{\kms}{\,\ensuremath{\rm{km\,s}^{-1}}}
\providecommand{\cm}{\,\ensuremath{\mbox{cm}^{-2}}}
\providecommand{\A}{\,\ensuremath{\mbox{\AA}}}
\providecommand{\mA}{\,\ensuremath{\mbox{m\AA}}}
\providecommand{\kpc}{\,\ensuremath{\mbox{kpc}}}
\providecommand{\Mpc}{\,\ensuremath{\mbox{Mpc}}}
\providecommand{\eV}{\,\ensuremath{\mbox{eV}}}

\providecommand{\HI}{\ensuremath{\mbox{\ion{H}{1}}}}
\providecommand{\Lya}{\ensuremath{\mbox{Ly\,}\alpha}}
\providecommand{\Lyb}{\ensuremath{\mbox{Ly\,}\beta}}
\providecommand{\Lyab}{\ensuremath{\mbox{Ly\,}\alpha, \beta}}
\providecommand{\Lycd}{\ensuremath{\mbox{Ly\,}\gamma, \delta}}
\providecommand{\Lyc}{\ensuremath{\mbox{Ly\,}\gamma}}
\providecommand{\Lyd}{\ensuremath{\mbox{Ly\,}\delta}}
\providecommand{\NeVIII}{\ensuremath{\mbox{\ion{Ne}{8}}}}
\providecommand{\NeIX}{\ensuremath{\mbox{\ion{Ne}{9}}}}
\providecommand{\OI}{\ensuremath{\mbox{\ion{O}{1}}}}
\providecommand{\OIII}{\ensuremath{\mbox{\ion{O}{3}}}}
\providecommand{\OIV}{\ensuremath{\mbox{\ion{O}{4}}}}
\providecommand{\OVI}{\ensuremath{\mbox{\ion{O}{6}}}}
\providecommand{\OVII}{\ensuremath{\mbox{\ion{O}{7}}}}
\providecommand{\OVIII}{\ensuremath{\mbox{\ion{O}{8}}}}
\providecommand{\CI}{\ensuremath{\mbox{\ion{C}{1}}}}
\providecommand{\CII}{\ensuremath{\mbox{\ion{C}{2}}}}
\providecommand{\CIII}{\ensuremath{\mbox{\ion{C}{3}}}}
\providecommand{\CIV}{\ensuremath{\mbox{\ion{C}{4}}}}
\providecommand{\NII}{\ensuremath{\mbox{\ion{N}{2}}}}
\providecommand{\NV}{\ensuremath{\mbox{\ion{N}{5}}}}
\providecommand{\SiII}{\ensuremath{\mbox{\ion{Si}{2}}}}
\providecommand{\SiIII}{\ensuremath{\mbox{\ion{Si}{3}}}}
\providecommand{\SiIV}{\ensuremath{\mbox{\ion{Si}{4}}}}
\providecommand{\SII}{\ensuremath{\mbox{\ion{S}{2}}}}
\providecommand{\scHI}{\ensuremath{\mbox{\scription{H}{1}}}}
\providecommand{\scOVI}{\ensuremath{\mbox{\scription{O}{6}}}}
\providecommand{\scSiIII}{\ensuremath{\mbox{\scription{Si}{3}}}}
\providecommand{\scCIII}{\ensuremath{\mbox{\scription{C}{3}}}}
\providecommand{\tabHI}{\ensuremath{\mbox{\tabion{H}{1}}}}
\providecommand{\tabOVI}{\ensuremath{\mbox{\tabion{O}{6}}}}
\providecommand{\tabOI}{\ensuremath{\mbox{\tabion{O}{1}}}}
\providecommand{\tabOIII}{\ensuremath{\mbox{\tabion{O}{3}}}}
\providecommand{\tabOIV}{\ensuremath{\mbox{\tabion{O}{4}}}}

\providecommand{\tabCII}{\ensuremath{\mbox{\tabion{C}{2}}}}
\providecommand{\tabCIII}{\ensuremath{\mbox{\tabion{C}{3}}}}

\providecommand{\tabNII}{\ensuremath{\mbox{\tabion{N}{2}}}}
\providecommand{\tabNV}{\ensuremath{\mbox{\tabion{N}{5}}}}
\providecommand{\tabSiII}{\ensuremath{\mbox{\tabion{Si}{2}}}}
\providecommand{\tabSiIII}{\ensuremath{\mbox{\tabion{Si}{3}}}}
\providecommand{\tabSiIV}{\ensuremath{\mbox{\tabion{Si}{4}}}}
\providecommand{\tabSII}{\ensuremath{\mbox{\tabion{S}{2}}}}
\providecommand{\hden}{\ensuremath{n_{\mathrm H}}}

\providecommand{\loghden}{\ensuremath{\mbox{log}\,\hden}}
\providecommand{\NHI}{\ensuremath{N(\HI)}}
\providecommand{\NOVI}{\ensuremath{N(\OVI)}}

\providecommand{\logN}{\ensuremath{\mbox{log\,}N}}
\providecommand{\logNOVI}{\ensuremath{\mbox{log\,}N(\OVI)}}
\providecommand{\logNHI}{\ensuremath{\mbox{log\,N(\HI)}}}
\providecommand{\Wr}{\ensuremath{W_r}}
\providecommand{\K}{\,\ensuremath{\mbox{K}}}
\providecommand{\Oion}{\ensuremath{{\mathrm O}^{5+}}}
\providecommand{\ROVI}{\ensuremath{R_{\scOVI}}}
\providecommand{\bOVI}{\ensuremath{b_{\scOVI}}}
\providecommand{\bHI}{\ensuremath{b_{\scHI}}}
\providecommand{\bSiIII}{\ensuremath{b_{\scSiIII}}}
\providecommand{\bCIII}{\ensuremath{b_{\scCIII}}}
\providecommand{\pI}{\ensuremath{\mbox{paper~I}}}

\providecommand{\dex}{\,\ensuremath{\mbox{dex}}}
\providecommand{\OmegaOVI}{\ensuremath{\Omega_{{\rm O}^{5+}}}}
\providecommand{\zqso}{\ensuremath{z_{\mathrm qso}}}
\providecommand{\zabs}{\ensuremath{z_{\mathrm abs}}}

\shorttitle{Physical Conditions of low-$z$ \OVI\ Absorbers}
\shortauthors{Thom \& Chen}

\begin{document}

\title{A STIS Survey for \OVI\ Absorption Systems at $0.12 < z\lesssim
0.5$ II.: Physical Conditions of the Ionized Gas}

\author{C. Thom and Hsiao-Wen Chen} \affil{Dept.\ of Astronomy \& Astrophysics and Kavli
  Institute for Cosmological Physics \\ 
  University of Chicago, Chicago, IL, 60637, U.S.A. \\ 
  {\tt cthom, hchen@oddjob.uchicago.edu}}

\altaffiltext{1}{Based in part on observations with the NASA/ESA Hubble Space
Telescope, obtained at the Space Telescope Science Institute, which is operated
by the Association of Universities for Research in Astronomy, Inc., under NASA
contract NAS5--26555.}

\begin{abstract}

  We present a complete catalogue of 27 \OVI\ absorbers at low
  redshift ($0.12 < z < 0.5$) from a blind survey of 16 QSO echelle
  spectra in the HST/STIS data archive.  These absorbers are
  identified based only upon matching line profiles and the expected
  doublet ratio between the $\lambda\lambda\,1031, 1037$ transitions.
  Subsequent searches are carried out to identify their associated
  transitions.  Here we present all relevant absorption properties.
  By considering absorption components of different species which are
  well-aligned in velocity-space, we derive gas temperatures and
  non-thermal broadening values, $\bnt$. We show that in all 16 cases
  considered the observed line width is dominated by non-thermal
  motion and that gas temperatures are well below those expected for
  \Oion\ in collisional ionization equilibrium.  This result reaffirms
  previous findings from studies of individual lines of sight, but are
  at odds with expectations for a WHIM origin. At least half of the
  absorbers can be explained by a simple photoionization model. In
  addition, in some absorbers we find evidence for large variation in
  gas density/metallicity across components in individual absorbers.
  Comparisons of multiple associated metal species further show that
  under the assumption of the gas being photoionized by the
  metagalacitic background radiation field, the absorbing clouds have
  gas densities $<\hden> < -2.9$ and sizes $L > 1\kpc$. Finally, we
  compare our absorber selection with the results of other independent
  studies.

\end{abstract}


\section{Introduction}
\label{sec: introduction}
One of the current key questions in observational cosmology is the location of the missing
baryons. The {\it total} baryon content of the universe is well constrained, with various
measurements in relatively good agreement \citep[e.g.][]{burles-etal-01-BBN-omega,
  spergel-etal-03-WMAP-yr1, omeara-etal-06-deuterium}.  In the high-redshift universe, the \Lya\
forest dominates the baryon census \citep{rauch-etal-97-Lya-Omegab}, but in the present day
universe, only 1/3 of the baryons have been identified in known components
\citep{fukugita-peebles-04-mean-densities}.  Cosmological simulations indicate that up to 50\% of
the baryons exist in a warm-hot intergalactic medium (WHIM), in which gas is shock-heated to $\sim
10^5 - 10^7\K$ by the accretion onto large-scale structures, and remains hot owing to the low gas
density and inefficient cooling \citep{cen-ostriker-99-missing-baryons, dave-etal-01-WHIM-baryons,
  cen-ostriker-06-II-winds}. This gas is a result of accretion onto large-scale filamentary
structures, where cooling is inefficient due to the low densities, and (in some models) the
operation of large scale winds that shock-heat outflowing gas to $\sim10^6\K$
\citep[e.g.][]{cen-ostriker-06-II-winds}.  It is therefore important to identify this gas, and
constrain its contribution to the baryon fraction.

At the temperature range in question ($\logT = 5.0 - 7.0$), the best observational window for this
hot gas is X-ray absorption lines \citep[e.g.\ \OVII\,K$\alpha$, \OVIII\,K$\alpha$,
\NeIX\,K$\alpha$; ][]{gnat-sternberg-07-non-equilibrium-cooling}, but the resolutions of current
X-ray spectrographs (e.g.\ Chandra and XMM/Newton) are an order of magnitude too coarse to be
useful, and the sensitivities are more than an order of magnitude too low
\citep[e.g.][]{fang-etal-06-galactic-OVII}. The best tool currently available, therefore, is UV
absorption spectroscopy. The \OVI\ doublet ($\lambda\lambda\,1031.9261, 1037.617\A$) offers the best
hot gas tracer for a number of reasons: the transition has a large oscillator strength; Oxygen is a
relatively common metal; the abundance of the \Oion\ ion peaks in collisional ionization equilibrium
(CIE) at $\logT \sim 5.3$\footnote{It also has an ionization fraction greater than $10^{-2}$ up to
  $\logT \sim 5.7$; see e.g.\ \citet{gnat-sternberg-07-non-equilibrium-cooling}.}; the transitions
occurs longward of the Lyman limit; and absorption can be detected down to limiting column densities
$\logN \approx 13.5 (\Wr = 30\mA)$ with current instruments such as STIS.

If the absorption lines of an element are resolved and unsaturated, we can measure directly the
column density ($N$) and Doppler parameter ($b$) of the absorbing gas by voigt profile
fitting\footnote{In cases where multiple transitions of the same species are available, a
  curve-of-growth analysis can also be used.}.  The Doppler parameter is an oft-used measure of the
temperature of the gas, through the well-known relation $b^2 = \bnt^2 + 2kT/m$, where \bnt\ accounts
for non-thermal broadening of the line due to e.g.\ turbulence. The column-density, meanwhile, may
be used in conjunction with the path-length to derive the contribution to the cosmological mass density of
the O$^{5+}$ ions, \OmegaOVI\ \citep[see][hereafter \pI]{thom-chen-08-I-OVI-statistics}.

In the optical band, ground-based high resolution (echelle) spectra have made \OVI\ detection
possible at $2.3 \simlt z \simlt 3$ \citep[e.g.][]{norris-etal-83-OVI, carswell-etal-02-OVI,
  simcoe-etal-02-OVI-search}. The lower limit is set by the plummeting transmissivity of the
atmosphere to UV photons below $\sim 3500\A$, while the upper limit is set by confusion with the
\Lya\ forest. In the high-redshift regime, \Oion\ is predominantly photoionized by the UV background
radiation field, much the same as the \Lya\ clouds \citep{carswell-etal-02-OVI, simcoe-etal-04}. To
push to lower redshifts, space-based UV spectrographs are required. Using near-UV spectra from the
Faint Object Spectrograph on-board HST, \citet{burles-tytler-96-OVI-Omega_b} identified 12 \OVI\
doublets with $W_{r}(1031) > 0.21\AA$ in the range $0.5 \simlt z \simlt 2$ and derived a
cosmological mass density $\OmegaOVI\,h \geq 7\times10^{-8}$, providing the first constraint on the
mass density of this highly ionized gas.

In the low-redshift universe ($z \simlt 0.5$), the \OVI\ doublet transition remains in the far-UV
($\lambda_{obs} \simlt 1600\A$).  The terminated Far Ultraviolet Spectroscopic Explorer \citep[{\it
  FUSE}; ][]{moos-etal-00-FUSE} could probe \OVI\ absorption out to $z \sim 0.15$ with good
signal-to-noise, but at a resolution of only $\sim 20\kms$
\citep[e.g.][]{danforth-etal-05-OVI_baryon_census}. This compares poorly with the thermal line width
of \OVI\ ($\sim 3\kms$ at $\logT = 4.0$ and $\sim 10\kms$ at $\logT = 5.0$).  The (currently
suspended) Space Telescope Imaging Spectrograph \citep[{\it STIS}; ][]{woodgate-etal-98-STIS}
on-board the Hubble Space Telescope ({\it HST}) offers a resolution of $\sim 6-7\kms$ and its data
are useful for studies of \OVI\ absorption at slightly higher redshifts than FUSE ($0.12 < z < 0.5$).

Early work on \OVI\ absorption in the low-$z$ universe typically concentrated on single lines of
sight, due to the paucity of data. Single-sightline or single-absorber analyses have been conducted
on the various lines of sight, beginning with the QSO H\,1821$+$643
\citep{tripp-etal-98-Lya-galaxies, tripp-etal-00-H1821+643, tripp-etal-01-H1821+643_z0.1212,
  oegerle-etal-00-H1821+643-FUSE}. As more data become available, further sightline were analysed:
PG0953$+$415 \citep{savage-etal-02-PG0953+415}, PG1259$+$593 \citep{richter-etal-04-PG1259+593},
PG1116$+$215 821\citep{sembach-etal-04-PG1116+215}, PKS0405$-$123
\citep{prochaska-etal-04-PKS0405-12}, HE0226$-$4110 \citep{lehner-etal-06-HE0226-4110} and
PKS\,1302$-$102 \citep{cooksey-etal-08-PKS1302-102}. Now, with a database of UV data
available\footnote{More data are likely soon to be available with the upcoming HST servicing
  mission.}, statistical approaches have become possible \citep{danforth-etal-05-OVI_baryon_census,
  danforth-etal-06-OVI_survey, thom-chen-08-I-OVI-statistics, tripp-etal-08-OVI,
  danforth-shull-08-OVI}.

This is the second in a series of papers reporting the results of our search for \OVI\ absorption
systems in the STIS E140M archive. Unlike other searches, we employ a {\it blind} search for \OVI\
doublets, which is independent of {\it a priori} knowledge of the presence of other transitions such
as \Lya. In \pI\ we reported results on the statistics of the \OVI\ absorbers. The major results
were: a) a measurements of the number of absorbers per unit redshift, $d\,{\cal N}(W\ge 30\,\mA)/dz
= 10.4 \pm 2.2$; b) a measurement of the cosmological mass density of the \Oion\ gas, $\OmegaOVI\,h
= (1.7 \pm 0.3) \times10^{-7}$; c) $<5$\% of \OVI\ absorbers originate in underdense regions that do
not show a significant trace of \HI; d) \HI\ column densities of \OVI\ absorbers span more than 5
orders of magnitude, and a moderate correlation exists between \NHI\ and \NOVI; and e) the number
density of \OVI\ absorbers along a given line of sight appears to be inversely correlated with the
number density of \HI\ absorbers. In this paper we present our catalogue of \OVI\ absorbers upon
which the results of \pI are based. We also address the physical conditions of the \Oion\ bearing
gas. The nature of the IGM, the \OVI\ absorbers in particular, and their relation to the WHIM is an
area that has seen much recent progress. At least two other groups have contemporaneously reported
results of similar analyses. We refer the interested reader to the works of
\citet{tripp-etal-08-OVI} and \citet{danforth-shull-08-OVI}. Specifically, see
\citet{tripp-etal-08-OVI} Sec~\ref{sec: not_recovered} for comments on the differences between both
works.

We recall the description of our search and selection technique from \pI\ in Sec~\ref{sec: data},
presenting the full table of absorbers. We discuss individual lines-of-sight, and present the
measured quantities for each system, in Sec~\ref{sec: los}. In Sec~\ref{sec: not_recovered} we
discuss those systems reported in \citet{tripp-etal-08-OVI} that are not accepted by our selection
criteria. The physical properties of the ionized gas selected via \OVI\ absorption are discussed in
Sec~\ref{sec: properties}. Sec~\ref{sec: discussion} contains a summary and concluding remarks.

\section{Data and Catalogue}
\label{sec: data}
\subsection{STIS Data}
Our data were drawn from the STIS data archive\footnote{CALSTIS v2.23 (2006 Oct 06)}. The data and
search technique is described in \pI\ (see in particular, Sec~2.1 for details). For completeness, we
repeat parts of that description here. We chose all data with sufficient signal-to-noise ratio ($S/N
\geq 5$ per pixel), and resolution, as to be able to detect weak ($\Wr > 30\mA$) \OVI\
absorbers. This selection yielded 16 lines of sight with STIS E140M data. Table~\ref{tab:
  journal_observations} describes these lines of sight.

\OVI\ absorbers were selected on the basis of the equivalent width ratio of the doublet lines
alone. In order not to bias our search by the presence of other transitions, we do not consider
other associated lines until a later stage. This differs from the traditional technique, which
relies on {\it a priori} knowledge of absorber positions (usually from \Lya), and then searches for
possibly associated species.  We began our search by Hanning smoothing the spectra, and identifying
all deviations $> 1.5\,\sigma$ from the continuum level. A gaussian profile was fit to each feature,
with the width restricted to $ 6 < \sigma < 300\kms$, where the lower limit is taken from the
spectrograph resolution, and the upper limit from consideration of the line width distribution of
known \OVI\ systems \citep[e.g.][]{heckman-etal-02, danforth-etal-05-OVI_baryon_census}. An
equivalent width was determined by directly integrating the data, with integration limits determined
from the gaussian width. All features with $<2\,\sigma$ significance were rejected.

We consider each feature a putative \OVI\,1031 line and fit a doublet absorption model to the data
to determine whether the spectrum is consistent with the presence of both \OVI\,1031, 1037
lines. The doublet model requires both transitions to have the same line width, and have a line
strength ratio of $2:1$ (i.e.\ the ratio of the oscillator strength--wavelength product,
$f\,\lambda^2$, for the two transitions). Each system was also visually inspected to determine
whether the spectrum is consistent with the presence of a doublet. We accepted candidates if (a) the
\OVI\,1031 member has $> 3\,\sigma$ significance, and; (b) the ratio of rest-frame line
strengths---$\ROVI \equiv W_r(1031)/W_r(1037)$---lies between $1-\sigma_{\ROVI}$ and
$2+2\,\sigma_{\ROVI}$.

Finally, we visually identified other species associated with each absorber, typically searching for
transitions of the ions H$^{0}$, Si$^{+, 2+, 3+}$, S$^{+, 2+}$, O$^{0, 2+, 3+}$, C$^{+, 2+}$, N$^{+,
  4+}$ and Ne$^{7+}$. The {\it vpfit}\footnote{http://www.ast.cam.ac.uk/\~{ }rfc/vpfit.html}
software was used to fit Voigt profiles to all components of all detected species in each
absorber. {\it vpfit} convolves the voigt profile with a gaussian line-spread function (LSF), whose
width is set by the instrument resolution. The STIS LSF has significant broad wings in some
configurations. We have tested that this difference does not affect our measurements of $N$ and $b$
(i.e. the differences are much smaller than the error in the measured values).  The number, position
and initial values for the absorption components was assessed initially by eye.  We performed a
minimum-$\chi^2$ analysis that includes multiple components; new components were added and fit
iteratively until either the normalized $chi^2$ did not decrease, or the newly added component
became ill-constrained (error-bars for the best-fit parameters were greater than the best-fit values)
by the data. For \OVI\ and \HI, this process is typically facilitated by the presence of multiple
transitions. Section~\ref{sec: los} has details of the component structure for each absorber.  The
fits were used to evaluate the total column density for each absorber, and are given in
Table~\ref{tab: ovi_absorbers}. Fit results for the individual components that comprise each
absorber are given in Tables~\ref{tab: 3C249.1_transitions}--\ref{tab: Ton28_transitions}. In our
fitting, we employed the latest version of the standard atomic data distributed with {\it
  vpfit}. These data are primarily from the compilation of \citet{morton-03}, with some more recent
updates included.

\subsection{A Catalogue of Random \OVI\ Absorbers at $0.12 < z < 0.50$}

The final catalogue of \OVI\ doublet systems is given in Table~\ref{tab: ovi_absorbers}. The table
lists the line-of-sight and redshift of the absorber, typically the redshift of the strongest \OVI\
component (columns 1 \& 2). Rest-frame equivalent widths ($W_r$), and errors ($\sigma_{W_{r}}$), are
reported in units of \mA\ (columns 3--6), for both lines of the \OVI\ doublet. The ratio of
equivalent widths, \ROVI, and associated error are listed in columns (7) and (8). Finally, columns
(9) and (10) give the total \OVI\ column density for the absorber, which is the sum of the
individual components. Individual component fitting results are given in the following section.

\placetable{tab: ovi_absorbers}

\section{Individual Lines of Sight}
\label{sec: los}
For each absorber in the following sub-sections, we present the results of our profile fitting in
the associated figures and tables. Spectra are unbinned, and profile fits are overlayed in solid
(blue). The error spectrum is plotted as a solid line at the bottom of each panel (red), also
unbinned. The dot-dashed (green) lines indicate the continuum and zero flux levels. Component
positions are marked above the spectra with the (red) solid ticks. The tables list, for each
absorber along each line of sight, the ion, component position in redshift ($z$) and velocity offset
($v$) (from the absorber redshift), and the profile fit Doppler parameter ($b$) and column-density
(\logN), and their associated errors. The flags denote upper (U) and lower (L) limits, or uncertain
features (Z). Note that errors are not reported for upper and lower limits.

\subsection{3C\,249.1}
The QSO 3C\,249.1 lies at $\zqso = 0.3115$, and we searched for \OVI\ absorption along the line of
sight from $0.122 < \zabs < 0.2885$. We identify a single \OVI\ doublet system along this line of
sight.

\begin{figure}
  \epsscale{0.6}
  \plotone{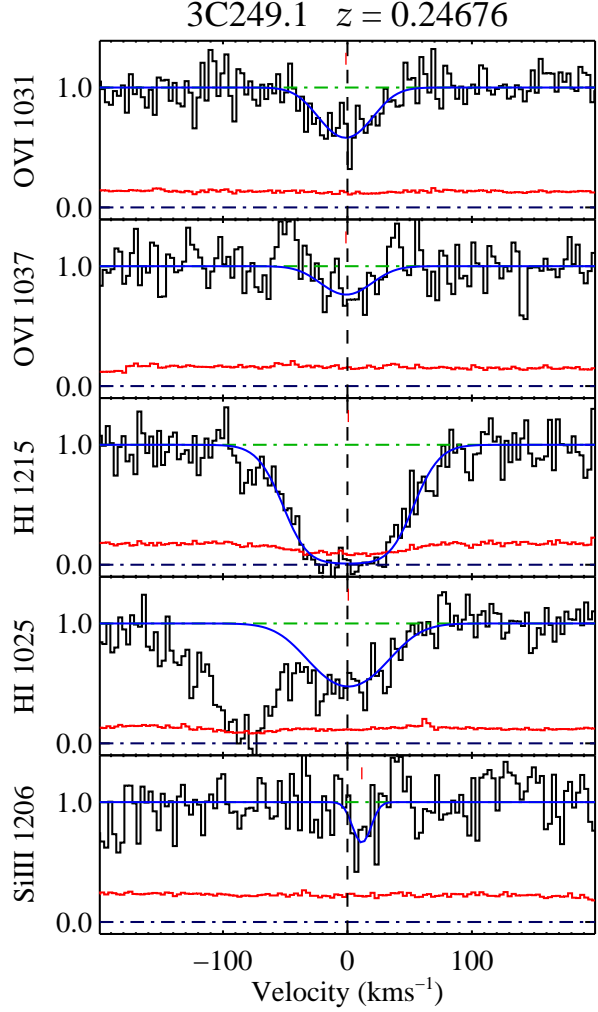}
  \caption{3C\,249.1 $z = 0.24676$---\OVI\ and strong \HI\ are detected in this system. \Lyb\ is blended
    with what we tentatively assign as \Lya\ line at $z = 0.0517$. There is a weak and uncertain feature
    which may be \SiIII\,1206.}
  \label{fig: 3C249.1_z0.24676}
\end{figure}

\noindent $\zabs = 0.24676$ (Figure~\ref{fig: 3C249.1_z0.24676}; Table~\ref{tab:
  3C249.1_transitions})---This system shows a clear detection of both \OVI\ lines, strong \Lyab, and
tentative \SiIII\,1206. The \OVI\,1037 line shows warm pixels on the very edges of the line wings,
which do not affect our line fits.  The blue wing of the \Lyb\ line partially blends with another
strong line, which we tentatively identify as \Lya\ at $z = 0.0517$. We cannot check this
assignment, as the corresponding \Lyb\ line for this putative assignment is below our wavelength
range. In the unblended region, the model fit to the data is good.  The \Lyc\ line is on the edge of
the Galactic \Lya\ trough, and was excluded from our fits. The \SiIII\ line is very weak and offset
from the \OVI\ and \HI\ absorption ($\Delta\,v = 11\pm3\kms$); its identification is thus
uncertain. With the well aligned \OVI\ and \HI\ components, we derive $\bnt=26.1\kms$ and $\logT =
4.7$; see Sec~\ref{sec: gas_temperature} for details.

\subsection{3C\,273} 
The STIS data for 3C\,273 ($\zqso = 0.1583$) are of excellent quality, but offer only a short
path-length over which to detect \OVI\ doublet systems ($0.1144 < \zabs < 0.1417$). The lowest
redshift absorber in our sample, at $\zabs = 0.12003$ is detected along this line of sight.

\begin{figure}
  \epsscale{1.0}
  \plotone{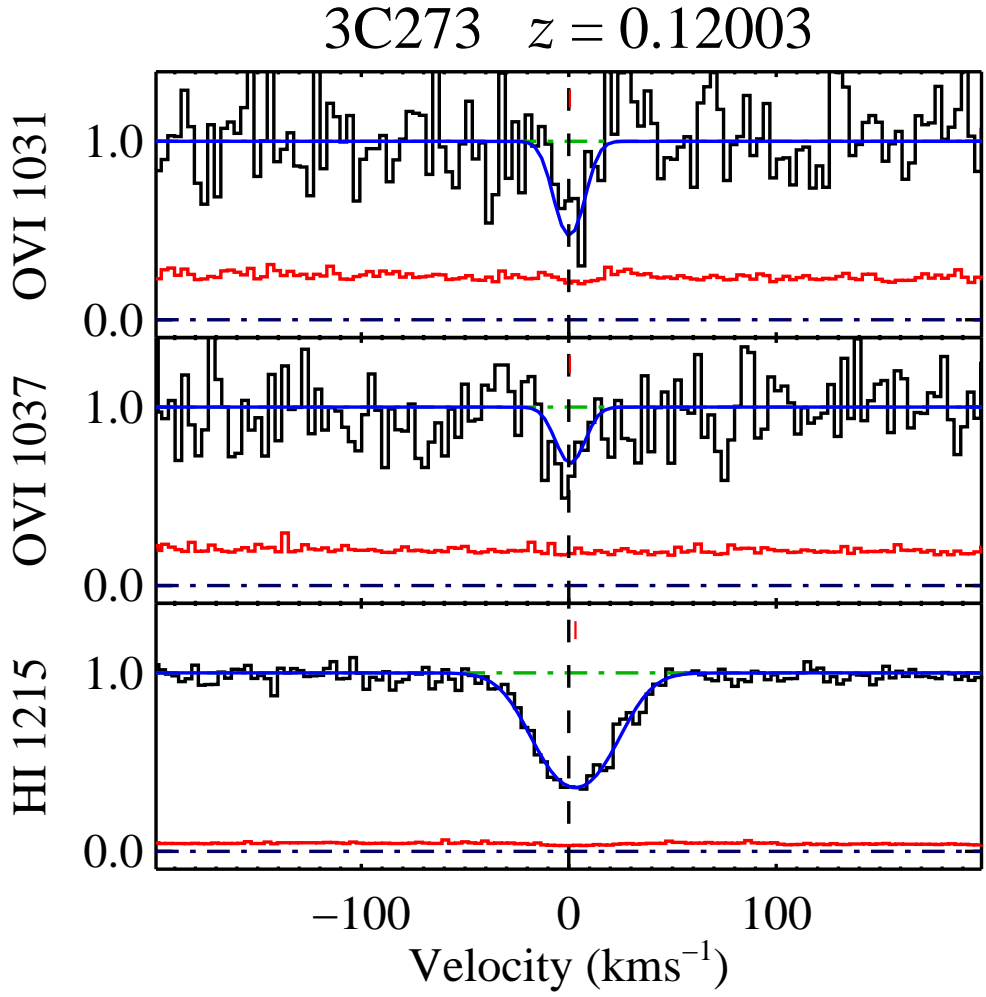}
  \caption{3C\,273 $z = 0.12003$---This weak system is detected in only \OVI\ and \Lya. The \OVI\ lines
    are mismatched in their strengths, but neither can be weak \Lya, since they both lie to the blue
    of the $z=0$ \Lya\,1215 line.}
  \label{fig: 3C273_z0.12003}
\end{figure}

\noindent $\zabs = 0.12003$ (Figure~\ref{fig: 3C273_z0.12003}; Table~\ref{tab:
  3C273_transitions})---This narrow, weak, single-component absorber is the lowest redshift absorber
in our sample, and is detectable at such a low wavelength ($\lambda_{obs} = 1156\A$) only due to the
high quality data for the 3C\,273 sightline. For both \OVI\ lines, while noisy, the data show
corresponding profiles, and the model fits are satisfactory for both lines. The line strengths from
direct integration of the data, are mismatched (the equivalent width ratio is $\sim 1.0 \pm 0.4$),
but neither line can be weak \Lya\ (the system is blueward of the Galactic \Lya\ line).  Of the \HI\
lines, only \Lya\ is present in our data at such low redshifts, and the line is unsaturated and well
fit by a single component. Due to the simple structure, we are able to derive $\bnt = 6.6\kms$ and
the gas temperature, $\logT = 4.5$; see Sec~\ref{sec: gas_temperature} for details.

\subsection{3C\,351.0}
3C\,351.0 lies at $\zqso = 0.3716$, giving a usable path-length for detecting \OVI\ systems $0.1309
< \zabs < 0.3483$. We detect only a single \OVI\ doublet along this line of sight.

\begin{figure}                 
  \epsscale{0.6}
  \plotone{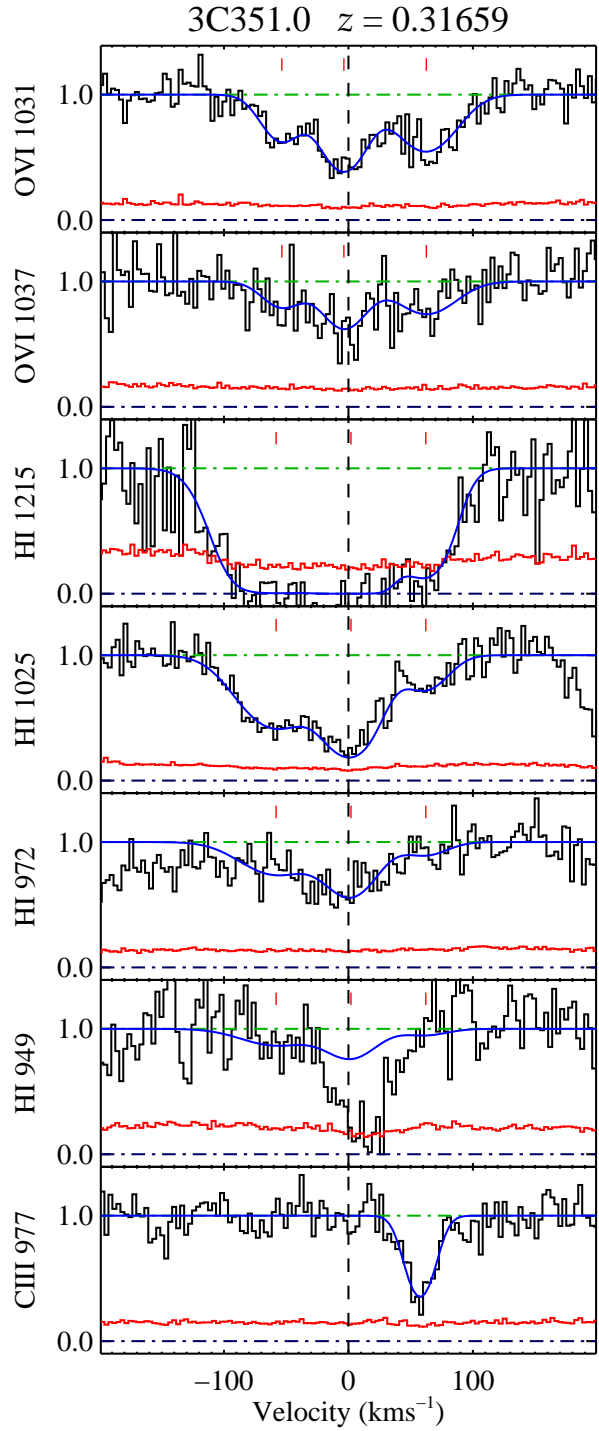} 
  \caption{3C\,351.0 $z = 0.31659$---Three obvious \OVI\ components are seen in this
    absorber. Corresponding \Lya\ absorption is heavily saturated, but \Lyb\ and \Lyc\ allow us to
    derive \HI\ column densities and positions. The \Lya\ components are well matched in position
    with the \OVI\ absorption. No associated metals are detected in our wavelength range.}
  \label{fig: 3C351.0_z0.31659}  
\end{figure}

\noindent $\zabs = 0.31659$ (Figure~\ref{fig: 3C351.0_z0.31659}; Table~\ref{tab:
  3C351.0_transitions})---This system shows a complex structure, with three well-defined \OVI\
components present, and corresponding \HI\ profiles.  The \Lya\ line is saturated, but the \Lyb\
line shows three components that are well aligned with the \OVI.  The weak \Lyc\ line is noisier and
less well fit by this \HI\ model; the \Lyd\ line is contaminated by Galactic \SII\,1250 absorption.
No \CIII\,977 is observed; the strong, putative \CIII\,977 line which aligns with the red-most \OVI\
component is \HI\,937 at $z = 0.37193$, associated with the QSO host. As all three \OVI\ components
are well-aligned with the three \HI\ components, we attempt to derive the gas temperature and
non-thermal broadening. For the components at $v = -3, -53\kms$ we derive $\bnt = 23.0, 19.5\kms$
and $\logT = 4.1, 4.8$ respectively. For the component at $v = +62\kms$, we cannot find a solution
since $\bOVI > \bHI$; at the $1\,\sigma$ level this component is consistent with a system whose
Doppler parameter is entirely dominated by non-thermal broadening. Sec~\ref{sec: gas_temperature}
contains further discussion of this case.

\subsection{H\,1821+643}
\label{subsec: H1821+643}
The H\,1821+643 ($\zqso = 0.297$) line has been studied extensively in terms of intervening
absorbers.  \citet{tripp-etal-98-Lya-galaxies} used GHRS and galaxy redshifts to study the \Lya\
absorbers. \citet{tripp-etal-00-H1821+643} followed up with STIS observations, focusing on \OVI\
absorption, complemented by \citet{oegerle-etal-00-H1821+643-FUSE} with FUSE observations. The STIS
data allow us to search for \OVI\ absorbers between $0.1144 < \zabs < 0.2741$. We uncover four
absorbers at $\zabs = 0.22496, 0.22638, 0.24532, 0.26656$.

\begin{figure*}                 
  \epsscale{1.0}
  \plottwo{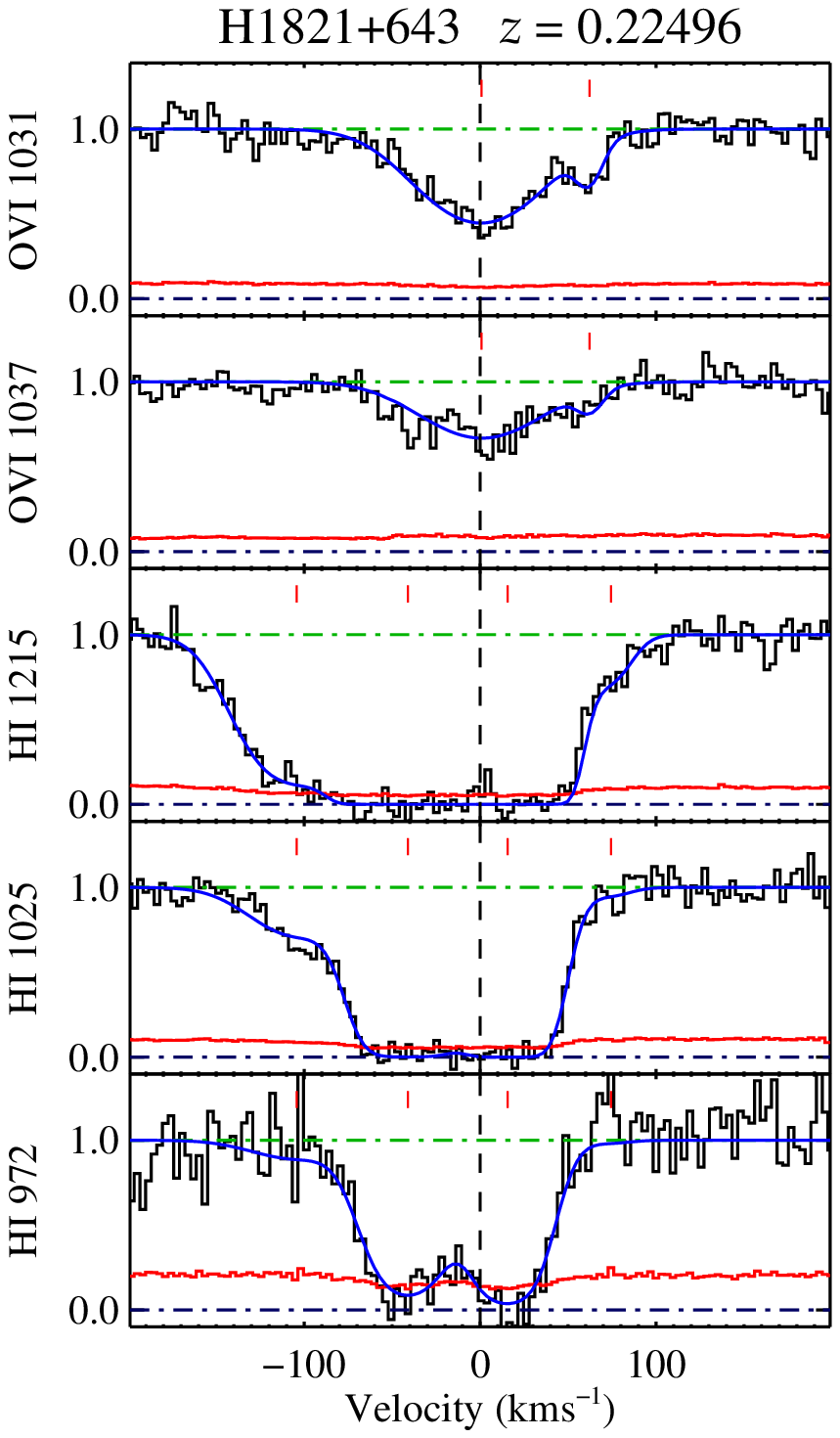}{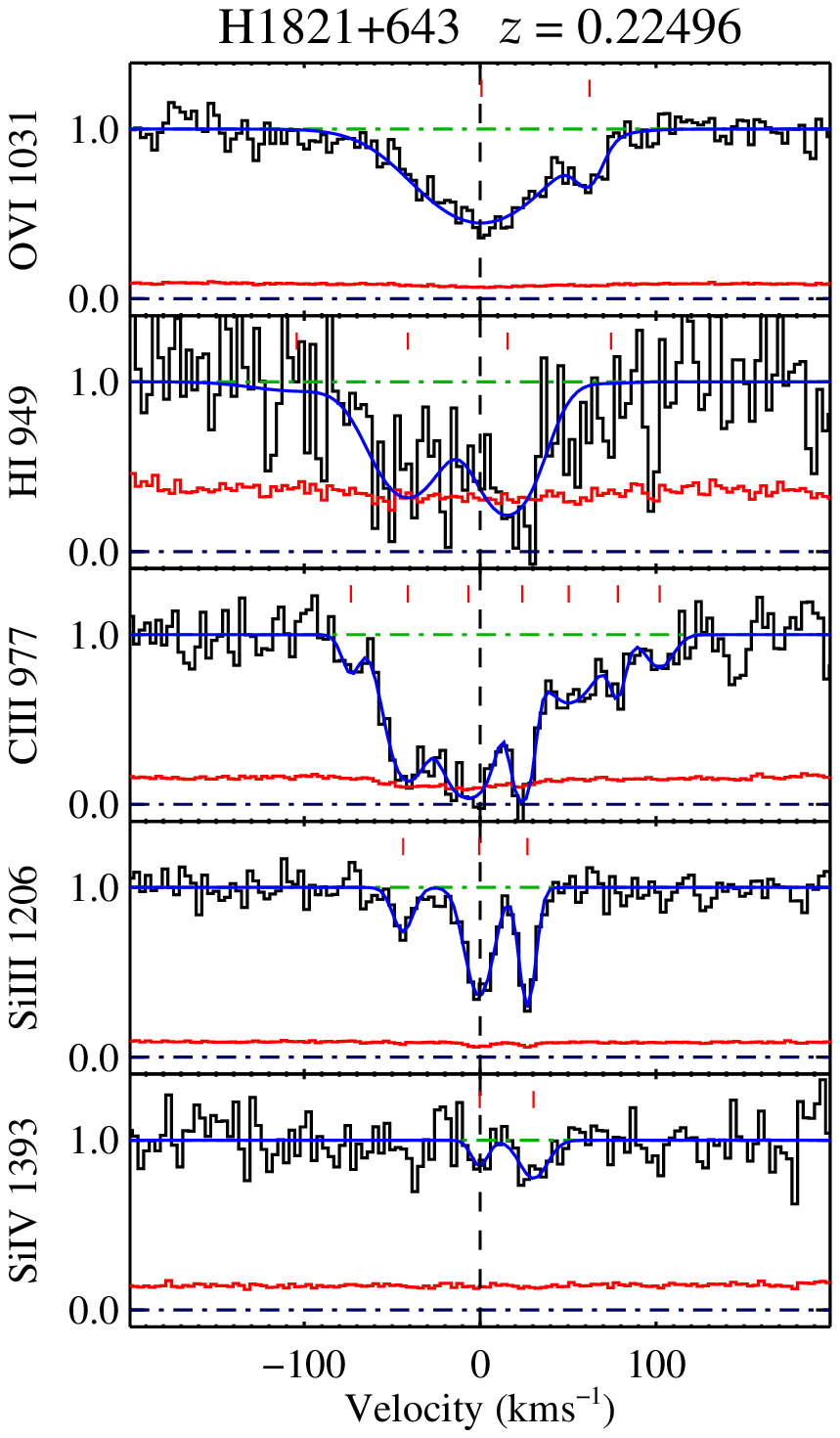}  
  \caption{H\,1821+643 $z = 0.22496$---A strong main \OVI\ absorption component and weaker, offset
    component are both obvious in this system. \Lya\ is heavily saturated, but higher order Lyman
    lines are available. Strong \CIII\ and \SiIII\ can also be seen. }
  \label{fig: H1821+643_z0.22496}  
\end{figure*}

\noindent $\zabs = 0.22496$ (Figure~\ref{fig: H1821+643_z0.22496}; Table~\ref{tab:
  H1821+643_transitions})---This absorber and the next (H\,1821+643; $\zabs = 0.22638$) are
separated by only $\sim350\kms$, but the two absorbers appear to be physically distinct systems (as
opposed to components of the same absorption system).  The \OVI\ absorption consists of a broad,
strong component, with a very weak, narrow component on the red edge. The associated \Lyab\ lines
are strongly saturated, while the \Lycd\ lines show some saturation. Three separate \HI\ components
are included in the fit, but \NHI\ is a lower limit; the two strong components are saturated in the
three lowest order Lyman lines, while the \Lyd\ line is very noisy. The \CIII\,977 profile is
similarly complex, with several possible weak components evident around the three main, strong
components, (at least two of which show signs of saturation).  There are three well-defined but
unsaturated \SiIII\,1206 components corresponding to the strong \CIII\,977 absorption lines. The
$v=0\kms$ component of \SiIII\ and the $v = -6\kms$ \CIII\ absorption component, while aligned with
the main \OVI\ absorption, cannot arise in the same gas phase, since e.g.\ $\bSiIII \ll \bOVI$, and
we require $0.8 < \bSiIII/\bOVI < 1.0$. These limits are discussed in more detail in Sec~\ref{sec:
  gas_temperature}.  At least two weak \SiIV\,1393 components are also present, which align with the
two strongest \CIII\ components, but are too weak to be detected in the weaker \SiIV\,1402 doublet
transition.

\begin{figure}                 
  \epsscale{0.8}
  \plotone{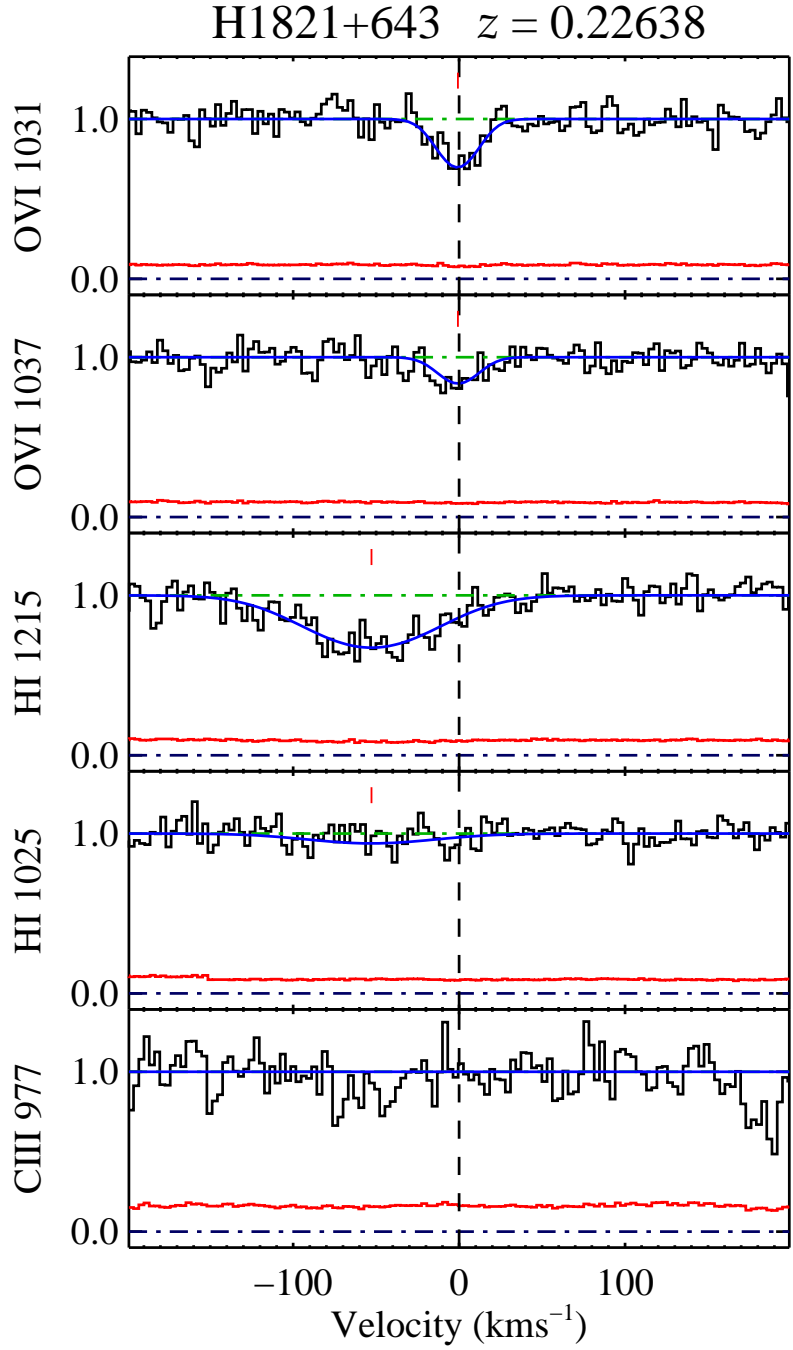} 
  \caption{H\,1821+643 $z = 0.22638$---This system comprises weak \OVI\ absorption, and very broad \HI
    absorption offset from the \OVI\ by $-53\kms$.  This absorber is only $\sim350\kms$ from the
    system at $z=0.22496$.}
  \label{fig: H1821+643_z0.22638}  
\end{figure}

\noindent $\zabs = 0.22638$ (Figure~\ref{fig: H1821+643_z0.22638}; Table~\ref{tab:
  H1821+643_transitions})---The weak absorber at $z=0.22638$ is separated from the strong system at
$z=0.22496$ by only $\sim350\kms$. The \OVI\ doublet is well fit by a weak, narrow, single component
absorption profile. \Lya\ absorption at $v = -53\kms$ shows a similar single component profile.
There is some evidence of weak \CIII\,977 at the same velocity as the \Lya, but we could not obtain
a satisfactory fit, and better quality data are needed to confirm this claim.
  
\begin{figure}                 
  \epsscale{0.6}
  \plotone{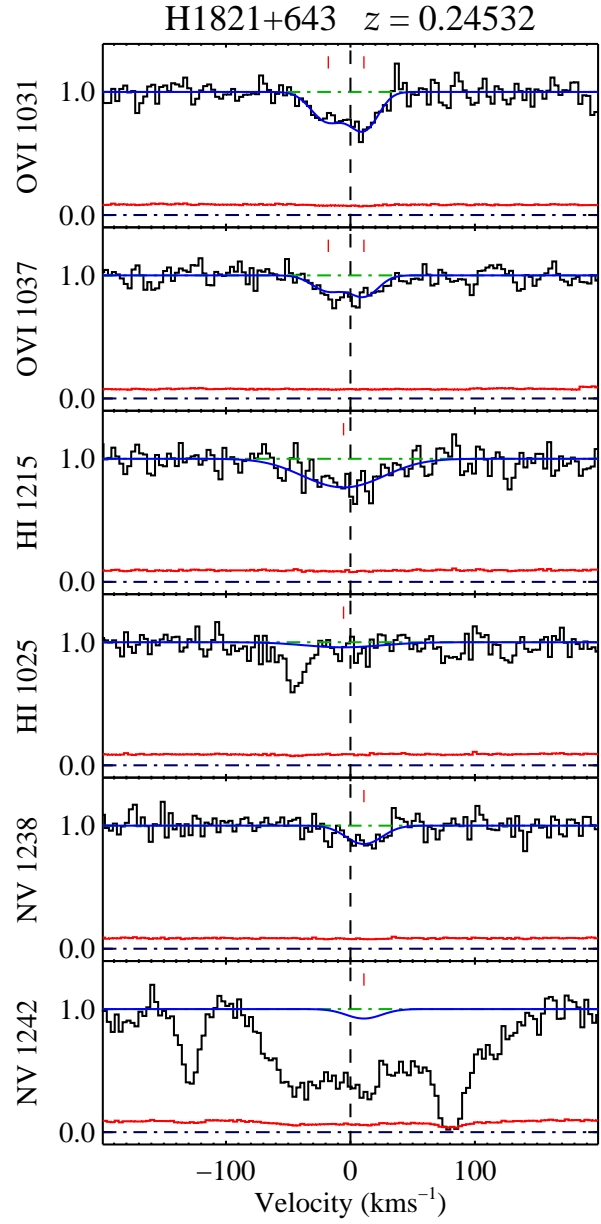} 
  \caption{H\,1821+643 $z = 0.24532$---Two weak \OVI\ components and a corresponding very weak \HI\
    absorber are seen in this system. \NV\,1238 is possibly present, but the \NV\,1242 line is lost
    in the Galactic \CIV\,1548 line.}
  \label{fig: H1821+643_z0.24532}  
\end{figure}

\noindent $\zabs = 0.24532$ (Figure~\ref{fig: H1821+643_z0.24532}; Table~\ref{tab:
  H1821+643_transitions})---There are two weak \OVI\ components in this system, with a single weak,
broad \HI\ component. There is possible \NV\ present, corresponding the red-most \OVI\ component, but
the weak \NV\,1242 line is totally obscured by Galactic \CIV\,1548, and we regard this
identification as uncertain.

\begin{figure}                 
  \epsscale{1.0}
  \plotone{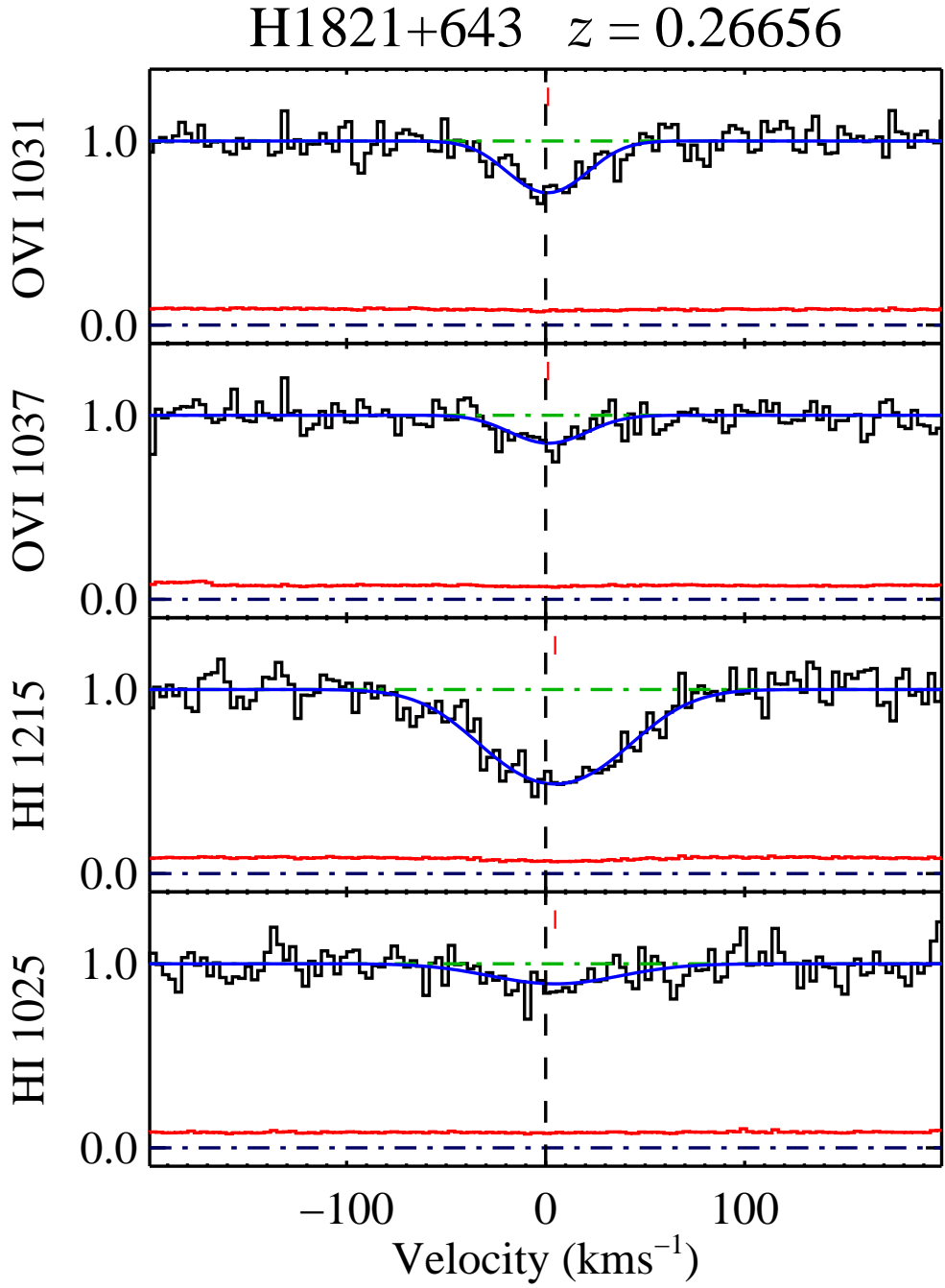} 
  \caption{H\,1821+643 $z = 0.26656$---This weak system consists of a single \OVI\ and \HI\ component,
    both well-aligned.}
  \label{fig: H1821+643_z0.26656}  
\end{figure}

\noindent $\zabs = 0.26656$ (Figure~\ref{fig: H1821+643_z0.26656}; Table~\ref{tab:
  H1821+643_transitions})---This absorber is relatively simple, with well-aligned, single-component
\OVI\ and \HI\ lines. The \OVI\ and \HI\ line centroids differ by only 4\kms, and the good alignment
permits us to derive gas temperature and non-thermal broadening of $\logT = 4.9$ and $\bnt =
24.4\kms$ respectively. Sec~\ref{sec: gas_temperature} has the details of this derivation.

\subsection{HE\,0226$-$4110}
The STIS spectrum of HE\,0226$-$4110, at $\zabs = 0.495$, allows us to search for intervening \OVI\
absorbers in the interval $0.1154 < \zabs < 0.4707$. We identified \OVI\ doublets at $\zabs =
0.20702, 0.32639 0.34034, 0.35525$. This line of sight has also been studied by
\citet{savage-etal-05-HE0226-4110-NeVIII}, who focused on the $\zabs = 0.20702$ system, and
\citet{lehner-etal-06-HE0226-4110} who studied the full path length using STIS and FUSE data.

\begin{figure*}                 
  \epsscale{1.0}
  \plottwo{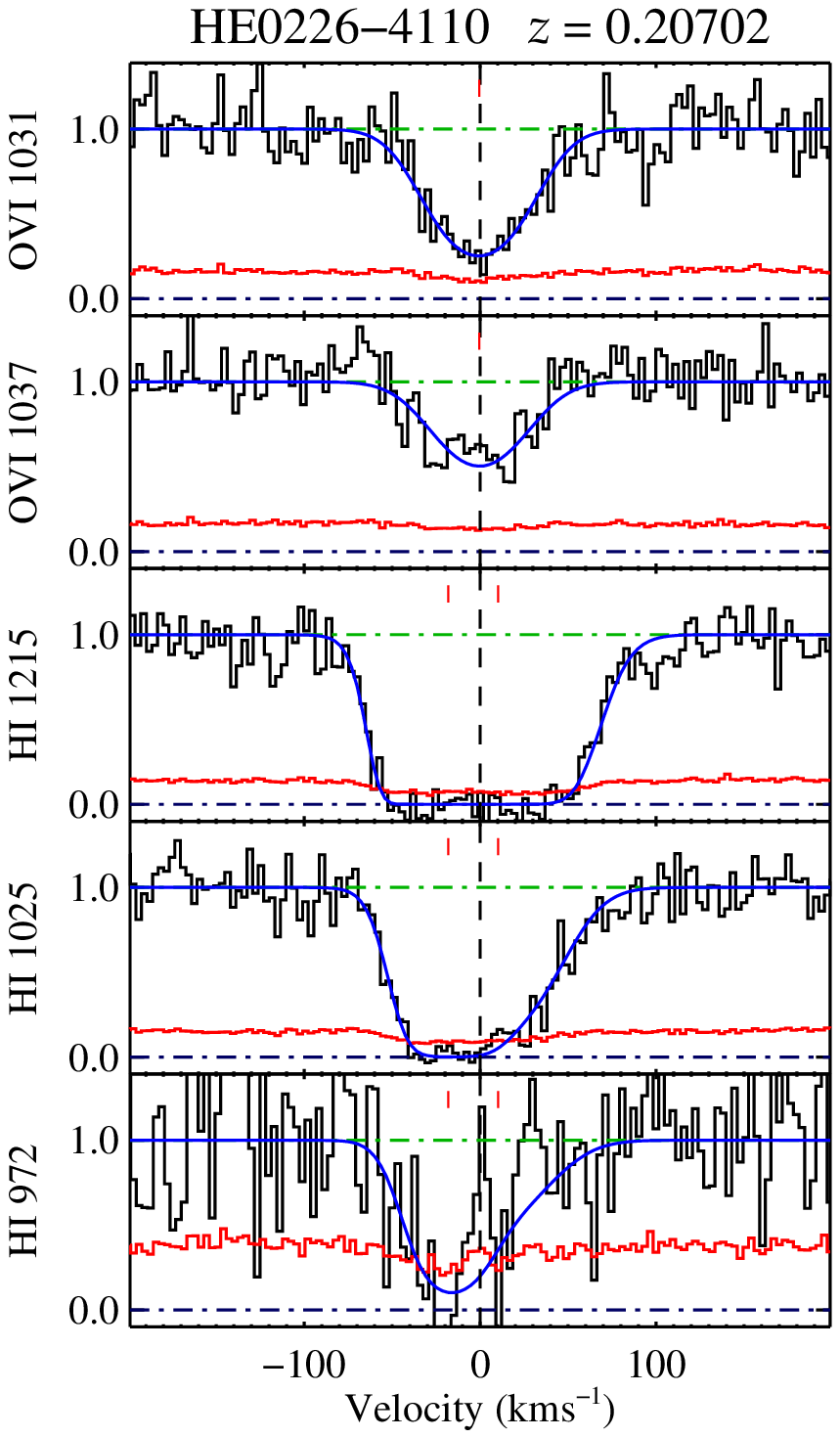}{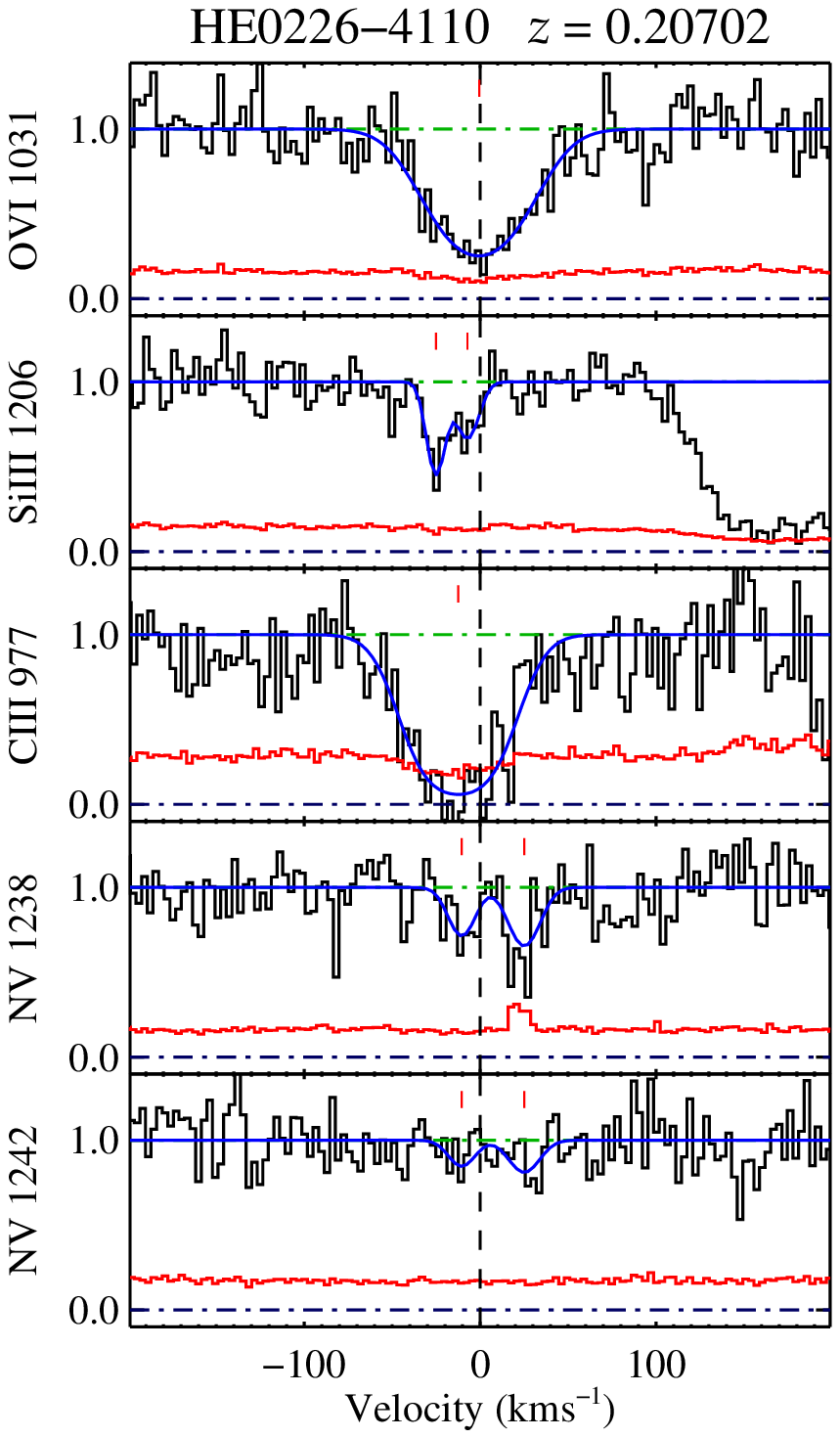} 
  \caption{HE\,0226$-$4110 $z=0.20702$---This strong \OVI\ absorber is also seen in saturated \Lyab\
  absorption. The \Lyc\ line is contaminated by hot pixels, making the \NHI\ determination
  uncertain. We see associated \SiIII, \CIII\ and possibly \NV\ absorption.}
  \label{fig: HE0226-4110_z0.20702}  
\end{figure*}

\label{subsec: HE0226-4110}
\noindent $\zabs = 0.20702$ (Figure~\ref{fig: HE0226-4110_z0.20702}; Table~\ref{tab:
  HE0226-4110_transitions})---The strong absorber at $\zabs = 0.20702$ has associated \HI,
\CIII\,977, \SiIII\,1206 and \NV. The \OVI\,1037 transition differs from the single-component
structure seen in \OVI\,1031, which is probably a result of bad pixels in the \OVI\,1037 region. We
measure only a lower limit on \NHI; the \Lya\ and \Lyb\ transitions are saturated, while the \Lyc\
line is contaminated by hot pixels.  \SiIII\,1206 has two components blueward of the fiducial \OVI\
position, which roughly correspond to the saturated \CIII\,977 line.  The \NV\ lines are weak, and
uncertain. \citep{savage-etal-05-HE0226-4110-NeVIII} detected \NeVIII\ aligned with the \OVI\
absorption in FUSE data. They show this system is likely collisionally ionized, which is consistent
with the single broad-component in \OVI\ that we observe.

\begin{figure}                 
  \plotone{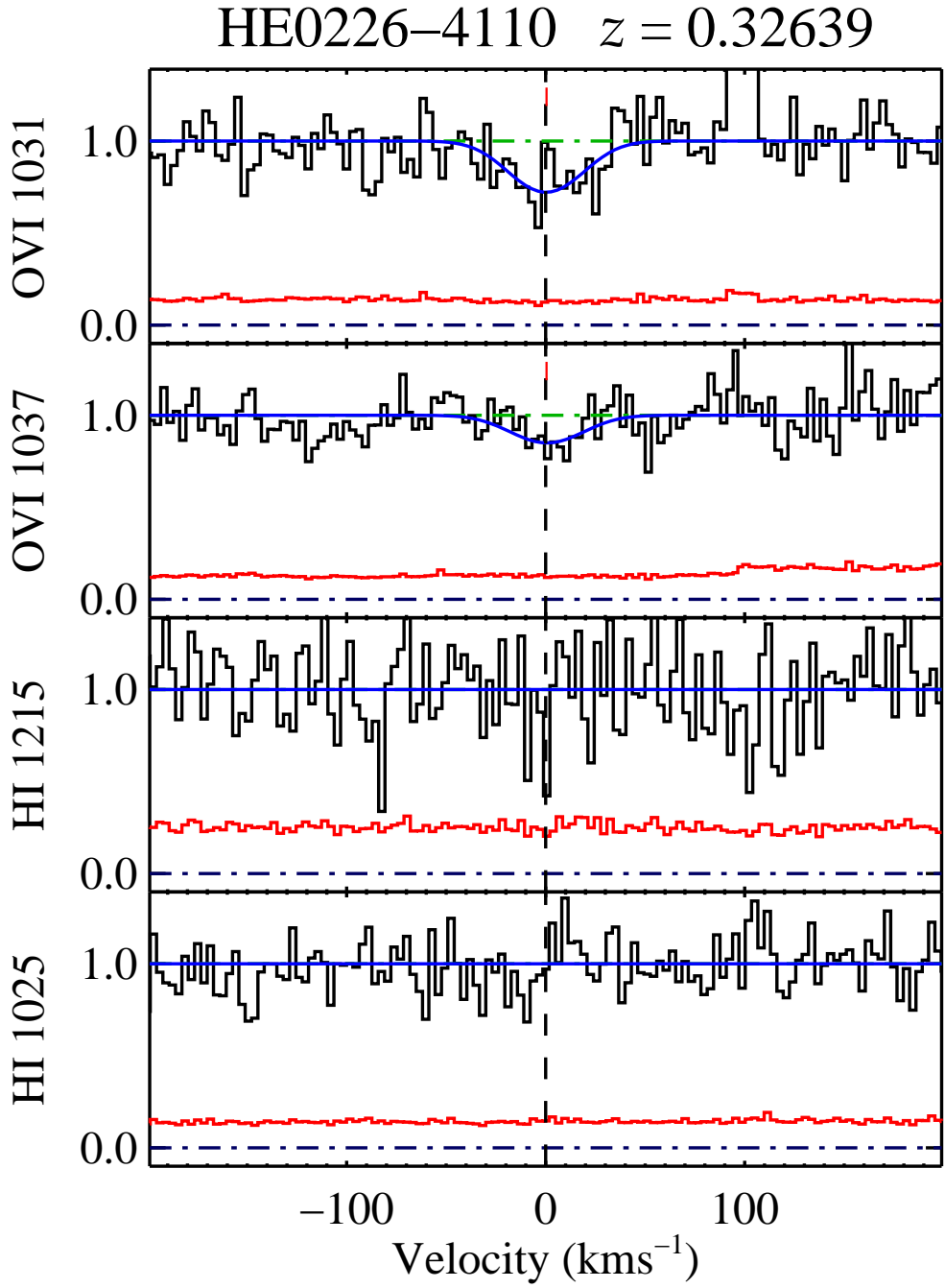} 
  \caption{HE\,0226$-$4110 $z=0.32639$---This system is weak and uncertain. It is the only system we
    detect without corresponding \HI\ absorption. No other transitions are present and correspond
    to the \OVI\ absorption.}
  \label{fig: HE0226-4110_z0.32639}  
\end{figure}

\noindent $\zabs = 0.32639$ (Figure~\ref{fig: HE0226-4110_z0.32639}; Table~\ref{tab:
  HE0226-4110_transitions})---We tentatively identify an \OVI\ doublet at $\zabs = 0.32639$ as the
only system in our sample that shows no sign of \HI\ absorption. Both \OVI\ lines are well fit by a
single doublet model, while the position of any putative \Lya\ line is in the red-most portion of
the STIS wavelength coverage, making the data noisier than the \OVI\ region, as can be seen in
Figure~\ref{fig: HE0226-4110_z0.32639}. Fixing the redshift and expected Doppler parameter from the
\OVI\ profile, we set an upper limit on the \HI\ column density $\NHI < 12.5$. No other transitions
are present in this system. We note that \citet{lehner-etal-06-HE0226-4110} do not report
identifications for either lines (see e.g.\ their Figure~3).  We also note that the \OVI\,1037 line
is detected at low significance (only $2\,\sigma$), and emphasize that caution is required
interpreting this system as \HI\ free. We suggest that further observations would be very valuable
to confirm this system, and whether it is \HI\ free.

\begin{figure}                 
  \epsscale{0.7}
  \plotone{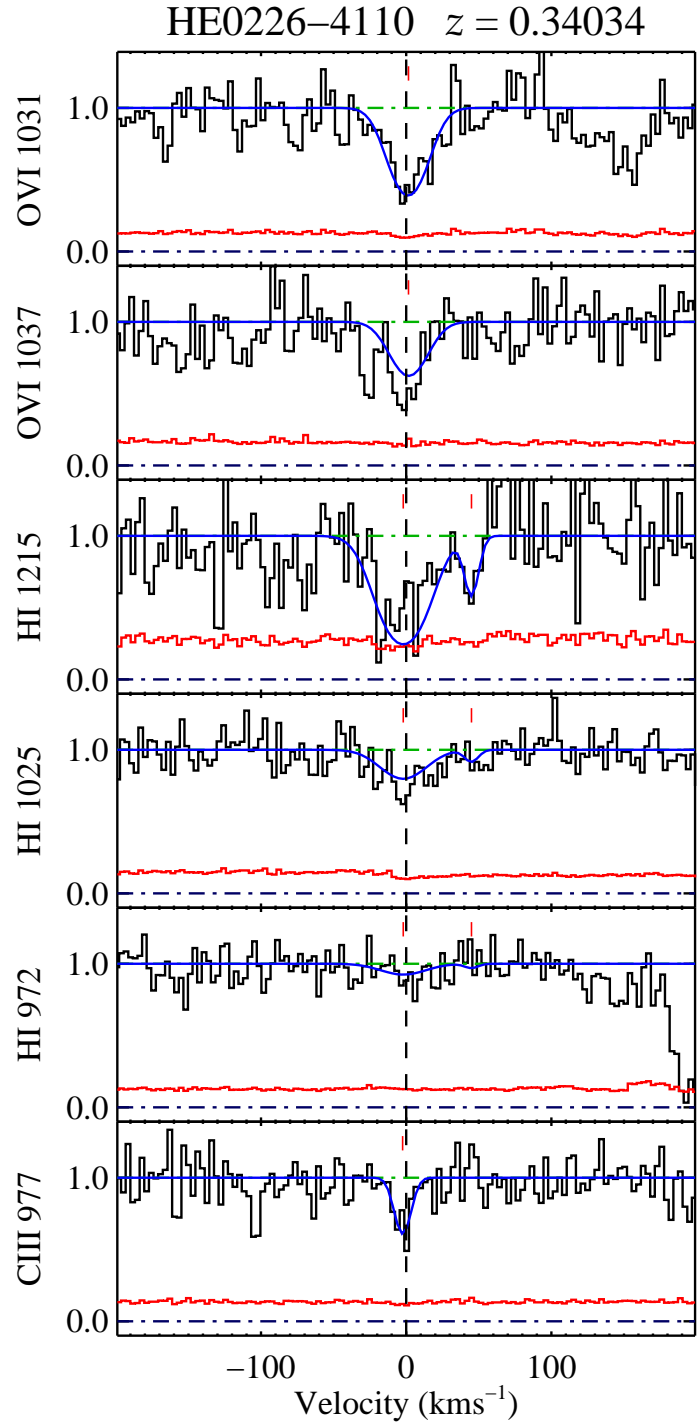} 
  \caption{HE\,0226$-$4110 $z=0.34034$---The \OVI\,1037 line in this system is contaminated by an
    unidentified metal line. \HI\ is present, but noisy, and the fit is poor. Weak \CIII\ is also
    detected.}
  \label{fig: HE0226-4110_z0.34034}  
\end{figure}

\noindent $\zabs = 0.34034$ (Figure~\ref{fig: HE0226-4110_z0.34034}; Table~\ref{tab:
  HE0226-4110_transitions})---The \OVI\,1037 line in this system is contaminated by an unidentified
metal line at $v = -26\kms$, which is excluded from the fit. The \Lya\ line aligns well with the
\OVI\ absorption, but is in a poor-quality region of the spectrum, and the agreement between the
\Lya\ and \Lyb\ profiles is poor.  A weak, narrow \CIII\,977 line is also detected, which aligns
well with the \OVI\ and main \HI\ component. The Doppler parameter of the \CIII\ line, \bCIII, is
significantly smaller than that of the \OVI\ component. This may be explained in several ways: the
\CIII\ line is not real---possible but unlikely, given its precise alignment with the \OVI\ and \HI\
positions; the \OVI\ Doppler width is over-estimated, either by unresolved components, or due to
noise in the profile; finally, the \OVI\ and \CIII\ absorption may arise in physically distinct gas
clouds.  For the well-aligned \OVI\ and \HI\ components, we calculate $\logT = 4.0$ and $\bnt =
16.5\kms$. Sec~\ref{sec: gas_temperature} has the details of this calculation.

\begin{figure}                 
  \epsscale{1.0}
  \plotone{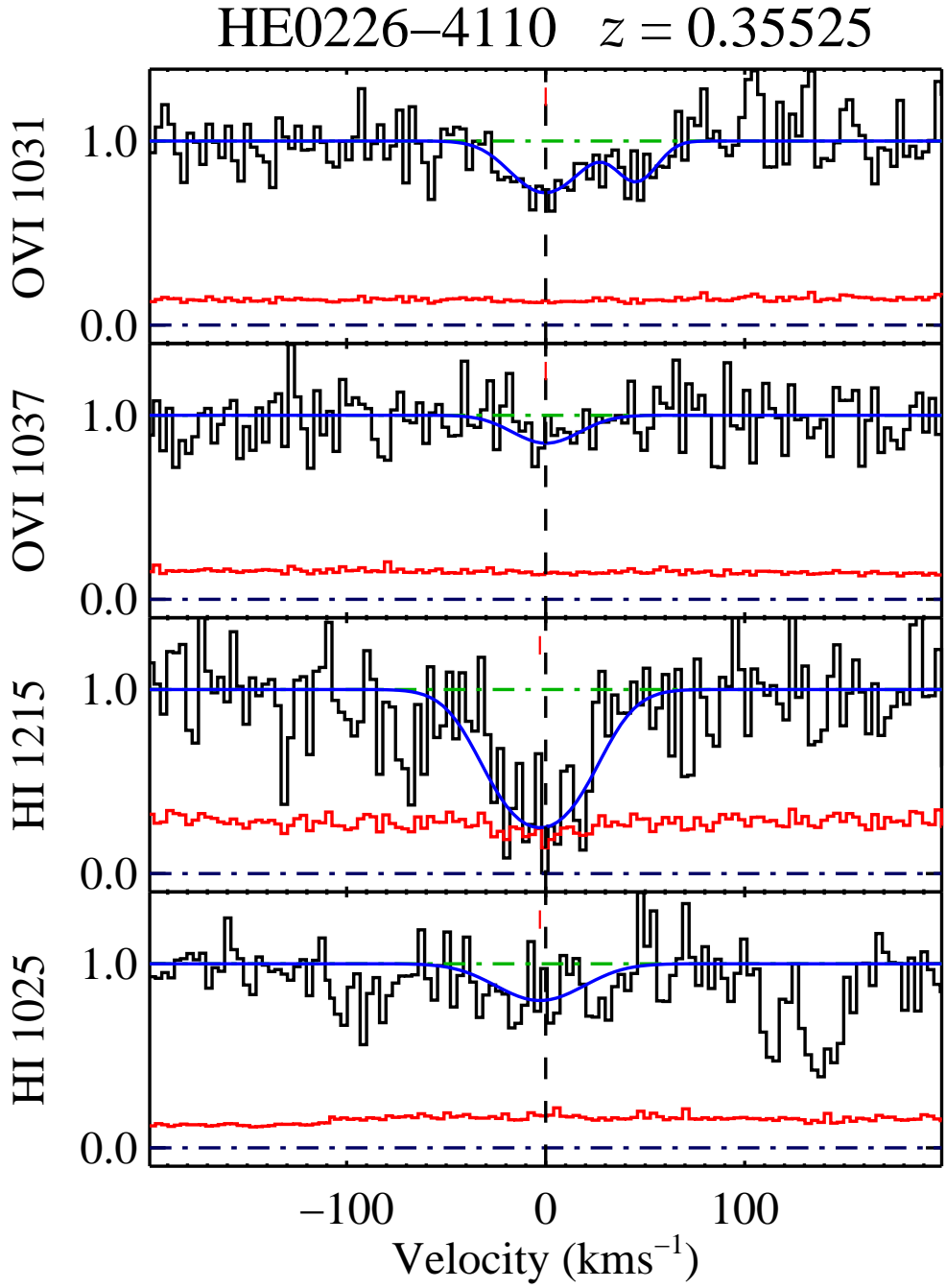} 
  \caption{HE\,0226$-$4110 $z=0.35525$---\OVI\,1031 in this system shows some evidence of contamination
    in the line wing, but this does not unduly affect our fits. Only associated \HI\ lines are
    detected; no metal lines are seen in our STIS data.}
  \label{fig: HE0226-4110_z0.35525}
\end{figure}

\noindent $\zabs = 0.35525$ (Figure~\ref{fig: HE0226-4110_z0.35525}; Table~\ref{tab:
  HE0226-4110_transitions})---There is only weak \OVI\ and \HI\ at $\zabs = 0.35525$, with both
transitions precisely aligned. The \OVI\,1031 line shows a weak contaminating component or feature,
although it is too weak to be detected in the \OVI\,1037 transition if it is real. We fit this
component as a contaminating \HI\ line, and the resulting model is a good fit to both \OVI\
transitions. As with the $z=0.34034$ system, the \Lya\ profile is quite noisy, but the \HI\,1215,
1025 regions are fit to within the noise. The resulting \HI\ position is closely matched to the
\OVI\ redshift. Using this close match, we derive gas temperature $\logT = 4.3$ and non-thermal
broadening $\bnt = 22.2\kms$; see Sec~\ref{sec: gas_temperature}.

\subsection{HS\,0624+6907}
The QSO HS\,0624+6907 at $\zqso = 0.370$ offers a path-length for \OVI\ absorption $0.1222 < \zabs <
0.3464$. We detect two \OVI\ absorbers, at $\zabs = 0.31796, 0.33984$.

\begin{figure}                 
  \epsscale{1.0}
  \plotone{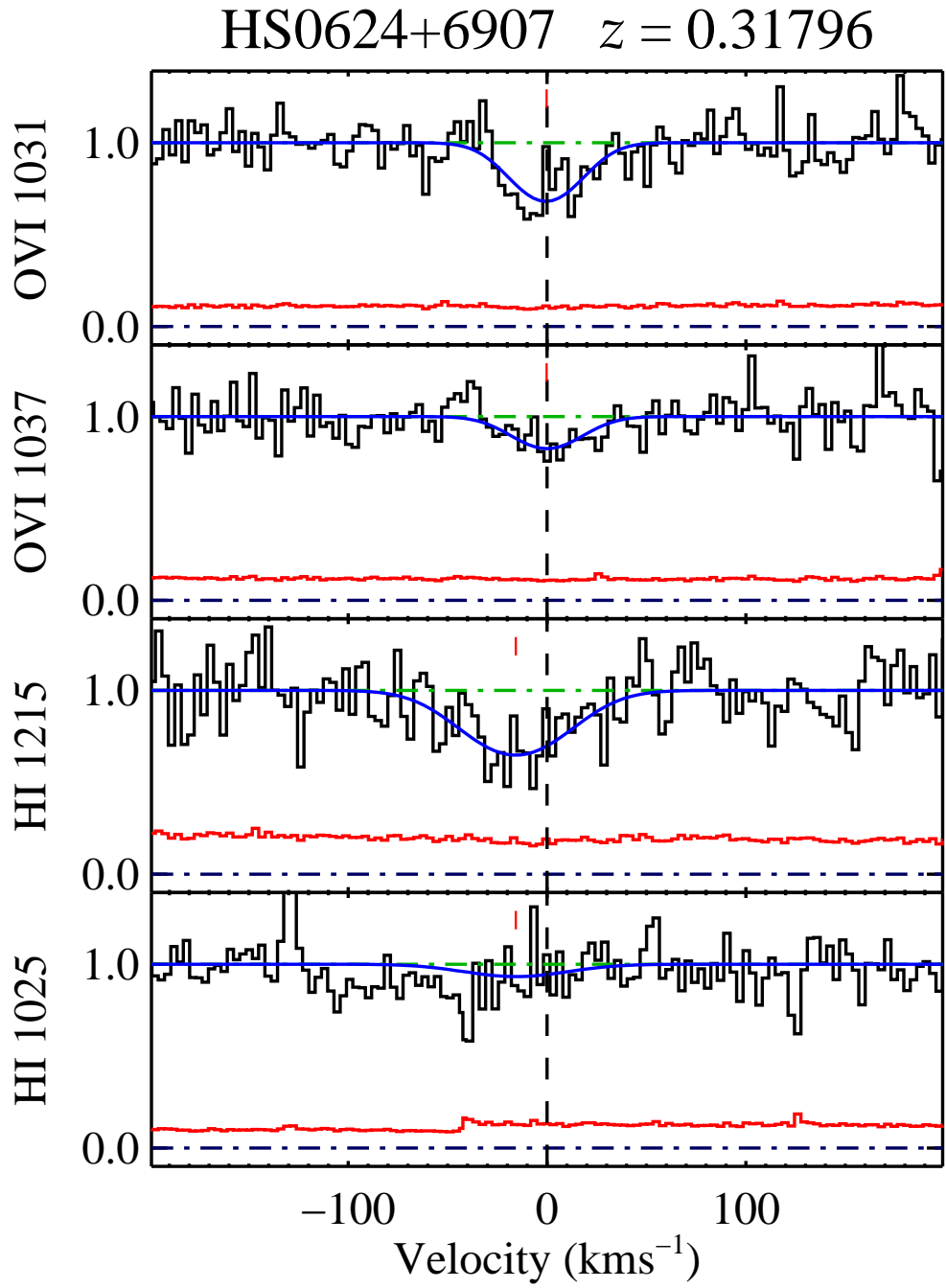} 
  \caption{HS\,0624+6907 $z=0.31796$---This weak absorber is detected only in \OVI\ and \Lya, with a
    15\kms\ offset between the two species.}
  \label{fig: HS0624+6907_z0.31796}  
\end{figure}

\noindent $\zabs = 0.31796$ (Figure~\ref{fig: HS0624+6907_z0.31796}; Table~\ref{tab:
  HS0624+6907_transitions})---This absorber is seen in only three lines---\OVI\,1031, 1037 and \Lya;
\Lyb\ is not present. The \Lya\ and \OVI\ positions are offset ($15 \pm 5\kms$).  The \OVI\,1031
line shows evidence of some warm or noisy pixels.

\begin{figure}                 
  \epsscale{0.5}
  \plotone{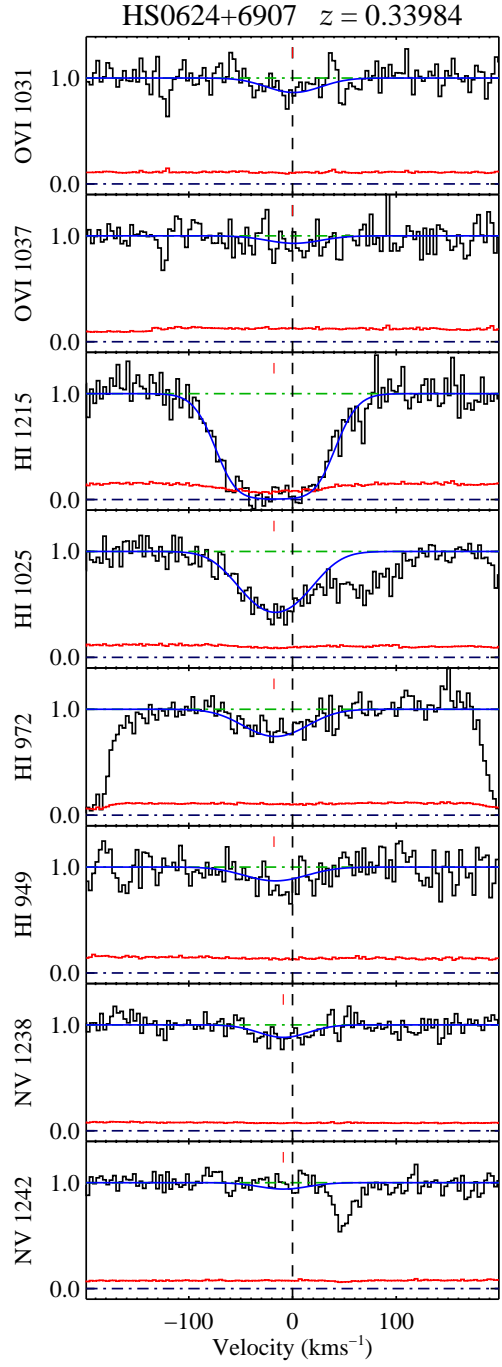} 
  \caption{HS\,0624+6907 $z=0.33984$---While the \Lya\ line is heavily saturated in this absorber,
    the \OVI\ lines are very weak. The \Lyb\ profile is partly blended, but the \Lyc\ line is free
    from contamination. No significant absorption from other species are detected.}
  \label{fig: HS0624+6907_z0.33984}  
\end{figure}

\noindent $\zabs = 0.33984$ (Figure~\ref{fig: HS0624+6907_z0.33984}; Table~\ref{tab:
  HS0624+6907_transitions})---The significance of the \OVI\ in this system is very weak. It meets
our formal definition of a significant \OVI\,1031 line, and an \OVI\,1037 profile that is consistent
with the \OVI\,1031 transition. We thus accept this system, noting that the \OVI\ doublet is
uncertain. Other transitions are clearly present, with strong \HI, and possibly weak \NV. The \Lya\
line is saturated and the fit is sub-optimal in the red wing, but good fits are possible to the
\Lyb, \Lyc\ and \Lyd\ lines. The \Lyb\ profile is partly blended with another line, whose
identification is uncertain---if this contaminant is \Lya, the expected \Lyb\ line is at a low
enough wavelength that it would not be detected in the noisy STIS data. The blended portion of the
spectrum is excluded from our fits. There is a very weak, broad line at the expected position of
\NV\,1238, but is too weak to be confirmed in the \NV\,1242 transition.

\subsection{PG\,0953+415}
Along the line of sight to PG\,0953+415 ($\zqso = 0.239$), we can detect \OVI\ absorption in the
range $0.1144 < \zabs < 0.2163$. A single system is detected, at $\zabs = 0.14232$, which has been
previously discussed by \citet{tripp-savage-00-PG0953+415_z0.1423}.

\begin{figure}                 
  \epsscale{0.8}
  \plotone{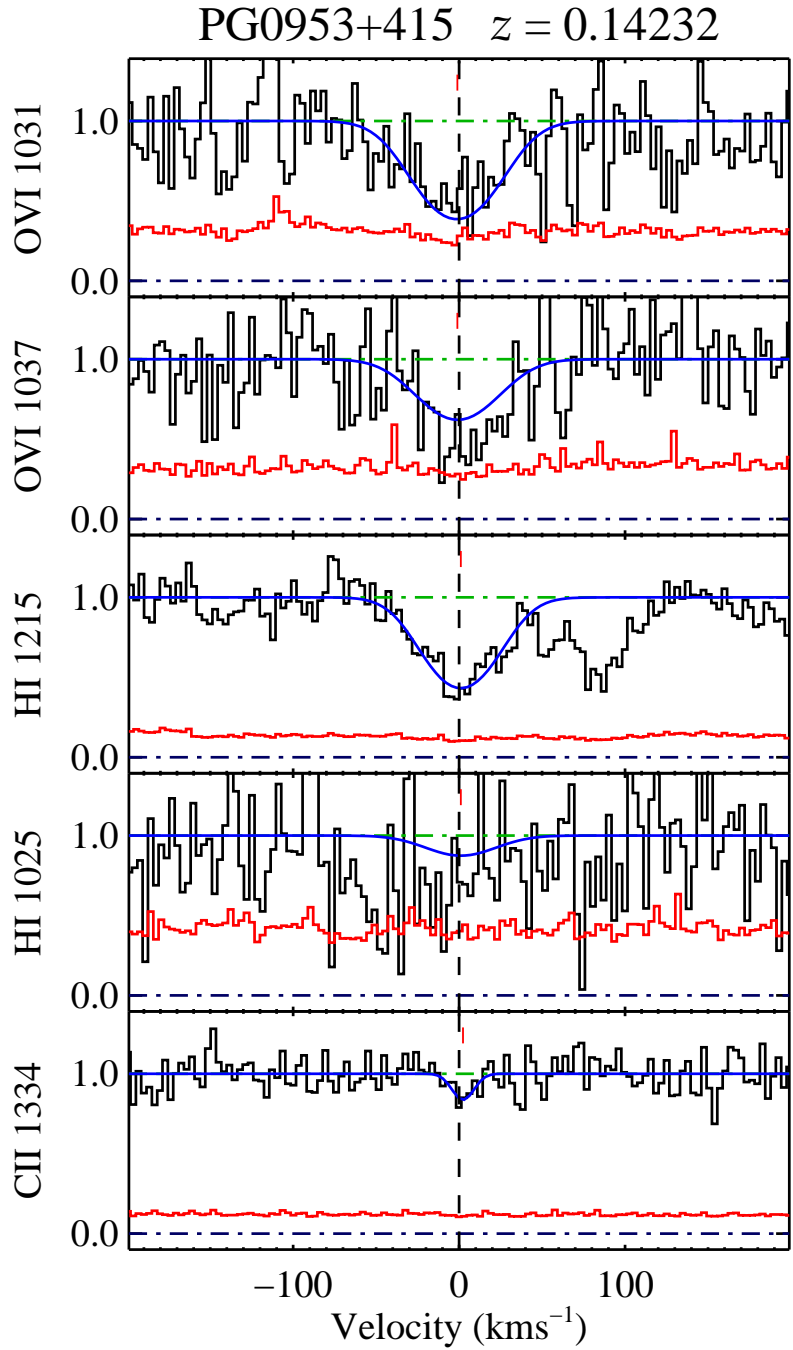} 
  \caption{PG\,0953+415 $z=0.14232$---Strong \OVI\ is present in this system, although the \OVI\,1037
    line is stronger than expected. \HI\ is well fit in the \Lya\ line, and is not affected by the
    blending seen to the red of the line. \Lyb\ in this absorber is contaminated by \HI\,949 line
    from the QSO host at $z=0.2335$. \CII\,1334 may be present, but is extremely weak and we flag it
    as uncertain.}
  \label{fig: PG0953+415_z0.14232}  
\end{figure}

\noindent $\zabs = 0.14232$ (Figure~\ref{fig: PG0953+415_z0.14232}; Table~\ref{tab:
  PG0953+415_transitions})---This strong \OVI\ absorber falls in a noisy part of the STIS spectrum
of PG\,0953+415. The \OVI\,1037 line is stronger than expected, based on the \OVI\,1031 line
strength, but this may be due to noise in the spectrum. \Lya\ is present, and well aligned with the
\OVI, but we are unable to obtain the gas temperature, \logT, and \bnt, since $\bOVI > \bHI$; see
Sec~\ref{sec: gas_temperature} for further discussion. Due to noise in the \Lyb\ portion of the
spectrum, and blending of the \Lyb\ profile with the QSO host \HI\,949 line at $z=0.2335$, we fit
only the zero-velocity \HI\ component. There is a hint of \CII\,1334, but it is very weak and
uncertain. This system has been discussed by \citet{tripp-savage-00-PG0953+415_z0.1423}.

\subsection{PG\,1116+215}
The PG\,1116+215 ($\zqso = 0.1765$) sightline has a single \OVI\ absorber detected at $\zabs =
0.13847$ from the available path-length between $0.1144 < \zabs < 0.1536$.

\begin{figure*}                 
  \epsscale{1.0}
  \plottwo{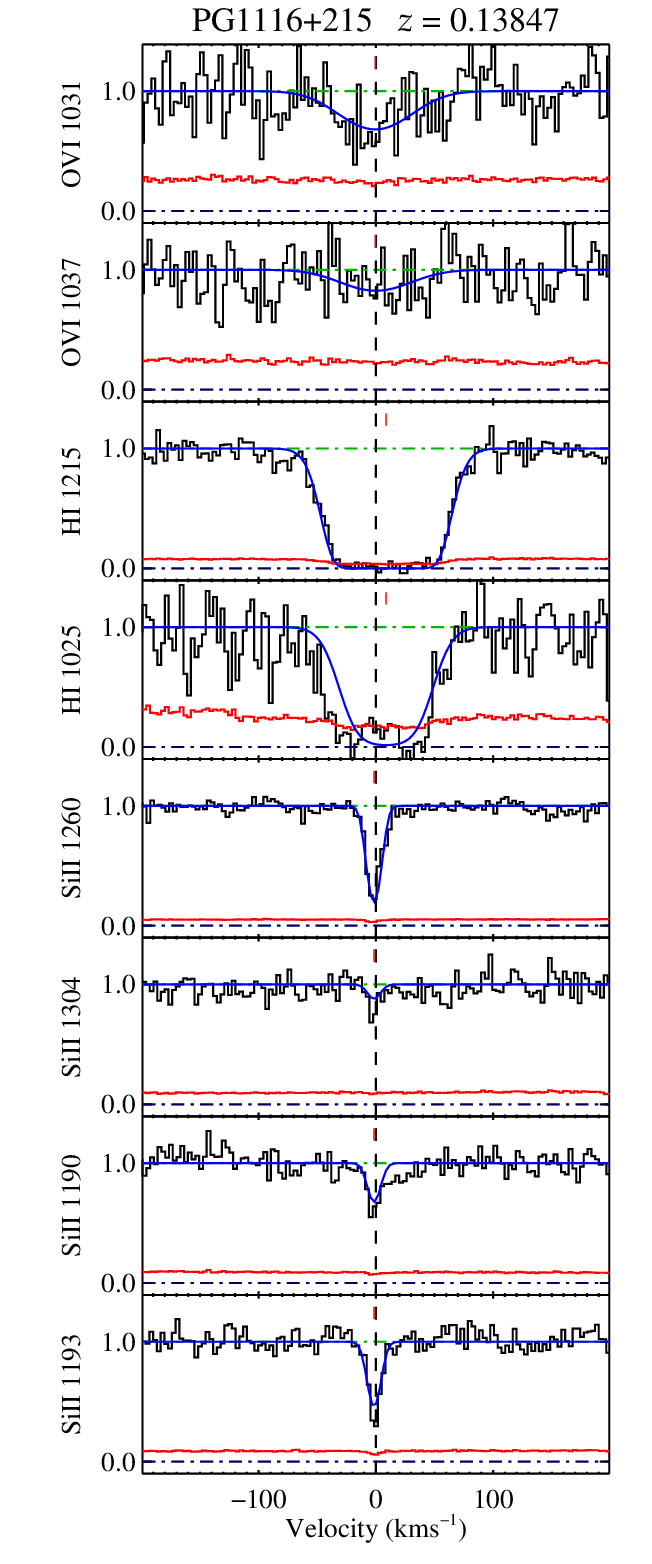}{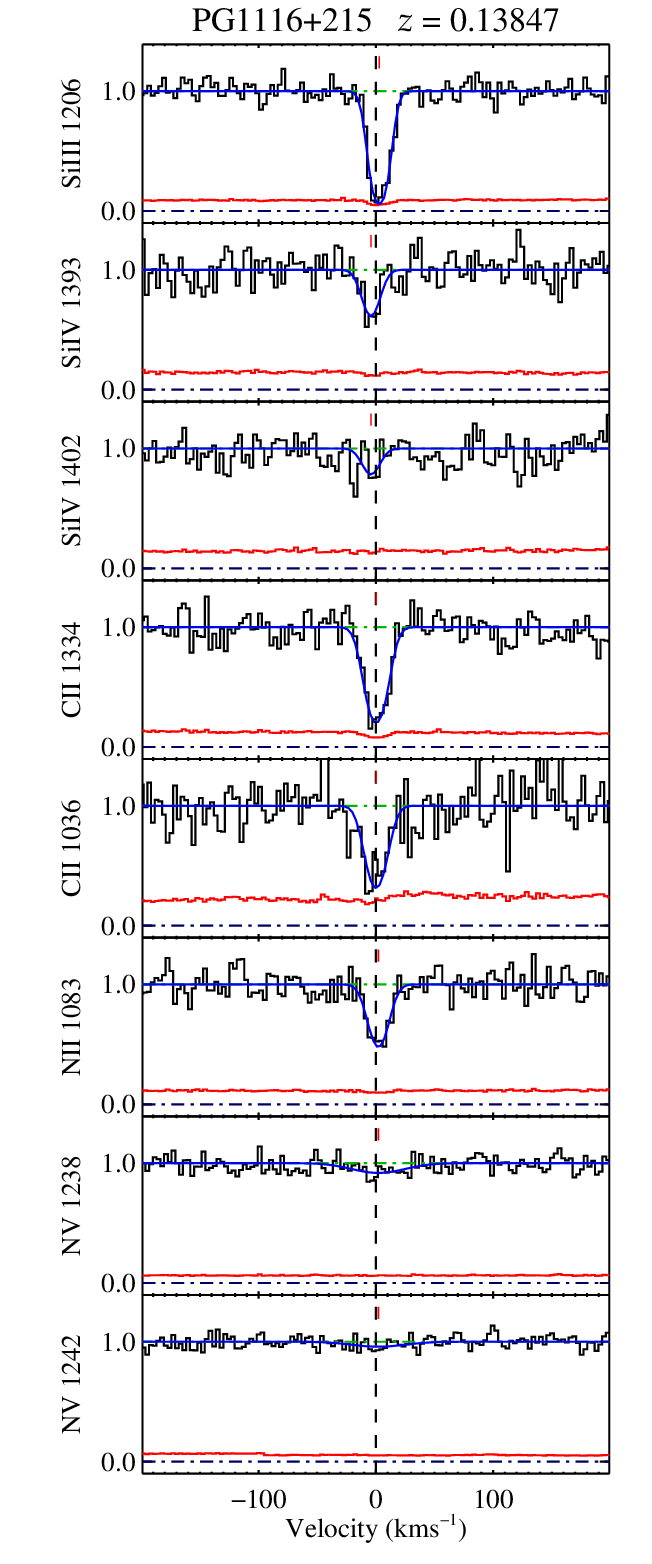} 
  \caption{PG\,1116+215 $z=0.13847$---While the \OVI\ transitions in this absorber are quite noisy, we
    have good data for \Lyab, both of which are saturated. We cannot obtain satisfactory fits to
    \Lya\ and \Lyb\ simultaneously. \SiII, \SiIII\, \SiIV, \NII\ and \CII\ are all clearly detected,
    often in multiple transitions (e.g.\ \SiII\ is detected in the 1190, 1193, 1260 and 1304
    transitions).}
  \label{fig: PG1116+215_z0.13847}  
\end{figure*}

\noindent $\zabs = 0.13847$ (Figure~\ref{fig: PG1116+215_z0.13847}; Table~\ref{tab:
  PG1116+215_transitions})---Previously analysed by \citet{sembach-etal-04-PG1116+215} this absorber
has noisy \OVI\ due to the low STIS efficiency at the observed wavelength ($\lambda_{\scOVI\,1031} =
1174.8\A$). Strong, saturated \HI\ absorption is observed.  Well-aligned metal lines of \SiII,
\SiIII, \SiIV, \CII, \NII\ and possibly \NV, are all observed. The \Lyab\ lines in our STIS data are
heavily saturated, and we cannot simultaneously provide a good fit to both \HI\ lines.  Using a
curve-of-growth analysis and the weak Lyman-limit, \citet{sembach-etal-04-PG1116+215} measured
$\logNHI \sim 16.2$ for this absorber. They also concluded that photoionization models at a single
ionization parameter, or collisionally ionization models at a single temperature, cannot explain all
the metal lines observed.

\subsection{PG\,1216+069}
The STIS data for PG\,1216+069 ($\zqso = 0.3313$) allow us to search for \OVI\ doublets in the range
$0.1309 < \zabs < 0.3078$. We detect only a single intervening \OVI\ absorber, at $\zabs = 0.28232$.

\begin{figure*}                 
  \epsscale{1.0}
  \plottwo{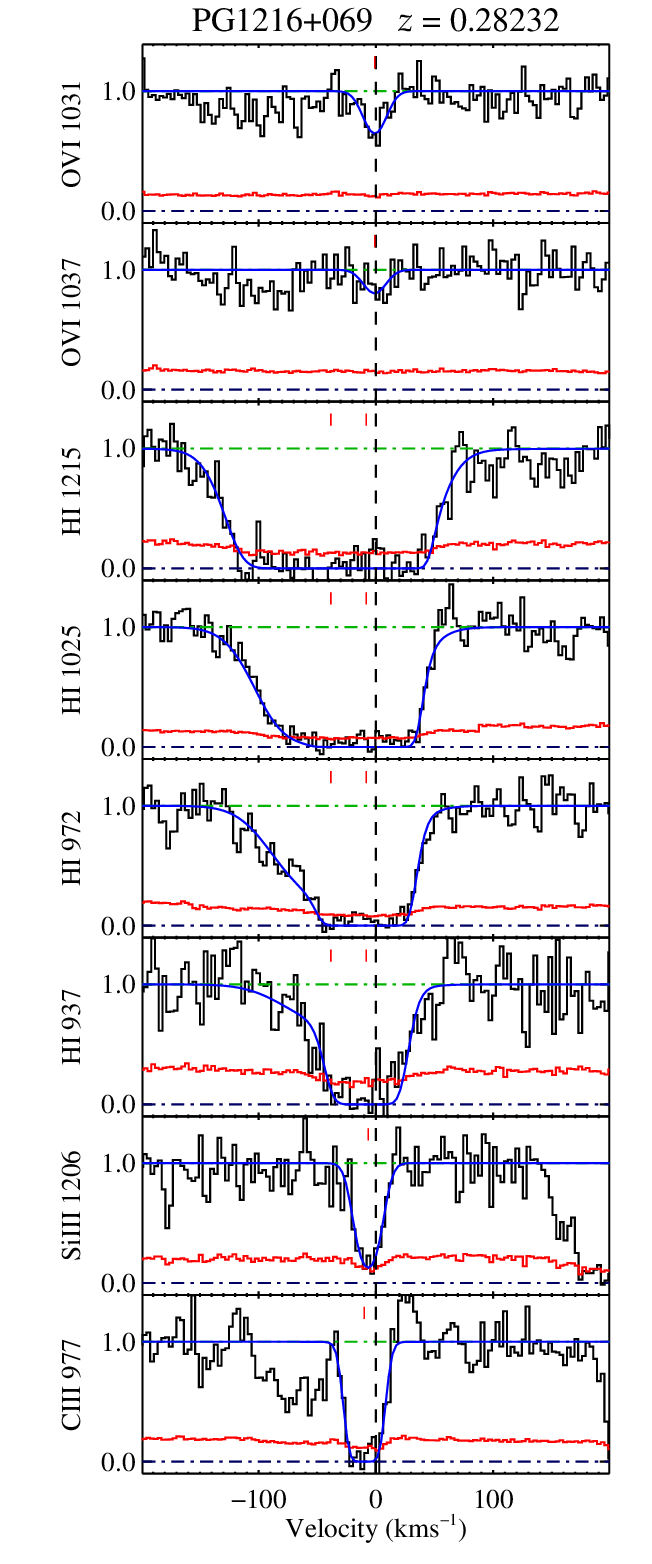}{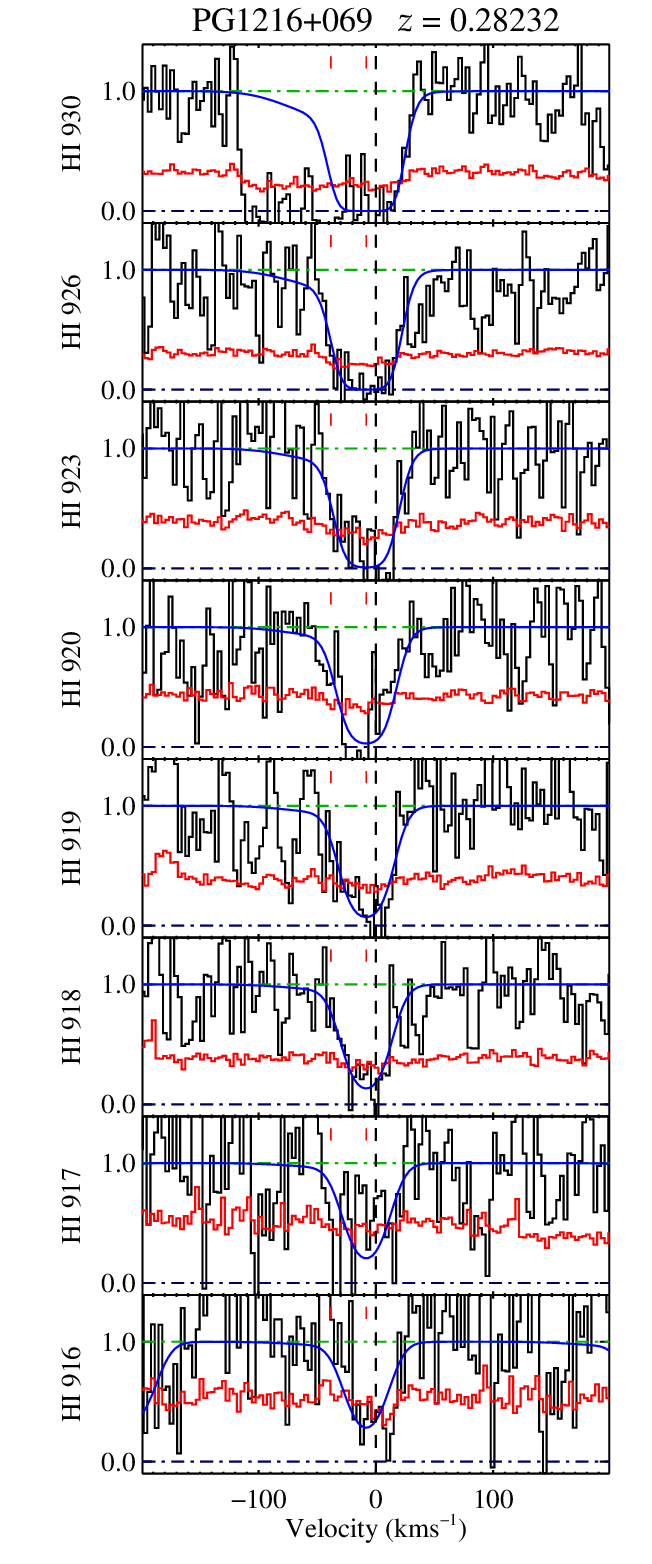}
  \caption{PG\,1216+069 $z=0.28232$---This system is another instance of very strong, saturated \HI,
    strong low ion metal species, but only weak \OVI. All the Lyman series lines in our spectrum are
    saturated, or are too noisy to be useful. \CIII\,977 is also saturated. It is not clear whether
    the absorption to the blue of the central \CIII\ component is \CIII\ associated with blue \HI\
    components, or a contaminating line.}
  \label{fig: PG1216+069_z0.28232}  
\end{figure*}

\noindent $\zabs = 0.28232$ (Figure~\ref{fig: PG1216+069_z0.28232}; Table~\ref{tab:
  PG1216+069_transitions})---The only \OVI\ system we detect towards PG\,1216+069, this absorber has
very weak \OVI\ absorption, along with very strong, saturated, multi-component \HI, allowing us to
set only a lower limit on \logNHI. We also detect saturated \CIII\,977, and strong \SiIII\,1206. In
the \CIII\ region, we fit only the zero velocity component. Other components may be {\it bona fide}
\CIII\ associated with the blue \HI\ component, but this is not clear.  Precise knowledge of the
\HI\ component positions would be valuable for determining this.
    
\subsection{PG\,1259+593}
The good quality STIS data for the line-of-sight PG\,1259+593 ($\zqso = 0.4778$) allow us to search
for weak absorbers over a large path-length $0.1144 < \zqso < 0.4533$. We detect two \OVI\ systems
at $\zabs = 0.21950, 0.25981$. This sightline has also been studied by
\citet{richter-etal-04-PG1259+593}.

\begin{figure}                 
  \epsscale{0.5}
  \plotone{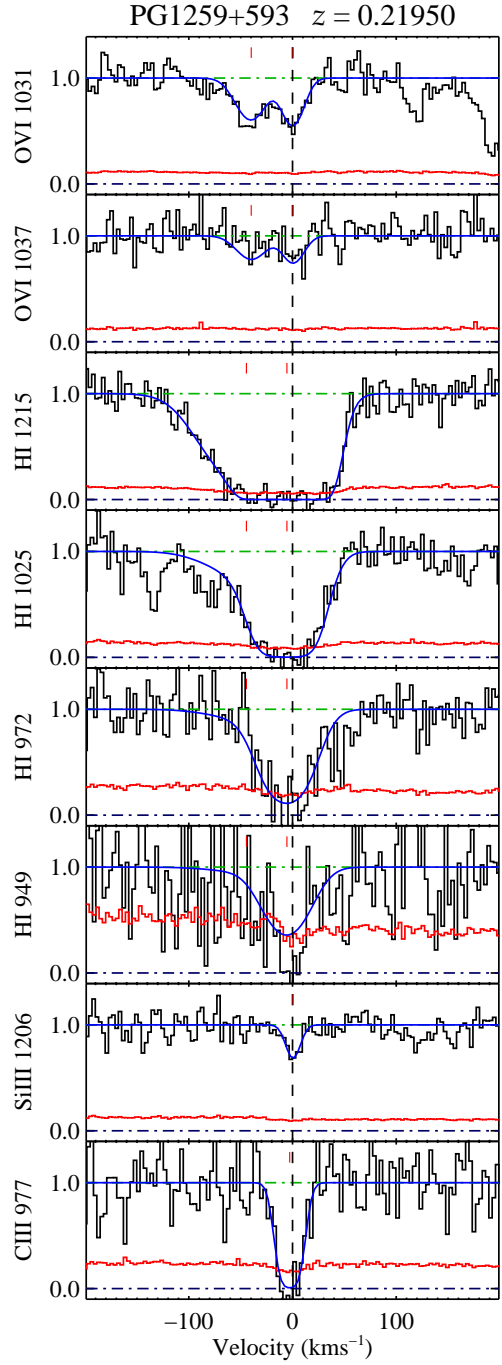} 
  \caption{PG\,1259+593 $z=0.21950$---Multi-component \OVI\ and \HI\ are both present in this system,
    with the \HI\ lines saturated. The blue wing of the \Lyb\ line blends with Galactic
    \SII\,1250. \CIII\ and \SiIII\ are both detected; \CIII\ is saturated, while the \SiIII\ line is
    weak. Both species align with the $v=0\kms$ component.}
  \label{fig: PG1259+593_z0.21950}  
\end{figure}

\noindent $\zabs = 0.21950$ (Figure~\ref{fig: PG1259+593_z0.21950}; Table~\ref{tab:
  PG1259+593_transitions})---This absorber contains two principle \OVI\ components, both of which
are well resolved. The blue component suffers from several bad pixels, which were excluded from the
fitting regions.  The \HI\ components are aligned with the \OVI, although the stronger component is
saturated even in the noisy, higher order \HI\,949 transition. The blue edge of the \Lyb\ profile
blends with Galactic \SII\,1250, but does not affect our fitting. We detect weak \SiIII\,1206 and
saturated \CIII\,977, both aligned with the zero velocity \OVI\
component. \citet{richter-etal-04-PG1259+593}, who studied all the systems on this sightline,
measured $\logNHI = 15.2$.
 
\begin{figure}                 
  \epsscale{0.8}
  \plotone{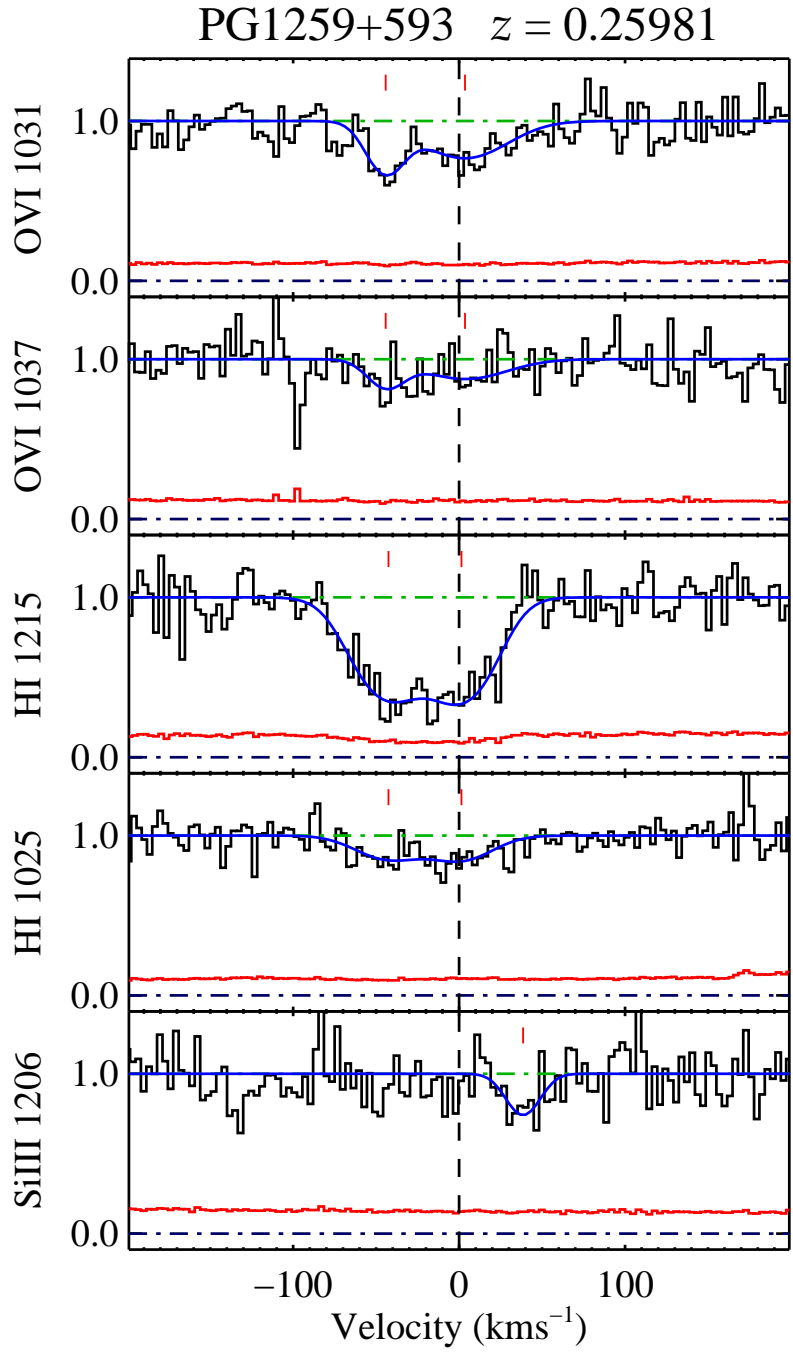}
  \caption{PG\,1259+593 $z=0.25981$---This double-component absorber shows only two \OVI\ and \HI\
    components.}
  \label{fig: PG1259+593_z0.25981}  
\end{figure}

\noindent $\zabs = 0.25981$ (Figure~\ref{fig: PG1259+593_z0.25981}; Table~\ref{tab:
  PG1259+593_transitions})---The multi-component absorber at $z=0.25981$ has two main \OVI\
components: a narrow component at $-44\kms$ separation from the broad zero-velocity component. The
best fit \HI\ profile has two components, well aligned with the \OVI\ components. We discount the
reality of a putative weak \SiIII\,1206 transition at $\sim +40\kms$. We attempted to derive \logT\
and \bnt\ for both components (Sec~\ref{sec: gas_temperature}), but were only successful only for
the \OVI/\HI\ pair at $-44\kms$. The zero-velocity \OVI/\HI\ component pair has $\bOVI > \bHI$, and
is discussed further in Sec~\ref{sec: gas_temperature}. For the component at $v = -44\kms$, we
measure $\bnt = 13.9\kms$ and $\logT = 4.5$.

\subsection{PHL\,1811}
The PHL\,1811 ($\zqso = 0.1917$) sightline contains a single \OVI\ absorber at $\zabs = 0.15786$ in
the available range $0.1144 < \zabs < 0.1690$. \citet{jenkins-etal-05-PHL1811_z0.0809} have studied
this sightline, focusing on the LLS at $\zabs = 0.0809$.

\begin{figure}                 
  \epsscale{1.0}

  \plotone{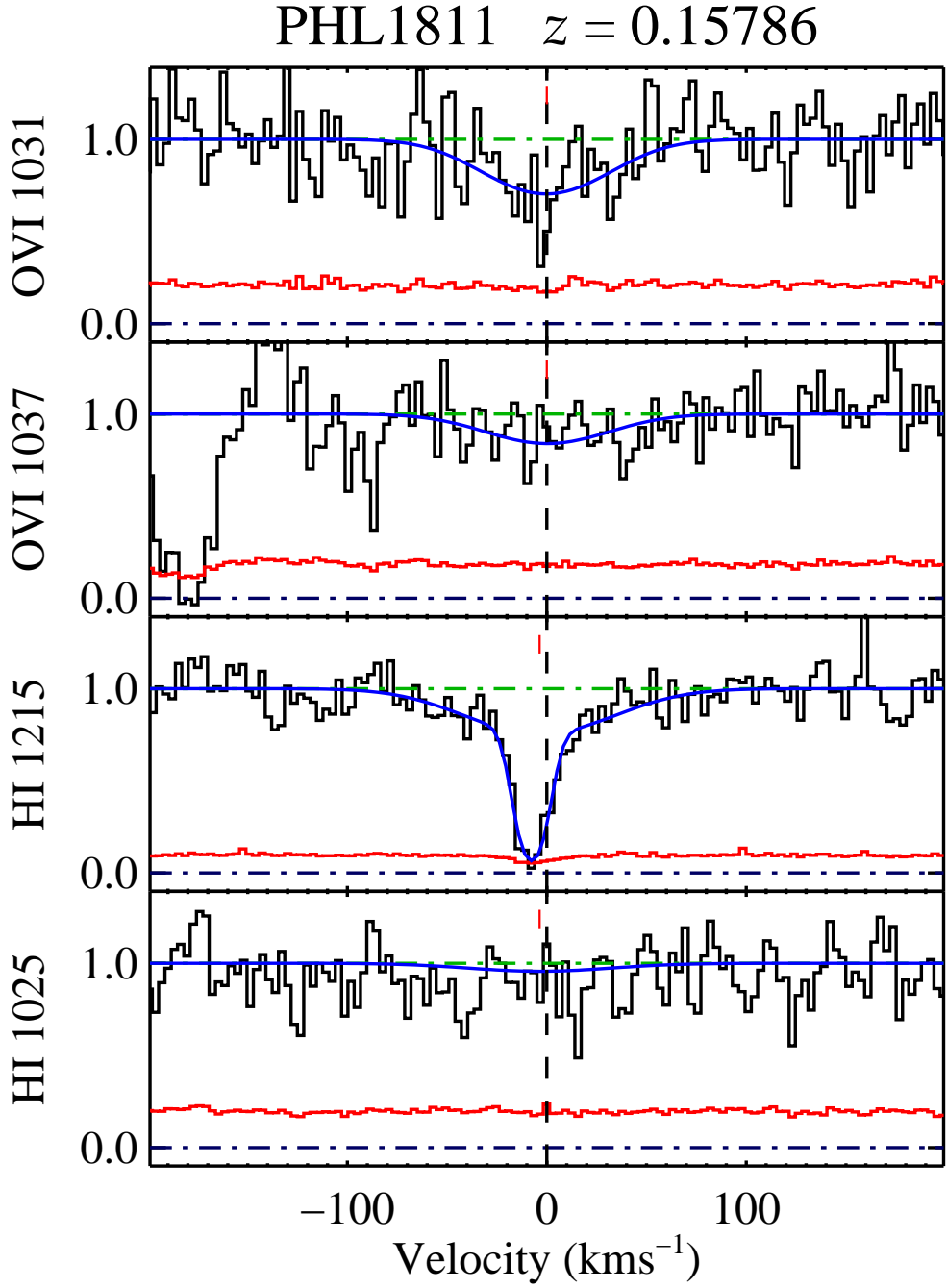} 
  \caption{PHL\,1811 $z=0.15786$---Both \OVI\ and \HI\ in this system are weak. The putative narrow
    \HI\ component is {\it not} \HI, but rather \OI\,1302 in the $z=0.0809$ Lyman-limit system. We
    include this line in the fit, detecting only broad, weak \HI\ that is too weak to be detected in
    the \Lyb\ transition.}
  \label{fig: PHL1811_z0.15786}  
\end{figure}

\label{sec: phl1811}
\noindent $\zabs = 0.15786$ (Figure~\ref{fig: PHL1811_z0.15786}; ---This \OVI\ system is very
weak. The $\OVI\,1031$ line is obvious, but the $\OVI\,1037$ line is detected at only $\sim
2\,\sigma$. Nevertheless, the data show a profile very consistent with the stronger line of the
doublet, and the fit to both lines is good. Further, this system is blueward of the Galactic \Lya\
line, so the \OVI\,1031 line cannot be ascribed to \Lya\ absorption. The \HI\ in this system is
harder to quantify. There are two components at the expected position for \Lya\ at this
redshift. Both are well fit, but the narrow component is conspicuous given the broad nature of the
\OVI. Our fits are consistent with those of \citet{tripp-etal-08-OVI}, but further investigation
reveals the narrow component to be \OI\,1302 in the $z=0.0809$ Lyman-limit system (LLS)
\citep{jenkins-etal-05-PHL1811_z0.0809}.  Our fit parameters for the narrow component agree well
with measurements of \citeauthor{jenkins-etal-05-PHL1811_z0.0809}.  Further, there is no
corresponding \Lyb\ feature, which would be expected if the narrow component were \HI. The broad
\Lya\ component is too weak for \Lyb\ absorption to be detectable. The broad \OVI\ and \HI\
components are well aligned, and we derive $\logT = 5.1$ and $\bnt = 40.5\kms$ (Sec~\ref{sec:
  gas_temperature}). This is the only system for which we derive a temperature greater that
$10^5\K$, and even at this temperature, \OIII\ and \OIV\ dominate the ionization state of Oxygen by
several orders of magnitude in CIE models. Both species, however, have absorption lines which are
shortward of the Lyman limit for the $z=0.0809$ LLS, and so cannot be observed.

\subsection{PKS\,0312$-$77}
The data for the PKS\,0312$-$77 ($\zqso = 0.223$) allow a path-length for our doublet search $0.1241
< \zabs < 0.1999$. We detect the strongest \OVI\ absorber in our survey in these data, at $\zabs =
0.20275$, which is part of a partial LLS.

\begin{figure*}                 
  \epsscale{1.0}
  \plottwo{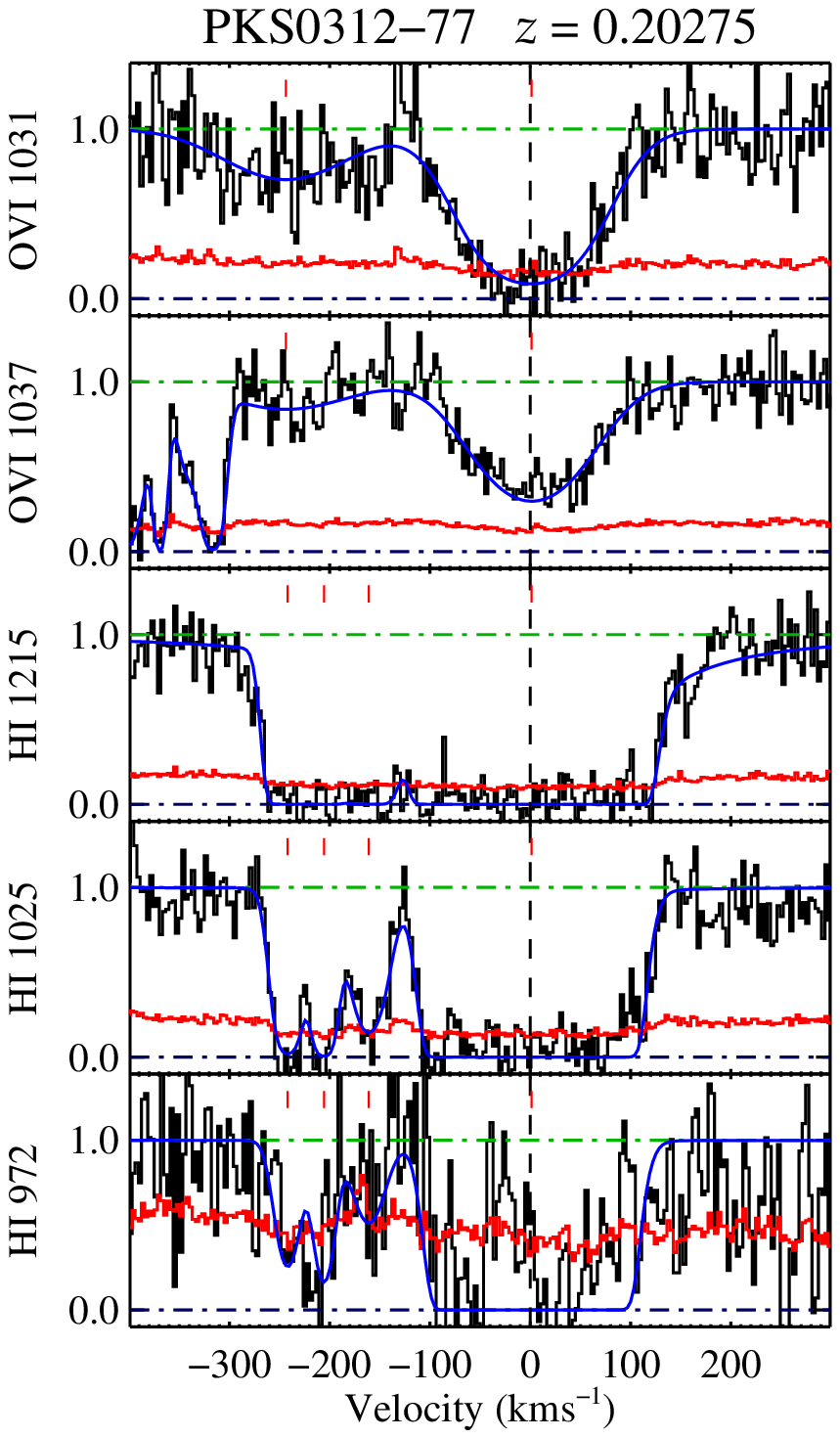}{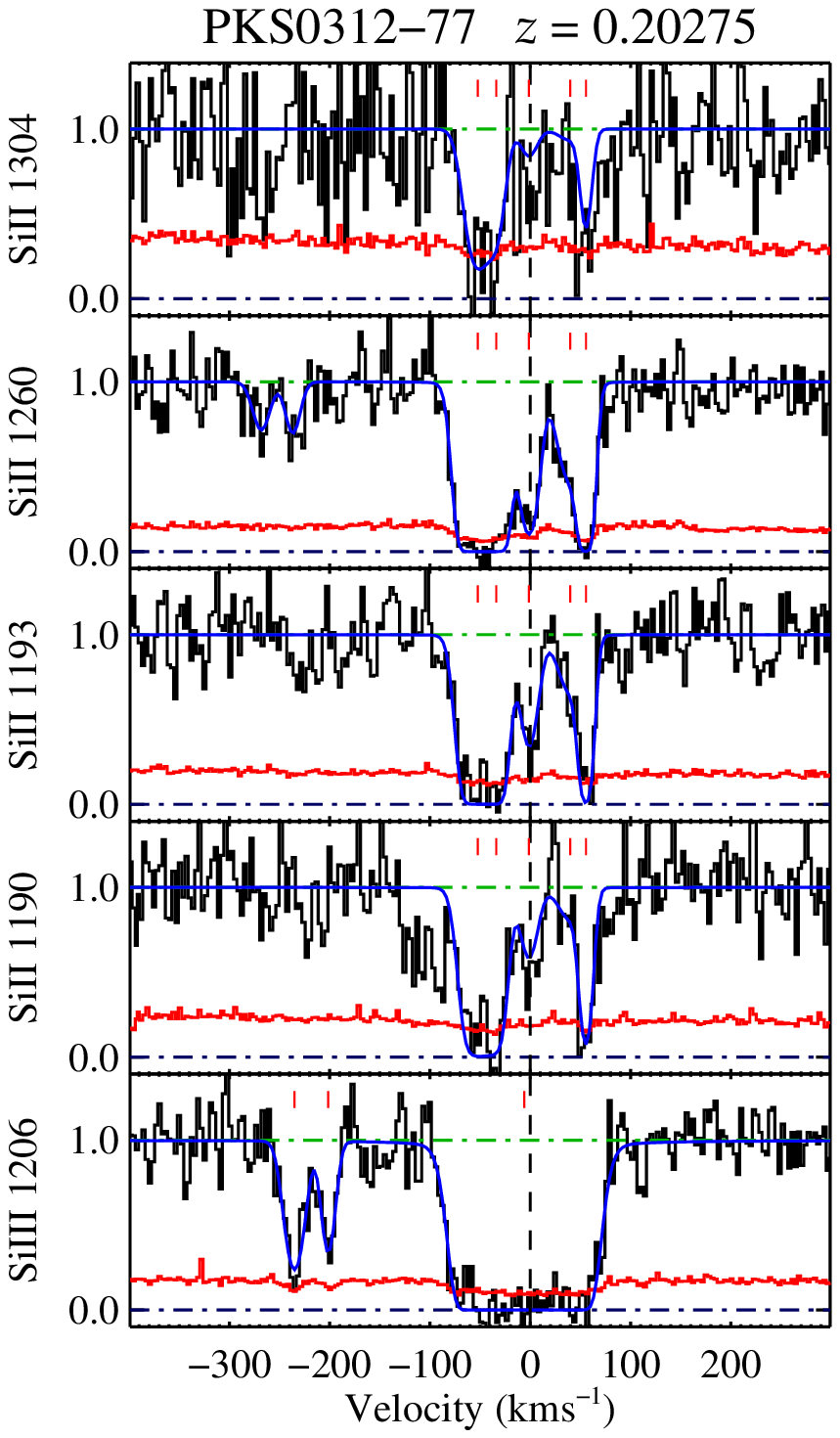}
  \caption{PKS\,0312$-$77 $z=0.20275$---This Lyman-limit system shows strong, saturated \OVI\, \HI,
    \SiII, \SiIII\, \SiIV and \CIII. We also see weak \NV. We are not able to resolve multiple
    absorption components in the main \OVI\ absorber, but their presence is strongly suggested by the
    complicated component structure exhibited in other transitions.}
  \label{fig: PKS0312-77_z0.20275}  
\end{figure*}

\noindent $\zabs = 0.20275$ (Figure~\ref{fig: PKS0312-77_z0.20275}; Table~\ref{tab:
  PKS0312-77_transitions})---The STIS spectrum of the sightline toward PKS\,0312$-$77 was obtained
under proposal id 8651 (PI Kobulnicky). This line of sight contains the strongest \OVI\ absorber of
our entire sample: a partial Lyman Limit System at $\zabs = 0.20275$ with $15.9 < \logNHI < 18.4$,
which is confirmed by a brief inspection of a FUSE spectrum. This system has two groups of
components, each having a complex structure. The main component group has strong, broad \OVI\
absorption and covers the range $-100 \simlt v \simlt 100\kms$, while the weaker group shows
evidence of weak \OVI\ ($-300 \simlt v \simlt -100\kms$).

Fitting for this absorber was complicated by the fact that many components are saturated,
particularly in stronger transitions such as \Lyab, \CIII\,977 and \SiIII\,1206. In general, we used
weaker, low ion transitions (typically \NII\ and \CII) to fix the redshifts of the components
(excluding \OVI), and initialized fits with fixed redshift, and initial $b$-value and column density
guided by weaker absorbers. Degeneracies and poorly constrained parameters naturally result from
fitting saturated absorbers; these are indicated by large formal parameter errors, and the fit
values are typically indicative of lower limits only.

\begin{figure}                 
   \epsscale{1.0}
 \plottwo{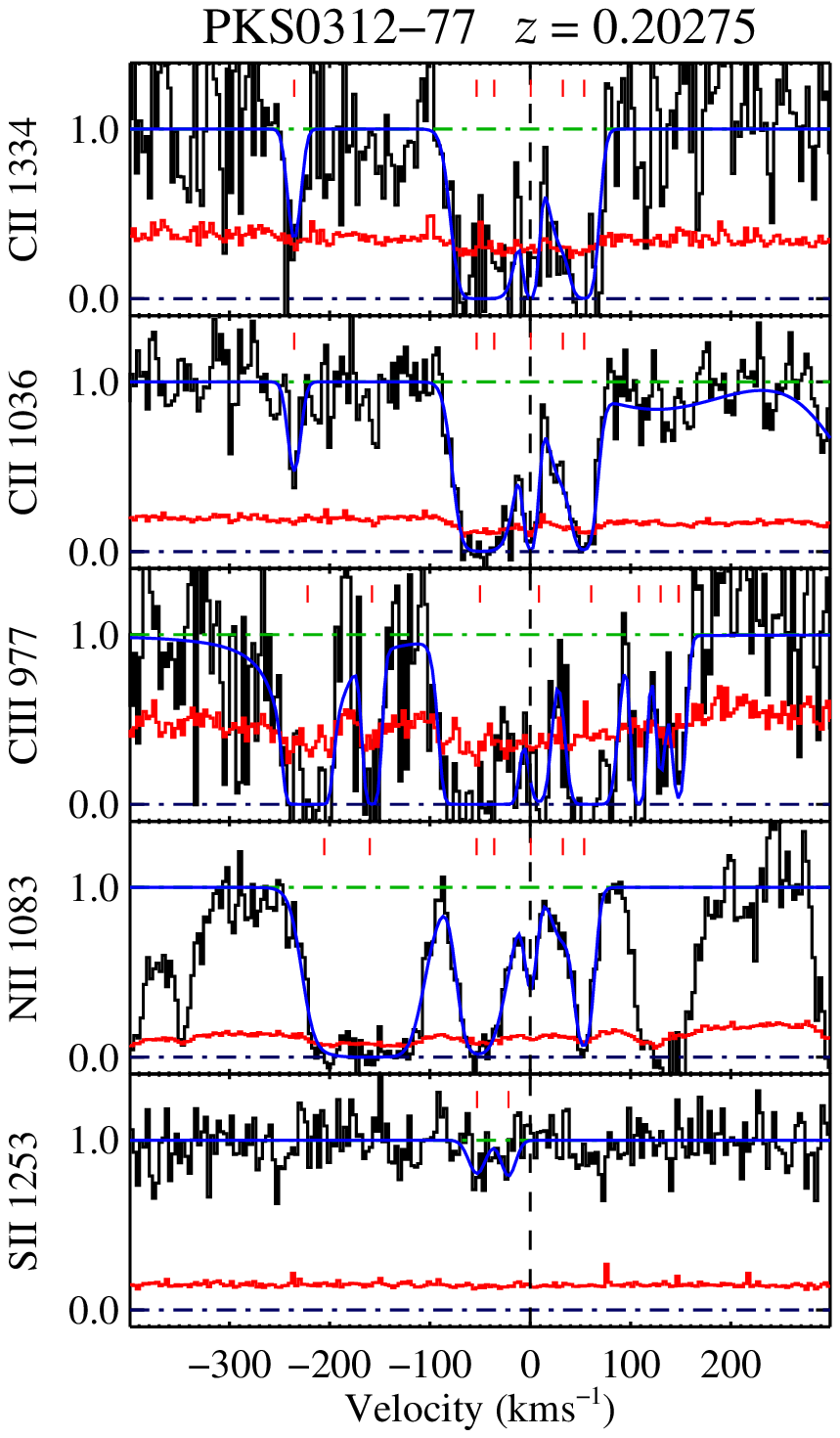}{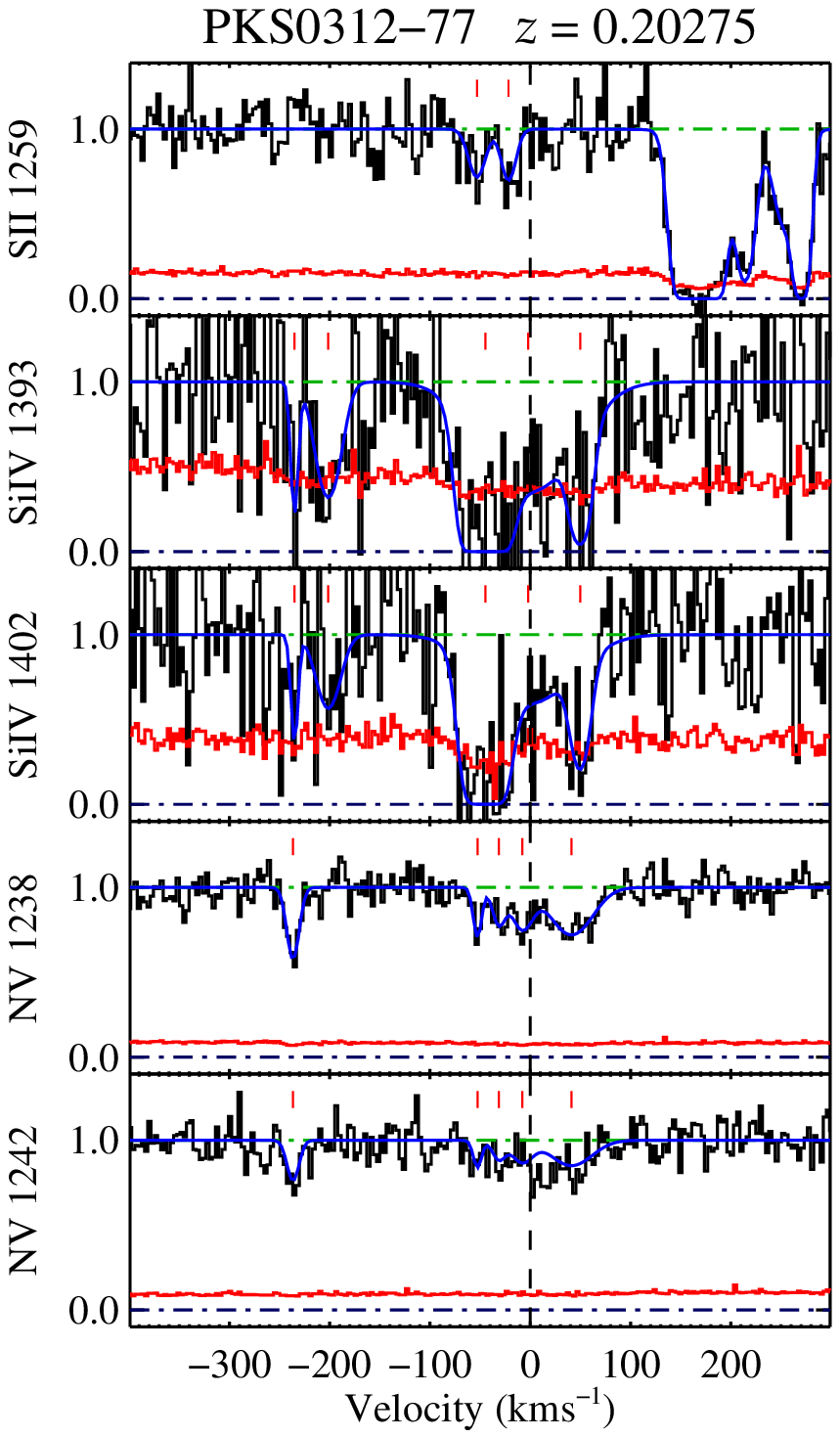} 
  \caption{Further transitions of the PKS\,0312$-$77 $z=0.20275$ system, as for Figure~\ref{fig: PKS0312-77_z0.20275}.}
  \label{fig: PKS0312-77_z0.20275_2}  
\end{figure}

For the main component group, we fit each species separately. Five components were first identified
from low the ion transitions \CII\,1036, \NII\,1038 and \SiII\,1193, 1260, which all show the same
structure. These redshifts were fixed and fits performed. The \OVI\, line, while probably requiring
more than one component for the measured breadth, shows little evidence of individual component
structure, and a single component provided adequate fits. Only three \HI\ transitions are present in
the STIS data, all of which are strongly saturated, resulting in mostly \NHI\ lower limits. \SiII\
transitions are well fit, except for \SiII\,1190, which shows possible contamination from Galactic
\CI*\,1190 at the red side, and Galactic \CI\,1190 at the blue side. The stronger \SiIII\,1206 and
\SiIV\ transitions show saturation, with data quality at \SiIV\ deteriorating markedly due to
dropping STIS sensitivity (\SiIV\ is at the extreme red end of the wavelength coverage).  \SII\
shows weak absorption with a different profile than other low ions---only the two strongest
components in the main body of the absorber are present; better quality data would be valuable to
confirm that this detection is correct, since the weaker \SII\,1259 transition is barely detected.
For \CII, the $\lambda\,1036$ line is well fit, but the $\lambda\,1334$ line suffers from the same
decrease in sensitivity as \SiIV, and is poorly fit, as is the strong \CIII\,977 transition. \NII\
is well fit; in the blueshifted part it is strongly saturated. Blueward of this saturated \NII\
absorption we see Galactic \OI\,1301 at $\sim -380\kms$, and at $\sim 140\kms$ we see Galactic
\SiII\,1304. Finally, \NV\ is present in multiple components, but very weak, and we use only the
stronger \NV\,1238 region for the fitting. Figure~\ref{fig: PKS0312-77_z0.20275_2} shows the predicted
profile of the \NV\,1242 region, based on the fits.
     
\subsection{PKS\,0405$-$12}
PKS\,0405$-$12 ($\zqso = 0.5723$) is the highest redshift QSO in our sample, allowing a path-length
to search for \OVI\ absorbers $0.1241 < \zabs < 0.5478$. We detect four \OVI\ systems, at $\zabs =
0.15597, 0.18291, 0.36333, 0.49514$. This line of sight has also been studied previously by
\citet{chen-prochaska-00-PKS0405-12_z0.167, prochaska-etal-04-PKS0405-12,
  prochaska-etal-06-PKS0405-12_galaxies}

\begin{figure*}                 
  \epsscale{1.0}
  \plottwo{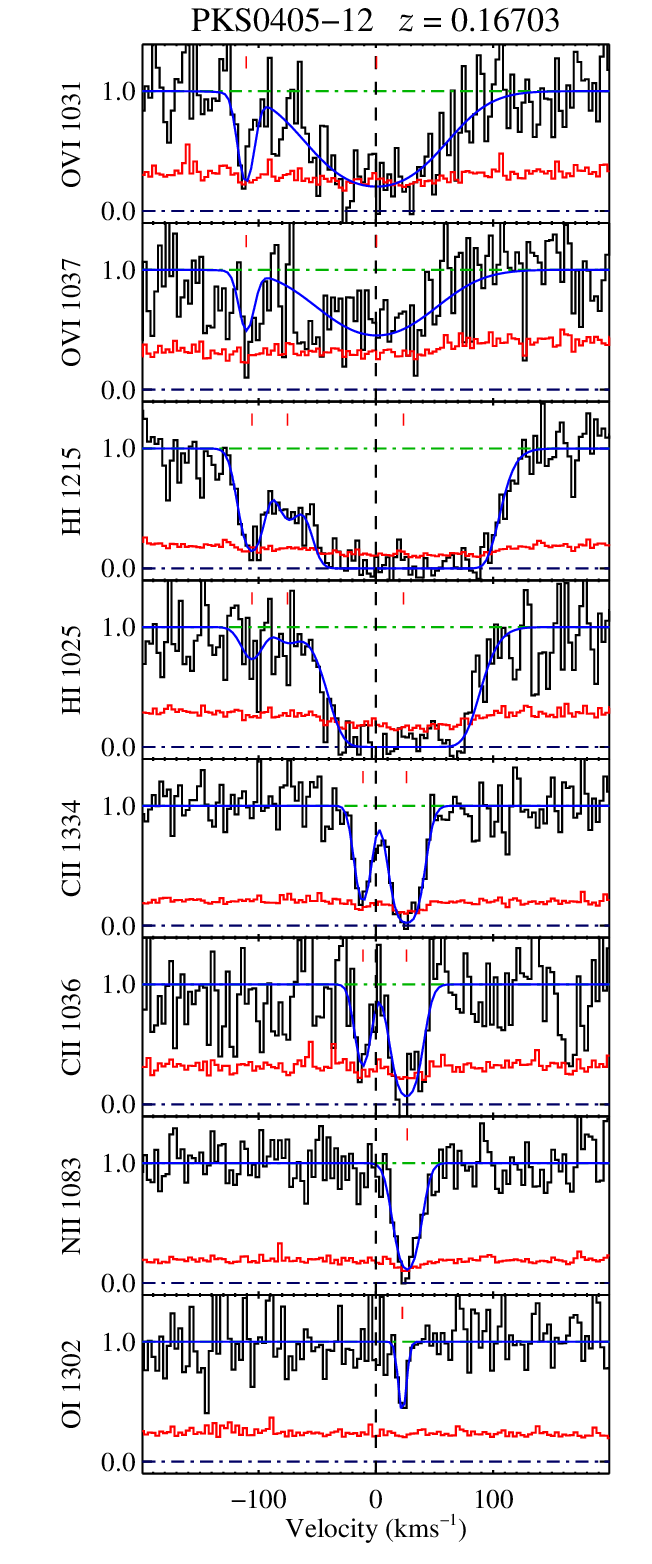}{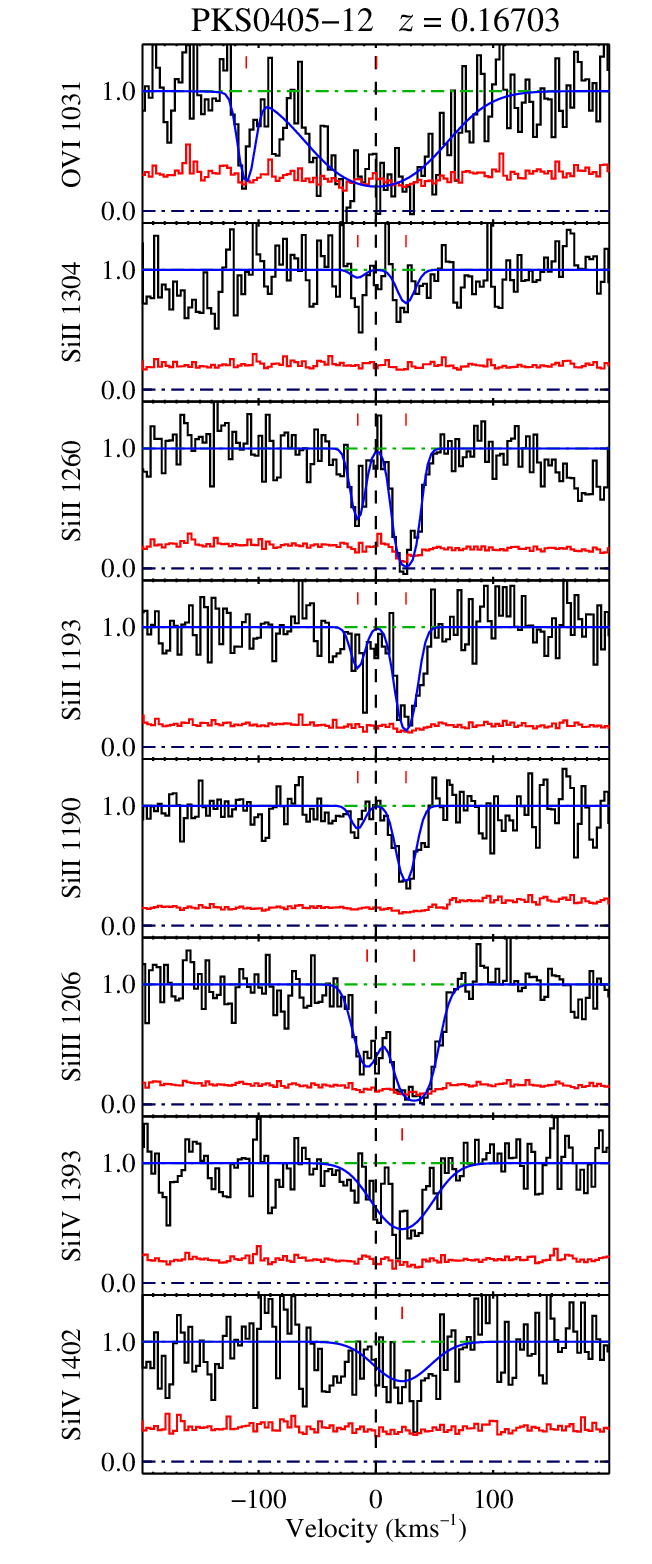}
  \caption{PKS\,0405$-$12 $z=0.16703$---This partial Lyman-limit system has strong \OVI\ and
    corresponding saturated \HI, and low ion metals with a complicated structure.}
  \label{fig: PKS0405-12_z0.16703}
\end{figure*}

\noindent $\zabs = 0.16697$ (Figure~\ref{fig: PKS0405-12_z0.16703}; Table~\ref{tab:
  PKS0405-12_transitions})---The PKS\,0405$-$12 sightline has been studied by
\citet{prochaska-etal-04-PKS0405-12}, and the partial Lyman-limit system at $z=0.167$ by
\citet{chen-prochaska-00-PKS0405-12_z0.167}. The system has strong, saturated \HI, and two \OVI\
components---one broad and strong; one weak and narrow. \citet{prochaska-etal-04-PKS0405-12}
estimate $\logNHI = 16.45$ from an analysis of the flux decrement shortward of the 912\A\
Lyman-limit in FUSE data for this sightline.

The STIS data exhibit strong lines of \SiII, \SiIII, \SiIV, \CII, \NII\ and \OI, some of which
appear saturated. In general, the low ion transitions are well aligned, and details are given in
Table~\ref{tab: PKS0405-12_transitions}.  Due to the heavily saturated \HI\ lines, only the weak
\OVI\ feature at $-110\kms$ can be matched well with \HI\ absorption components, and the resulting
parameters $\bnt = 7.0\kms$ and $\logT = 3.7$ are derived in Sec~\ref{sec: gas_temperature}.

\begin{figure}                 
   \epsscale{1.0}
 \plotone{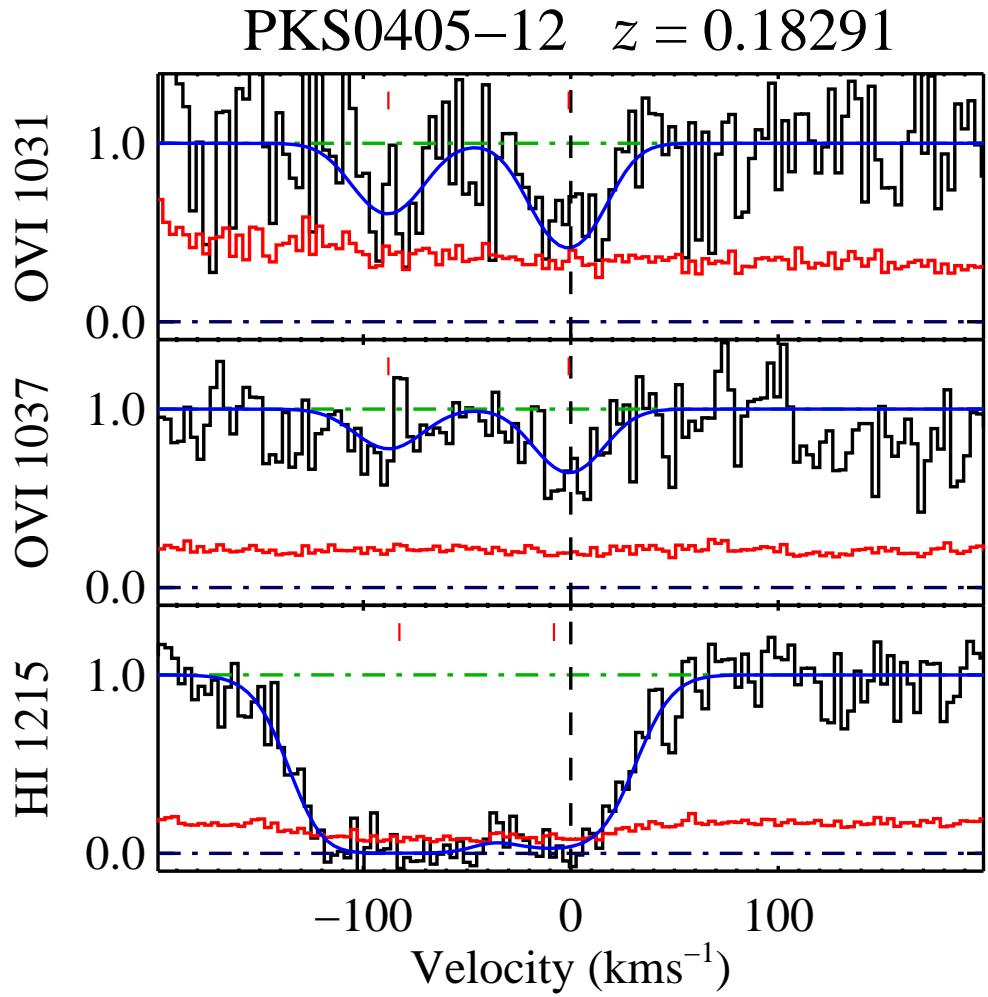} 
  \caption{PKS\,0405$-$12 $z=0.18291$---This is system has two strong \OVI\ components separated by some
    90\kms. The \OVI\,1031 data are quite noisy. Both \OVI\ components have corresponding \HI\
    absorption, but the \Lya\ line is heavily saturated and higher order lines are outside the
    wavelength coverage of our data. No metal absorption is seen.}
  \label{fig: PKS0405-12_z0.18291}
\end{figure}

\noindent $\zabs = 0.18291$ (Figure~\ref{fig: PKS0405-12_z0.18291}; Table~\ref{tab:
  PKS0405-12_transitions})---The absorber at $\zabs = 0.18291$ consists of two separate components,
separated by $\sim 90\kms$, both of which are reported in \citet{prochaska-etal-04-PKS0405-12}. In
our STIS data, we have only the \Lya\ line corresponding to these absorbers (\Lyb\ blends into the
Galactic \Lya\ trough), which is heavily saturated. We thus derive only lower limits to \logNHI,
consistent with \citet{prochaska-etal-04-PKS0405-12}.  The two \OVI\ components are also well fit,
despite the \OVI\,1031 line appearing on the red edge of the Galactic \Lya\ absorption.
     
\begin{figure}                 
   \epsscale{1.0}
 \plotone{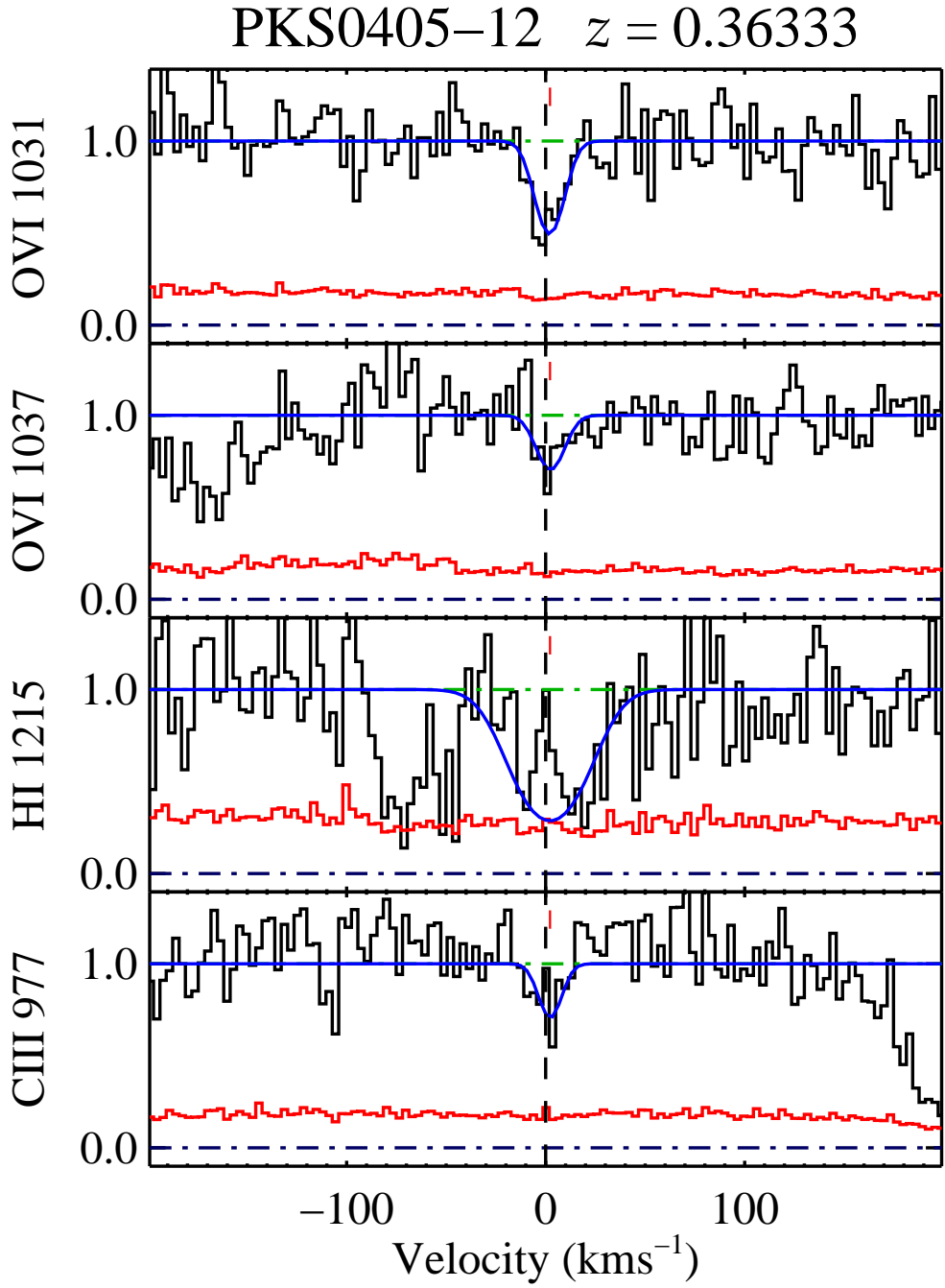} 
  \caption{PKS\,0405$-$12 $z=0.36333$---This system shows weak \OVI\ with a well-fit doublet. The \Lya\
    line shows contamination from bad pixels, and is also possibly blended with Galactic \CI*\,1657;
    \Lyb\ is not detected. Weak \CIII\ is also present.}
  \label{fig: PKS0405-12_z0.36333}  
\end{figure}

\noindent $\zabs = 0.36333$ (Figure~\ref{fig: PKS0405-12_z0.36333}; Table~\ref{tab:
  PKS0405-12_transitions})---This system has a well-fit \OVI\ doublet, and corresponding noisy \HI\
profile. \Lya\ is the only \HI\ transition we detect, but it is marred by hot pixels in the data, so
our resulting measurements are uncertain. Further, the \Lya\ line lies very close to the Galactic
\CI\,1656, blending with the fine structure \CI*\,1657.38 line.  Lacking the detection of \Lyb, we
cannot obtain reliable \logNHI. We also detect weak \CIII.
     
\begin{figure}                 
  \epsscale{0.6}
  \plotone{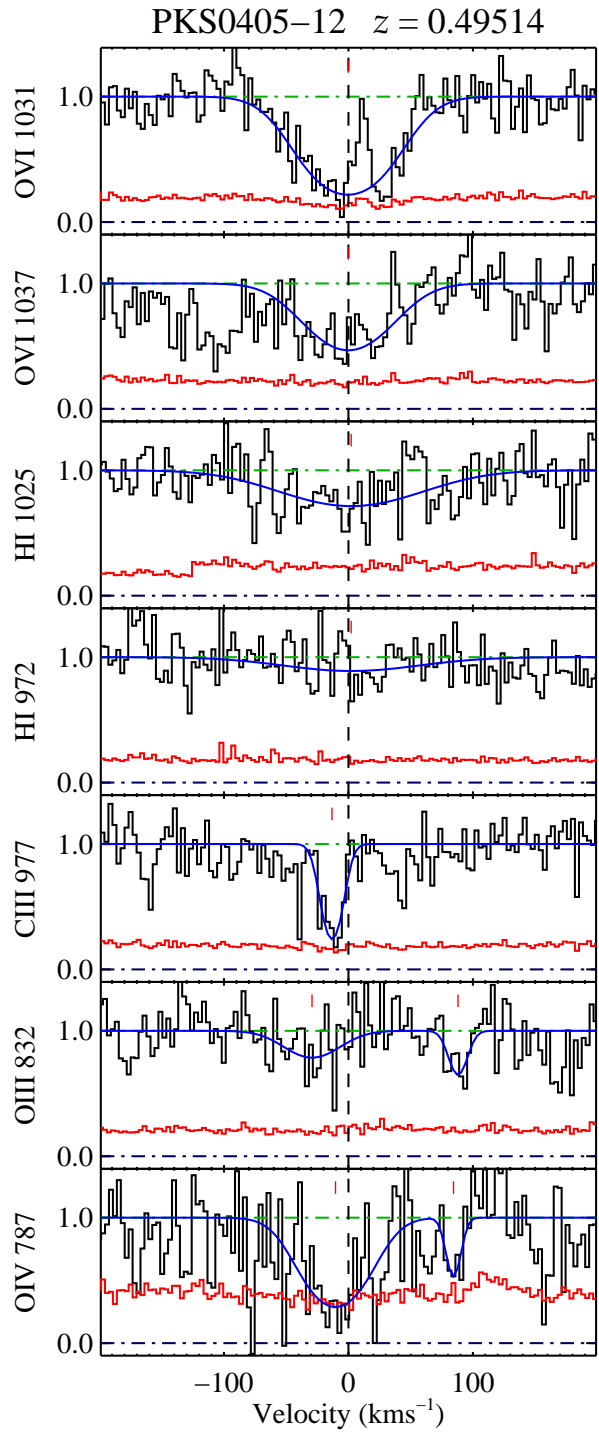} 
  \caption{PKS\,0405$-$12 $z=0.49514$---The highest redshift absorber in our sample, both \OVI\ lines
    are clearly detected, but both suffer from bad pixels, complicating the fitting. \Lya\ is
    redshifted out of our wavelength range, and \Lyb\ shows a very wide profile. Strong \CIII\ is
    detected, as are \OIII\ and \OIV.}
  \label{fig: PKS0405-12_z0.49514}  
\end{figure}

\noindent $\zabs = 0.49514$ (Figure~\ref{fig: PKS0405-12_z0.49514}; Table~\ref{tab:
  PKS0405-12_transitions})---\citet{prochaska-etal-04-PKS0405-12} present a brief analysis of the
absorber at $\zabs = 0.49514$. At such a high redshift, the \Lya\ line is redshifted out of the STIS
bandpass, and the \Lyb\ and \Lyc\ transitions are weak and noisy. Both \OVI\ lines are contaminated
by hot pixels. In the region of the \OVI\,1037 line, Galactic \CIV\,1538 falls at $\sim
-90\kms$. Inspection of the \CIV\,1542 line shows that it does not significantly affect the \OVI\
line. \CIII\ is detected at $v = 0 \kms$, as are offset \OIII\,832 and \OIV\,787. Both the \OIII\
and \OIV\ transitions also show a component at $\sim 90\kms$. While there is no associated \HI\ with
these components at $v = 90\kms$, the exact alignment of both transitions argues that the absorption
is real (as opposed to, say, weak \HI). Due the noise in the data, we could not obtain an acceptable
fit to the \HI\ lines and we were forced to fix the \HI\ doppler-parameter. Higher quality data
would be very valuable for this system, both to obtain an accurate \HI\ column density, and confirm
the assumed doppler parameter for \HI.

\subsection{PKS\,1302$-$102}
We searched the STIS data for PKS\,1302$-$102 ($\zqso = 0.2784$) for \OVI\ absorbers in the range
$0.1183 < \zabs < 0.2558$, detecting two close absorbers at $\zabs = 0.22565, 0.22744$. This
sightline has also been studied by \citet{cooksey-etal-08-PKS1302-102}.

\begin{figure}                 
  \epsscale{1.0}
  \plotone{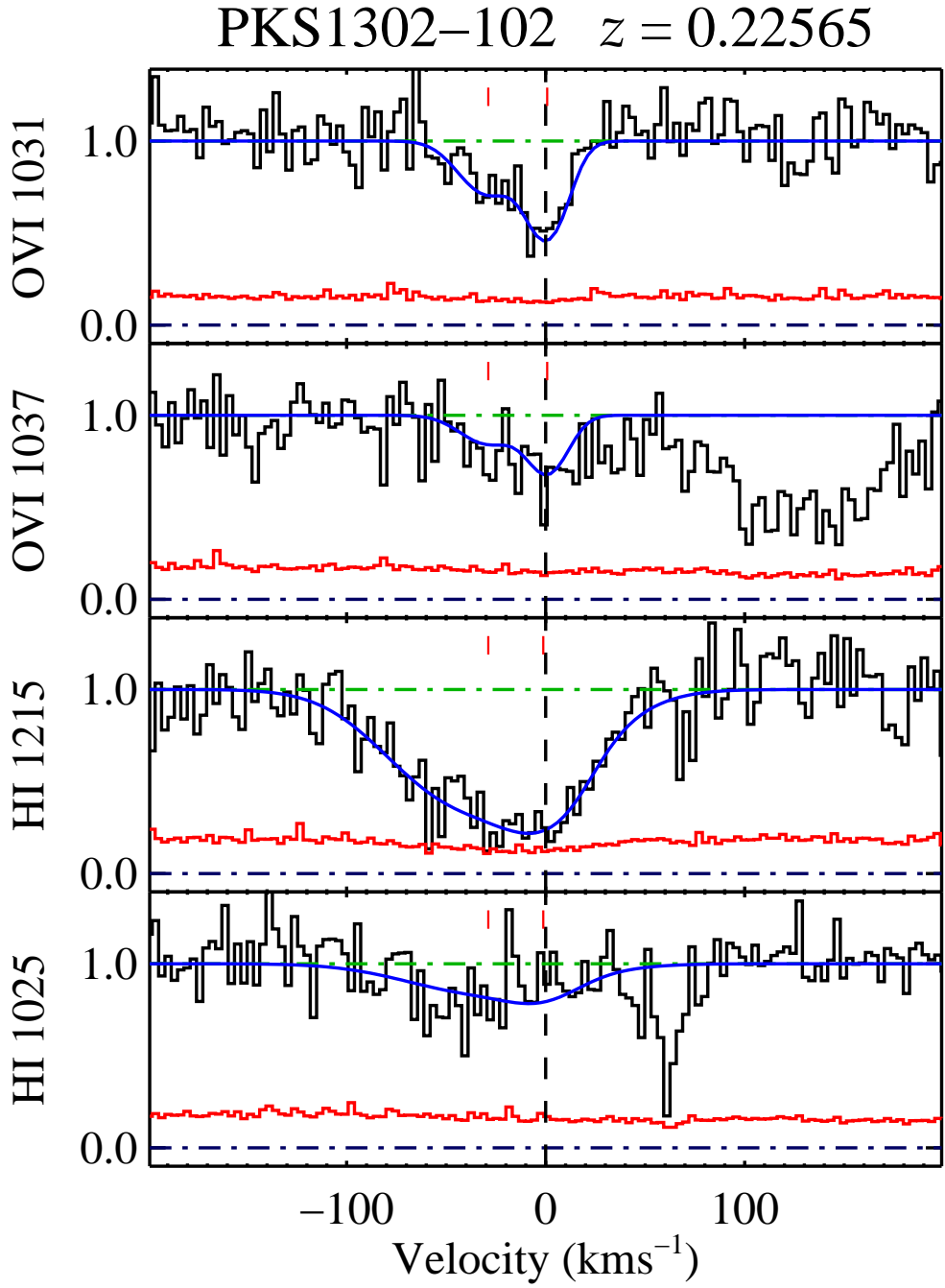} 
  \caption{PKS\,1302$-$102 $z=0.22565$---The lopsided, two-component structure seen in \OVI\ in this
    system is mirrored in the \HI\ absorption. No other transitions are detected.}
  \label{fig: PKS1302-102_z0.22565}  
\end{figure}

\noindent $\zabs = 0.22565$ (Figure~\ref{fig: PKS1302-102_z0.22565}; Table~\ref{tab:
  PKS1302-102_transitions})---As with the H\,1821+643 absorber pair, these absorbers at $\zabs = 0.22565$
and $\zabs = 0.22744$ along the PKS\,1302$-$102 sightline are separated by only $\Delta v = 440\kms$. Also
similar to the previous example, this absorber pair consists of a strong and weak system, with the
weaker system at a slightly higher redshift (although in this case, the difference is far less
dramatic). The system at $\zabs = 0.22565$ consists of two, well defined \OVI\ components separated by
28\kms. Both \OVI\ components have matching \HI\ absorption, detected primarily in \Lya. The two
strong lines at the expected positions of the \NV\ doublet are obviously unrelated \Lya\
absorbers. The temperature and non-thermal broadening can be determined for only the $v=0\kms$
component ($\logT = 4.4; \bnt = 12.5\kms$), since the \HI\ position for the component at $v=-28\kms$
has been fixed to that of \OVI\ in our fitting; see Sec~\ref{sec: gas_temperature} for details.

\begin{figure}                 
  \plotone{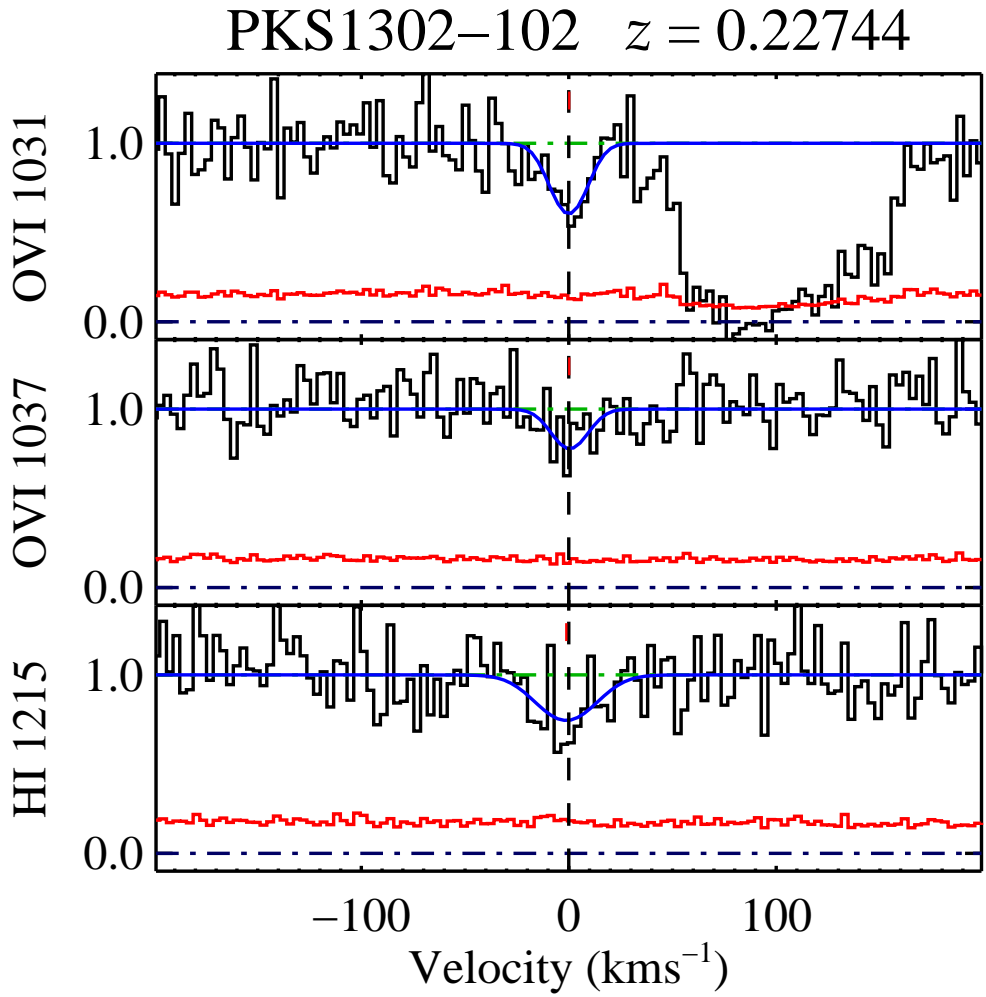} 
  \caption{PKS\,1302$-$102 $z=0.22744$---A mere 440\kms from the absorber at $z=0.22565$, this absorber
    is weaker than the lower redshift system. The \OVI\,1037 line is detected with low significance,
    and the corresponding \HI\ line is very weak. The strong line to the red of the \OVI\,1031 line
    is \Lya.}
  \label{fig: PKS1302-102_z0.22744}  
\end{figure}

\noindent $\zabs = 0.22744$ (Figure~\ref{fig: PKS1302-102_z0.22744}; Table~\ref{tab:
  PKS1302-102_transitions})---The weaker of the two close \OVI\ absorbers towards PKS\,1302$-$102, this
system has only very weak transitions of \OVI\,1031, 1037 and \Lya. The \OVI\,1031 line is close to,
but unaffected by, a strong \Lya\ absorption line at $z=0.4224$. Both species are well aligned, and
suitable for temperature analysis in Sec~\ref{sec: gas_temperature}. We derive $\logT = 4.2$ and
$\bnt = 10.7\kms$.

\subsection{Ton\,28}
The Ton\,28 ($\zqso = 0.3297$) line of sight hosts only one \OVI\ absorber at $\zabs = 0.27340$ in
the available path-length $0.1231 < \zabs < 0.3059$. 

\begin{figure}                 
  \epsscale{0.8}
  \plotone{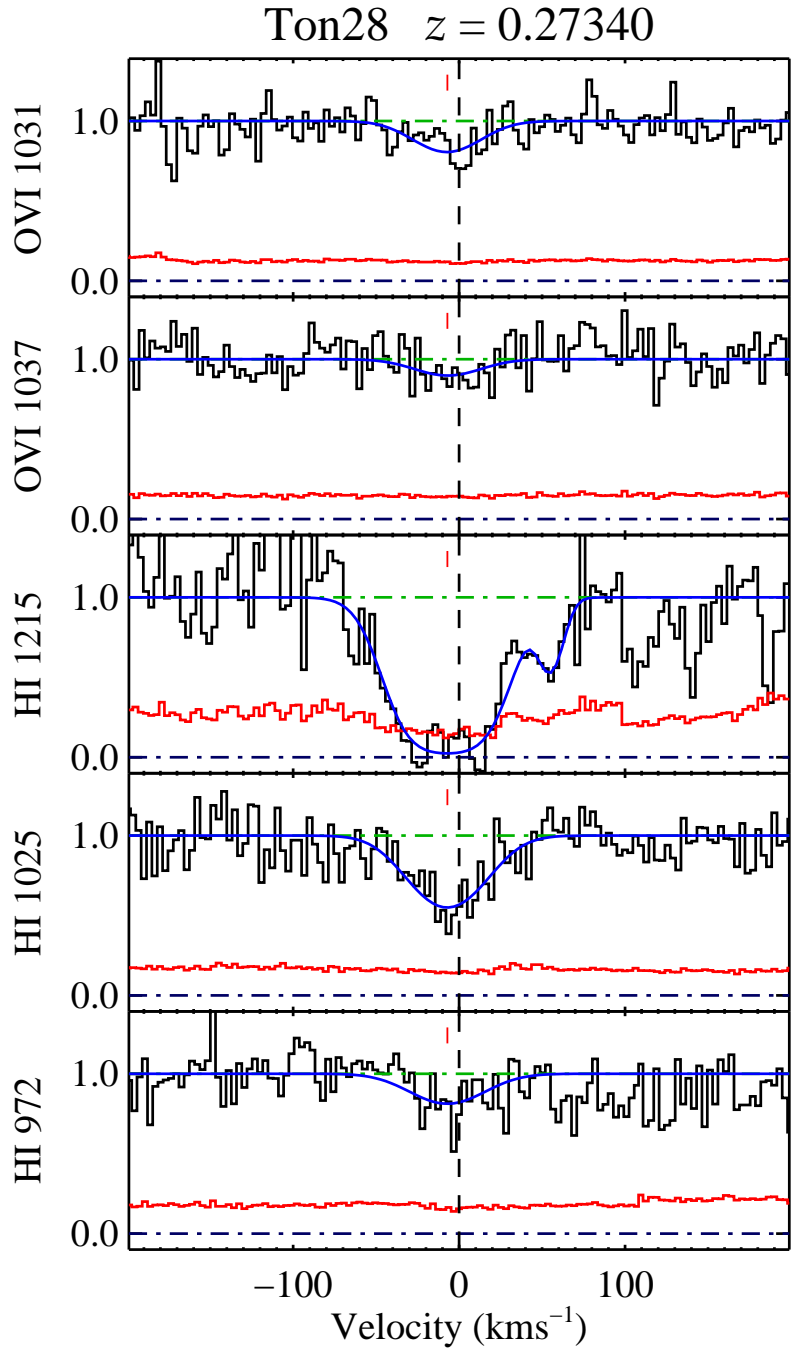} 
  \caption{Ton\,28 $z=0.27340$---Only very weak \OVI\ absorption is seen to correspond with the
    saturated \Lya\ in this system. \Lyb\ is present and unsaturated, as is \Lyc. The \Lya\ line is
    blended with Galactic \CIV\,1548, which we fit simultaneously.}
  \label{fig: Ton28_z0.27340}  
\end{figure}

\noindent $\zabs = 0.27340$ (Figure~\ref{fig: Ton28_z0.27340}; Table~\ref{tab:
  Ton28_transitions})---The system towards Ton\,28 contains strong \HI\ absorption, and only weak
\OVI. The \OVI\,1031 line is detected at $>3\,\sigma$ significance, while the \OVI\,1037 is only
$\sim 2.5\,\sigma$. The \Lya\ absorption line is saturated, and blended with Galactic
\CIV\,1548. Comparison of the Galactic \CIV\,1548 and 1550 lines shows that the full \Lya\ profile
is blended, not just the red wing (which is apparent in Figure~\ref{fig: Ton28_z0.27340}). The \Lyb\
and \Lyc\ lines allow us to obtain a good \HI\ fit, and we include the \CIV\ absorption component
that is obvious in the \Lya\ wing. The discrepancy between the \Lya\ data and fit profile is can be
ascribed to the \CIV\ blending. No other associated transitions are detected. The alignment of the
\HI\ and \OVI\ positions makes this system useful for our temperature analysis in Sec~\ref{sec:
  gas_temperature}. We have derived $\logT = 4.5$ and $\bnt = 20.2$.

\section{Comparison With Other Work}
\label{sec: not_recovered}
As noted in Sec~\ref{sec: introduction}, two similar contemporaneous studies have been conducted by
other groups \citep{tripp-etal-08-OVI, danforth-shull-08-OVI}. While differences exist in the
selection of \OVI\ systems in the different samples there is typically good agreement in measured
quantities for common systems.  Here we specifically address systems reported by
\citep{tripp-etal-08-OVI} that are not accepted by our selection criteria.  In general, this is a
difference between performing a {\it blind} search for the \OVI\ doublet, and looking for one of the
doublet lines in previously identified systems (e.g.\ \Lya\ absorbers). This associated \Lya\
technique accepts cases in which one or other of the \OVI\ doublet lines is masked by strong
absorption from ISM or IGM lines at a different redshift.  Other systems simply fall outside our
wavelength range or are identified in different data (e.g.\ the \OVI\ doublet is identified in FUSE
spectra).

We only consider absorbers from the \citet{tripp-etal-08-OVI} compilation that are within our
wavelength range, but are not recovered in our search.  Table~\ref{tab: unconfirmed_systems}
summarizes these systems, and we comment on each system below. The \OVI\,1031 (top) and 1037
(bottom) spectral regions for each system are shown in Figures~\ref{fig:
  OVI_rejected_1}~\&~\ref{fig: OVI_rejected_2}.  Note that \citet{danforth-shull-08-OVI} have also
recently conducted a similar survey using both STIS and FUSE data. Finally, where our sample has
common systems with the samples of \citet{tripp-etal-08-OVI} and \citet{danforth-shull-08-OVI}, we
see no marked differences in the measured total column densities, or the component $b$-values and
column densities where similar component structures are fitted.

\subsection{3C\,351.0}
\noindent $z = 0.21811$---This system shows broad absorption at the expected position of both \OVI\
doublet members, but is rejected because acceptable fits could not be obtained; the line strengths
of the putative doublet members are inconsistent at the $3-4\,\sigma$ level, with \OVI\,1037
stronger than \OVI\,1031. There is a putative \Lya\ absorber at this redshift, but its velocity is
inconsistent with the \OVI\ position ($\Delta\,v \sim 50\kms$). \CIII\ is obscured by Galactic
\SiII\,1190 and no \SiIII\ is detectable. It is possible that the \OVI,1037 line in this system is a
broad \Lya\ absorber \citep[see e.g.\ ][]{sembach-etal-04-PG1116+215}. We concur with
\citet{tripp-etal-08-OVI} that higher S/N observations would be very valuable to firmly establish
the nature of this system.

\noindent $z = 0.22111$---This system is not considered because the \OVI\,1031 position is clearly
blended with heavily saturated Galactic \SiII\,1260 absorption. \Lya\ and \SiIII,1206 are present at
the expected wavelengths; \CIII\,977 blends with \SiII\,1193 at $z = 0$.

\subsection{H\,1821+643}
\noindent $z = 0.12143$---There is strong \Lya\ absorption at $\sim -80\kms$ from the nominal
position of this system, but our STIS data are far too noisy to detect \OVI\ absorption ($\rm{S/N}
\approx 2-3$ in this region), which is reported by \citet{tripp-etal-08-OVI} in FUSE data. The
strong absorption at the \OVI\,1037 position for this redshift is \Lyd\ at z=0.225
\citep{tripp-etal-01-H1821+643_z0.1212}.

\noindent $z = 0.21331$---We initially considered this system, but rejected it due to mis-matched
line strength ratio. The \OVI\,1037 is stronger than the \OVI\,1031 line, and the strength ratio is
$R_{\scOVI} = 0.77 \pm 0.14$. As noted by \citet{tripp-etal-00-H1821+643}, the region of the
\OVI\,1037 line contains a strong Galactic \SII\,1259 at a velocity $v \simeq +100\kms$ with respect
to the absorber redshift, and a weaker Galactic \SII\,1259 feature at $\sim+30\kms$. This weaker
component partially blends with the putative \OVI\,1037 line. The weaker component is also seen in
\SII\,1253, and is likely associated with the intermediate-velocity cloud (IVC) in the Milky Way
known as the IV arch \citep{kuntz-danly-96-ComplexM}.

\subsection{HE\,0226$-$4110}     
\noindent $z = 0.42670$---The data show a strong \OVI\,1031 line in this candidate, but there is no
corresponding \OVI\,1037 line (the $\lambda\,1037$ position does contain a narrow absorption spike
due to noise, which is reflected in the error array). \Lya\ is redshifted out of the STIS bandpass
and there is no \Lyb. Galactic \SiIV\,1393 blends with the expected position of \CIII\,977, and
there is also no \SiIII\,1206. We conclude that this system is not real.

\subsection{PG\,1216+069}      
\noindent $z = 0.26768$---\OVI\,1037 blends with the heavily saturated \Lyb\ line from the absorber
at $z = 0.28232$, and hence this system cannot meet our criteria for a doublet search.

\subsection{PG\,1259+593}      
\noindent $z = 0.31972$---We do not detect the \OVI\,1037 line in this system, although there is a
single-pixel negative noise spike (which is reflected in the error array). A model based on the
putative \OVI\,1031 line is not consistent with the data in the $\lambda\,1037$ region. The red wing
of the \OVI\,1037 region blends with a weak ($\sim 20\mA$) unidentified line

\subsection{PG\,1444+407}     
\noindent $z = 0.22032$---The \OVI\,1031 line blends with Galactic \SII\,1259. The $z = 0$ \SII\
line exhibits the same structure in the \SII\,1253 transition, confirming its Galactic nature. Shallow,
broad \Lya\ may be seen, but the \OVI\,1037 is very weak and uncertain.

\subsection{PHL\,1811}         
\noindent $z = 0.13240$---At the observed wavelength of this system ($\lambda_{\rm obs} =
1168.5\A$), the spectrum is quite noisy. We detect a strong \OVI\,1031 line, and at higher
wavelengths the saturated, offset \Lya\ line is easily detected in higher S/N data. We do not,
however, see evidence of an \OVI\,1037 line, and thus reject this system.

\subsection{PKS\,0312$-$77}      
\noindent $z = 0.15890$---The data for the PKS\,0312$-$77 sightline is quite noisy, especially in the
low-wavelength region about the \OVI\,1031 position. We see a well-fit \OVI\,1031, but this same
profile bears no resemblance to the data in the expected region of \OVI\,1037. Multi component \Lya\
is also seen, but the lack of \OVI\,1037 absorption leads us to reject this system.

\noindent $z = 0.19827$---As with the lower redshift system described above, we detected \Lya\
(saturated in this case) and broad \OVI\,1031, but no evidence of \OVI\,1037. There is no
correspondence in the data between \OVI\,1031 and \OVI\,1037 regions, and the profile fit from the
\OVI\,1031 region is a poor fit to \OVI\,1037 region.

\subsection{PKS\,0405$-$12}      
\noindent $z = 0.36156$---Initially detected in our doublet search, and published by
\citet{prochaska-etal-04-PKS0405-12}, there is considerable mis-match between the \OVI\,1031 and
\OVI\,1037 regions. \Lya\ is uncertain, but there may be a weak line in the wing of the strong
(\NHI\ = 15.1) \Lya\ line at $z = 0.3608$. Due to the profile mismatch, we are unable to confirm
this system.

\subsection{PKS\,1302$-$102}     
\noindent $z = 0.19159$---This system is another example that we are unable to confirm due to lack
of \OVI\,1037 absorption.  A line at the position of \OVI\,1031 is present, as is saturated \Lya,
and possibly \SiIII\,1206. Given the strength of the apparent \OVI\,1031 line ($\sim 62\mA$), we
should expect to detect any \OVI\,1037 line, but no significant absorption exists, and the
\OVI\,1031 profile fit is a poor descriptor of the data in the \OVI\,1037 region. The independent
work of \cite{cooksey-etal-08-PKS1302-102} confirms this result; those authors report \Lya\ and
\OVI,1031, but do not detect \OVI,1037 (see their Table~3).

\subsection{Ton\,28}           
\noindent $z = 0.13783$---A strong line at the position of \OVI\,1031 is seen in this absorber, but
no corresponding \OVI\,1037 is seen. At such low wavelengths, however, this is hardly
surprising---our STIS spectra have ${\rm S/N} < 3$ per pixel at the position of \OVI\,1037. A
profile fit to the \OVI\,1031 line does not follow the data in the \OVI\,1037 region. \Lya\ is
detected as part of a stronger, saturated \Lya\ system.

\section{Physical Properties of Ionized Gas Selected by OVI Absorption}
\label{sec: properties}
Having identified a sample of \OVI\ absorbers, and related transitions, we now consider these
absorbers in detail. We begin by matching individual absorption components of \OVI\ and \HI\ within
an absorber. These components trace physically distinct gas clouds, and we consider first the global
properties of these components. With the sample of absorption components well-aligned in velocity
space, we derive the gas temperature and non-thermal broadening, and examine whether the
temperatures are consistent with a collisional ionization origin. We also consider the ionization
state of the \Oion-bearing gas using photoionization models. We close with a brief comment on the
lack of \OVI-only systems.

\subsection{\NOVI\ vs \NHI\ for Individual Components}
\label{subsec: components}
To consider any interpretation of the \Oion-bearing gas, we must determine whether the different
species observed in the absorbers arise from the same gas. It is well established from very high
resolution studies of the ISM that absorbers can be resolved into multiple (even many) individual
absorption components ($\Delta v < 1\kms$) \citep[e.g.][]{welty-etal-96-CaII-ISM,
  welty-etal-99-SN1987A}. Since these components will trace physically distinct gas clouds, it is at
this component level that we must assess the physical nature of the gas.  To define well-matched
components, we compare the positions of the \OVI\ and \HI\ components in each absorber. The
motivation for this matching is to constrain the temperature of the \Oion-bearing gas, a subject to
which we will return in the following sections.  In principle, highly ionized species such as \NV\
would be the ideal candidate to compare to \OVI\ in deriving physical conditions, in particular the
gas temperature. \NV\ peaks in CIE at $\logT = 5.25$ \citep{sutherland-dopita-93-gas-cooling}, so
occupies the same temperature range as \OVI. Conversely, \HI\ is orders of magnitude more abundant,
and so is much more likely to result in a good component match. We therefore attempt to match all
\OVI\ components with corresponding \HI\ components. If the two component positions are equal to
within their $2\,\sigma$ positional uncertainties (to a maximum of the STIS resolution---$6-7 \kms$)
we consider them related.

In Figure~\ref{fig: column_density_component_plots} we plot, for all well-matched components, the
\OVI\ column density as a function of the neutral hydrogen column density (c.f. Figure~19 of
\citealt{tripp-etal-08-OVI}).  It is not surprising that no points lie in the upper left part of
this plot since: they would require high metallicity {\it and} optimum ionization parameter, and; we
see little evidence of a class of systems lacking (or with weak) \HI\ (See Section~\ref{subsec:
  noHI}). The lower portion of the plot is empty due to a selection effect; in most of the survey,
our detection limit is $\Wr \gtrsim 30\mA$ which corresponds to $\logNOVI \gtrsim 13.5$.

A range of physical conditions may be used to explain any particular point on this plot, since the
ionization parameter and metallicity are degenerate. Nevertheless, it is worthwhile considering the
range of conditions which are allowed by the data. In this vein, we have overplotted two sets of
model photoionization curves: models for a variable ionization parameter at fixed metallicity; and,
varying metallicity at fixed \logU. The dot-dashed line shows our "fiducial" condition of gas at
peak ionization parameter and $1/10^{th}$ solar metallicity. Since \NOVI\ peaks for fixed density
and metallicity at $\logU = -0.2$, any increase or decrease in \logU\ will act to shift this line to
the lower right (i.e. \NHI\ will be larger at fixed \NOVI), an effect that is clear in
Figure~\ref{fig: photo_model}. What, then, of the points above this line?  Under the assumption of
photoionization equilibrium, we then infer that these systems {\it must} have higher metallicity
than our canonical 1/10$^{th}$ solar value. The dashed line shows the model curve for $\logU = -1.2;
\MH\ = 1.0$ and hence, at fixed $\MH\ = 1.0$, the data allow a range of ionization parameters of
$\sim 1\dex$.  If we instead consider the variation in the points as solely a function of
metallicity at fixed (optimal) ionization parameter, we obtain the bounds delimited by the crosses
and square-dots, and a range of metallicities of $\sim 1.7\dex$.

It is immediately apparent that the data do not follow the same global trend as the model curves
shown. Although we cannot determine a zero-point for the model curves, due to the degeneracy between
metallicity and \logU, the trend is nevertheless valid, and we should expect a variety of
metallicities and ionization conditions in the absorbers. The scatter in the observations of \NOVI\
vs \NHI\ can thus be understood as due to scatter in the metallicity and ionization state of the
gas\footnote{Other factors such as length scale, and varying redshift will also play a role here.}.

The case of the 3C\,351.0 absorber at $z = 0.31659$ (open squares) is interesting, since the \OVI\
and \HI\ clearly separate into three individual components. No other species are detected in the
STIS band. If we posit a physical connection, based on the proximity of the components, we might
expect similar conditions for each gas cloud. The three components are shown as open squares in
Figure~\ref{fig: column_density_component_plots}. It is clear that no single set of ionization and
metallicity conditions can explain all the components, yet the temperatures do not admit of a
collisional origin (see Sec~\ref{sec: gas_temperature}). Likely we are simply seeing, for a constant
flux of ionizing photons, density variations along the line of sight. By contrast, we would require
variations of $\sim 1.0\dex$ to explain the observations by metallicity variations alone.

\subsection{Gas Temperature}
\label{sec: gas_temperature}

In the previous discussion, we have assumed that the matched components arise from photoionized
gas. Is this assumption justified?  Having identified components of two different species that
likely arise from the same gas phase, we can use the Doppler parameters to derive an estimate of the
gas temperature, which are related through the relation $b^2 = \bnt^2 + 2kT/m$, where $\bnt^2$ is
the non-thermal component to the Doppler parameter (i.e. turbulence, bulk motion etc), and $k$, $T$
and $m$ are the Boltzmann constant, temperature and atomic mass as usual.  The two equations thus
determine both the gas temperature and the non-thermal line broadening, and this well-established
fact has been used in the past to determine gas temperatures of absorbers \citep[e.g.\
][]{chen-prochaska-00-PKS0405-12_z0.167, tripp-savage-00-PG0953+415_z0.1423}. It is also clear that,
in the absence of a turbulent component, the Doppler parameters will be related through the
square-root of the ratio of atomic masses. The distribution of \OVI\ and \HI\ $b$-values for
well-aligned absorption components is shown in Figure~\ref{fig:
  doppler_parameter_component_plots}. The solid lines bound two limiting cases. At one extreme, for
$\bnt = 0$, we have simply that $\bHI = 4\times \bOVI$. The upper boundary in Figure~\ref{fig:
  doppler_parameter_component_plots} shows the other limiting case, where $\bHI = \bOVI$. Formally,
we should not see absorbers above this limit, since that regime is non-physical. In practice, any
absorbers in this area imply either unresolved components, or that the \Oion\ and $\mbox{H}^0$ gas
is unrelated---either physically distinct\footnote{In which case the component alignment is purely
  coincidental.}, or in a separate phase of a multi-phase system.  While we have three components in
our sample that lie above this limit, all are consistent with $\bHI = \bOVI$ at the 1$\,\sigma$
level.

Gas in collisional ionization equilibrium with $T \simgt 2\times10^5$ implies $\bHI \simgt
60\kms$. It is immediately apparent that few of our components satisfy this criteria. Including a
non-thermal contribution to the $b$-value will further lower the implied temperature (or conversely,
increase the implied Doppler parameter). This simple argument indicates that collisional ionization
is not the dominant ionization mechanism.  It does not, however, rule out collisionally ionized
systems, since it is conceivable that weak, collisionally ionized systems may not contain a
detectable amount of neutral hydrogen (i.e. by considering {\it only} systems with well-aligned
\OVI/\HI\ components, we may be biased against systems in CIE).  To quantify this statement, we
examined with {\it Cloudy} (version 07.02.01) a simple model of gas in collisional ionization at
$\logT = 5.5$ and $1/10^{th}$ solar metallicity. For \OVI\ column densities in the range $\logN =
13.5 - 14.0$ the corresponding \HI\ column density is $\logN \simlt 13.0$, which will be difficult
to detect in our data. We note that while we see no evidence of a population of \OVI\ {\it
  absorbers} lacking \HI\ absorption, we certainly see {\it components} that are not well matched
with \HI.

Table~\ref{tab: ovi_components_temperature} gives the temperature and \bnt\ for the well-matched
components defined above. The first three columns identify the absorber and component. The measured
\HI\ and \OVI\ Doppler parameters are recalled here for reference. The derived gas temperature and
error are given, along with the non-thermal contribution to the Doppler parameter, and the resulting
thermal line-width for both \HI\ and \OVI. Finally, for comparison, we give the gas temperature we
would have derived from \bOVI\ had we assumed $\bnt = 0$, which is strictly a formal upper-limit.
Only one system lies above $\logT = 5.0$: the absorber towards PHL\,1811 at $z=0.15786$, and even in
this case, the data are ambiguous (see Sec~\ref{sec: phl1811}). It is clear, however, that for all
cases in which we can derive a gas temperature, the temperature is well below the $\logT = 5.5$
expected for CIE conditions. It may also be possible that we are observing gas that is radiatively
cooling out of equilibrium. \citet{gnat-sternberg-07-non-equilibrium-cooling} have shown that \OVI\
can be present at temperatures well below those expected in equilibrium models, since the cooling
time is much shorter than the recombination time. These non-equilibrium effects are only important
for enriched gas (solar metallicity or higher), however, and are unlikely to be a dominant effect
for our sample.

In three cases we could not determine \logT\ and \bnt, because $\bOVI > \bHI$; these cases are
listed in Table~\ref{tab: ovi_components_temperature} by ``...''. Note also, that these are the same
three components in Fig~\ref{fig: doppler_parameter_component_plots} which lie above the $\bOVI =
\bHI$ boundary. All are consistent with \bnt-dominated systems. Alternatively, this lack of
correspondence between the \HI\ and \OVI\ line widths may indicate that either the assumption of the
absorption being physically related is not correct for these systems, or there are unresolved
absorption components. In other words, {\it if} both absorption components arise from the same gas,
then we require that $1/4 < \bOVI/\bHI < 1$, since $m_{\rm O} = 16 \times m_{\rm H}$. The upper
limit comes from a state where the $b$-values are dominated entirely by \bnt (i.e. $\bHI = \bOVI$),
while the lower limit is the pure thermal case, and the b-values are simply related through the
square-root of the ratio of their respective atomic masses. Similar limits can be trivially written
down for all other pairs of species.

It is worth emphasizing, at this stage, that these measurements are dependent on several critical
factors: a) the resolution of the spectrograph must be sufficient to resolve the lines in question,
reinforcing our decision to rely solely on STIS E140M data with superior resolution (e.g.\ the FUSE
FWHM is $\sim 20-25\kms$, but our {\it median} \bOVI\ is $\sim 21\kms$); b) Further, the resolution
must be sufficient to resolve multiple absorption components and possible blending; c) the
non-thermal contribution to the Doppler parameter {\it must} be considered, since it dominates
$b_{tot}$ in most cases. As Table~\ref{tab: ovi_components_temperature} clearly demonstrates, if we
do not take this into account, most of our absorbers would be consistent with the coronal
temperature range $10^5 - 10^7\K$, and we could easily mis-interpret these as WHIM absorbers. This
point has also been made by \citep{tripp-etal-08-OVI}, while \citet{danforth-shull-08-OVI} derive
only upper limits to the temperature using \bHI.

\subsection{Ionization State of \Oion\ Bearing Gas}
\label{subsec: ionization}
Before considering the ionization conditions of the \Oion-bearing gas, we first inspect the results
of a typical photoionization model of low-density gas. We use similar models when considering
individual systems below, each tailored to the specific conditions of the individual absorber.
Figure~\ref{fig: photo_model} shows the ionization state for the most commonly observed species as a
function of the ionization parameter, $\logU \equiv {\rm log}\,(n_{\gamma}/\hden)$, where
$n_{\gamma}$ and \hden\ are the volume densities of ionizing photons (${\rm h}\nu > 13.6\eV$) and
hydrogen, respectively. The model was calculated for a slab of gas at with neutral hydrogen column
density $\logNHI = 15.0$ and metal abundance $1/10^{th}$ the solar value. The gas temperature was
constrained to be $\logT > 4.0$, although in practice this constraint has little
effect\footnote{Collisional ionization is not yet significant at these temperatures.}. Ionizing
radiation comes from an updated version of the \citet{haardt-madau-96-UV-background} spectrum
included with {\it cloudy}\footnote{Cloudy denotes this ``HM05''.}, at redshift $z = 0.25$ (the
median redshift of our sample). Since the spectrum of the UV background is fixed, and the relative
intensity is set at a fixed redshift, only changes in the density affect the ionization parameter.
Thus at high densities, there are fewer ionizing photons per atom, and lower ionization states
predominate. At lower densities (higher \logU), there are more ionizing photons for a given atom,
and higher ionization states may be ionized. Inspection of the curves in Figure~\ref{fig:
  photo_model} shows that \OVI\ can be observed over a wide range of conditions, especially if one
considers that metallicity may vary over a $1-2\dex$ range. Thus, unlike with CIE where \OVI\
occupies a very narrow temperature range, simply detecting \OVI\ is not enough to determine the
ionization conditions.

The column density ratios predicted by the {\it Cloudy} models are not sensitive to the adopted
value of \logNHI\ in the optically thin regime, although the absolute column densities would
obviously be lower for a lower gas column. A comparison between models with solar and $1/10^{th}$
solar metallicity shows that the column density ratios are unaffected by metallicity effects in the
regime $\logU > -4.0$, which corresponds to densities $\loghden < -1.6$.  This is well away
from $\logU = -0.2$, where \OVI\ peaks for photoionization models. None of the systems we consider
fall into this region of parameter space, so the adopted $\MH = -1.0$ for our models should not
affect our conclusions.

The detection of multiple metal transitions in an absorber offers a powerful diagnostic of the
ionization conditions of the gas, assuming that the metals responsible for the absorption are
co-spatial. By comparing the column density ratios for well-aligned transitions with predictions of
a model like that in Figure~\ref{fig: photo_model}, we can place limits on \logU; even saturated
lines can provide limits in \logU. With the ionization parameter determined, we obtain the average
volume density of the cloud and using the ionization fraction (from the Cloudy output) we can then
calculate a length scale for the cloud. These numbers are, however, dependent on the strength of the
UV background ionization field, which is uncertain. When relevant in the following analysis, we give
a range of parameters for a range of UV background intensities $J_{\nu, 912\,\mathrm{\AA}} = 2 - 7
\times 10^{-23}\,\mathrm{erg\,s^{-1}\,cm^{-2}\,Hz^{-1}\,sr^{-1}}$
\citep{shull-etal-99-UV-background, scott-etal-02-UV-background}.

We address several systems below. Many of these systems have already been studied in detail in terms
of their ionization conditions, characteristic sizes and/or average densities. In these cases, we do
not comment unless we have additional information to add, beyond what has been published. The
interested reader is encouraged to see the following references for more information:
\citet[][HE\,0226$-$4110 $z = 0.20702$]{savage-etal-05-HE0226-4110-NeVIII}; \citet[][HE\,0226$-$4110
$z = 0.34034$]{lehner-etal-06-HE0226-4110}; \citet[][PG\,0953+415 $z =
0.14232$]{tripp-savage-00-PG0953+415_z0.1423}; \citet[][PG\,1116+215 $z =
0.13847$]{sembach-etal-04-PG1116+215}; \citet[][PKS\,0405$-$12 $z = 0.16703,
0.36333$]{chen-prochaska-00-PKS0405-12_z0.167}.  Note that all our {\it Cloudy} models assume a
solar abundance pattern. If there are significant deviations from this pattern (which may be
particularly problematic for C or N), then we will deduce incorrect limits. Hence, the more metal
transitions we have for a system, the better our results will be.

\noindent H\,1821+643; $z=0.22496$---This system was first examined by
\citet{tripp-etal-00-H1821+643}, who report $N$ for \OVI\ absorbers on this sightline, but were
primarily interested in the cosmological mass density of \OVI, and did not discuss the ionization
condition of individual absorbers.  As noted in Sec~\ref{subsec: H1821+643} the \OVI\ and \SiIII\
Doppler parameters for the $v = 0\kms$ component are inconsistent with a single phase, since $\bOVI
\gg \bSiIII$ and $\bOVI \gg \bCIII$. For this $v=0\kms$ component, we also detect weak \SiIV, but no
\HI\ components are aligned with the \OVI\ position. The \SiIII\ to \SiIV\ column density ratio is
consistent with the \SiIII\ to \CIII\ column density limit, and $\logU \approx -2.1$ ($\loghden
\approx -3.5$). The weaker component at $v \simeq 25\kms$ also contains saturated \CIII, \SiIII\ and
\SiIV. This component also falls within the width of the \OVI\ line, but the centroids are
significantly different. The nearest \HI\ component to this low-ionization gas is at $v = 15\kms$,
and the relation between the two is unclear.  For this low-ion $v \simeq 25\kms$ component, \SiIII\
and \SiIV\ have the same column density, while log\,N(\CIII)/N(\SiIII) is $\geq 1.0\dex$, both of
which are consistent with $\logU \approx -1.4$ ($\loghden \approx -4.2$). There are no obvious
matches to the weak \OVI\ component at $v = 60\kms$. The mismatch of the broad \OVI\ and low ion
Doppler parameters, in addition to the multiple low-ion components falling within the \OVI\
absorption range, argues for a core-halo cloud model, with a hot gaseous halo traced by \OVI,
surrounding cooler gas traced by lower ionization species.

\noindent PG\,1216+069; $z = 0.28232$---\citet{tripp-etal-05-PG1216+069_NGC4261} were the first to
publish STIS data of PG\,1216+069, but concentrate on the DLA at $z = 0.00632$ associated with the
NGC\,4261 group, and do not consider this absorber in detail.  We detect \OVI\, \SiIII\ and
saturated \CIII.  The \CIII, \SiIII\ and \HI\ are offset from the \OVI\ position by $\Delta\,v =
6-9\kms$ to the blue, and its association is unclear. We measure $\log\,N(\CIII)/N(\SiIII) > 1.3$
which gives the limit $\logU > -2.6$ ($\loghden \simlt -2.9; L > 5\kpc$, where we have folded in the
range in $J_{\nu}$ into the quoted limits).

\noindent PG\,1259+593; $z = 0.21950$---Data for the PG\,1259+593 sightline was published by
\citet{richter-etal-04-PG1259+593}, who report the detection of \OIII\ in FUSE data, in addition to
\OVI, \SiIII\ and (saturated) \CIII\ metal lines seen in the STIS data. They conclude that the
absorber is a multi-phase medium, based on the inconsistency of the \SiIII\ data with an ionization
parameter derived using \OIII\ and \OVI. This conclusion is emphasised by our discussion above of
the limits placed by the ratio of Doppler parameters. In our STIS data we measure
$\bOVI/\bSiIII = 1.6 \pm 0.6$ which is inconsistent at the $1\,\sigma$ level with a single
gas phase, since $m_{\scOVI}/m_{\scSiIII} = 0.6$ i.e. for these species to arise from the same gas
phase, we must have $0.6 < \bOVI/\bSiIII < 1.0$, irrespective of the relative
contributions of \bnt\ and \btherm.

\noindent PKS\,0312$-$77; $z = 0.20275$---While the Lyman limit system at $z=0.20275$ in the sightline
towards PKS\,0312$-$77 shows many metal absorption lines, particularly those of low ions, none can be
unambiguously matched with the \OVI\ absorption. The main \OVI\ absorption component is very broad
compared to other ions; even the highly ionized species like \SiIV\ and \NV\ exhibit a complex
multi-component structure that is seen in the low ionization species. It is therefore likely that
\OVI\ arises in a different gas phase, or that there are unresolved \OVI\ absorption components.

\noindent PKS\,0405$-$12; $z = 0.49514$---The highest redshift \OVI\ absorber in our survey, this
absorber is fit with a single \OVI\ component. We detect \OIII\ and \OIV\ at $v = 90\kms$, but no
associated \OVI\ or \HI, and we do not consider this component in detail. At $v = -10\kms$, the data
clearly show \CIII\ and \OIV\,787. We discount the \OIII\,832 absorption at $v = -29\pm8\kms$ as too
far from the \OVI\ centroid. \citet{prochaska-etal-04-PKS0405-12} do not provide a full analysis of
this system, referring instead to a future paper which was not forthcoming. From our \OVI, \OIV\ and
\CIII\ measurements, we find $\logU \simeq -0.8 - -1.3$ ($\loghden \approx -4.0 - -4.5; L \approx 12
- 140\kpc$).

\subsection{Systems lacking \HI}
\label{subsec: noHI}
In a survey for \OVI\ absorbers, we might expect to find a class of systems that do not contain
associated \HI.  If large-scale galactic winds and superwinds shock-heat metals to a WHIM phase,
they may also ionize hydrogen to a level that is not detectable for weak systems (as in the above
discussion on CIE systems).  Such hot bubbles may exist even without strong winds
\citep{kawata-rauch-07-galactic-winds}. Since we conduct a {\it blind} search for the \OVI\ doublet,
we should be sensitive to the existence of such systems, down to our detection limit, which is
approximately 30\mA\ with good data (but see \pI\ for a better quantification of this, since not all
data are of equal quality). 

Given the above discussion, it may be somewhat telling that we find no strong evidence for such a
class of systems. Our survey contains two system that we suggest are \HI\ free. The $z=0.32639$
absorber toward HE\,0226$-$4110 (Figure~\ref{fig: HE0226-4110_z0.32639}) shows no evidence of \Lya\
or \Lyb\ absorption, although the \Lya\ region is quite noisy. The upper limit on \NHI\ for this
system implies $\log[NOVI/NHI] \simgt 1.1\dex$. The $z = 0.22638$ system toward H\,1821+643
(Figure~\ref{fig: H1821+643_z0.22638}) shows a single broad \HI\ component significantly offset
($\Delta\,v = -53\pm2\kms$) from the \OVI\ position. If we posit that this component could hide a
weak, aligned \HI\ component, and attempt to force-fit an \HI\ component aligned with the \OVI\
absorption, we obtain $\NHI < 12.3$, implying $\log[NOVI/NHI] \simgt 1.2\dex$. Such large column
density ratios are well outside the bounds of what is found for the well-aligned components ($-0.75
< log[\NOVI / \NHI] < 0.67$), and imply either strongly enriched gas (super-solar assuming
photoionization; See Figure~\ref{fig: column_density_component_plots}), or that the gas is
collisionally ionized.  Expanded samples of low-$z$ \OVI\ absorbers will be very valuable for
determining the fraction of such systems, and their potential as tracers of hot gas.

\section{Summary}
\label{sec: discussion}

We have presented a catalogue of \OVI\ absorbers selected from high-resolution HST/STIS echelle
data. Our selection technique followed a blind-search for the \OVI\,1031, 1037 doublet
feature. Relying solely on the presence of the \OVI\ doublet, and the ratio of doublet line
strength, we detect 27 \OVI\ absorption systems, independent of other transitions.  The statistics
of these absorbers were presented in \pI.  We note that 16 systems reported in a contemporaneous
work by \citet{tripp-etal-08-OVI} do not satisfy our selection criteria and therefore are not
included in our sample of 27 absorbers.  In cases where only one transition is found or the doublet
ratio appears to be inconsistent with model expectations, these authors include the presence of
other transitions, such as \Lya\ absorption, for justifying the identifications of \OVI.

In our absorbers, it is common to find multiple absorption components in any given system,
corresponding to physically distinct gas structures. By matching these absorption components from
different species, we can identify different transitions likely due to the same gas. Under this
assumption, we can then derive both the temperature of the gas, and the non-thermal contribution to
the Doppler parameter, \bnt. This analysis demonstrates that, for well-matched \OVI/\HI\ components,
gas temperatures are in the range $\logT = 3.7 - 5.0$, well below the temperature range at which
\OVI\ is expected to be found in collisional ionization equilibrium ($\logT = 5.5$). We thus advise
caution in identifying the \OVI\ absorbers with the hot WHIM gas that simulations predict will
contain a significant fraction of the baryons. Our finding based on all available absorbers in the
current HST/STIS data archive reaffirms previous findings from studies of individual lines of sight.
Future generations of X-ray spectrographs are likely to be necessary to solve this question
conclusively.

If galactic winds are the dominant source of \OVI\ formation, shock heating the gas to the WHIM
regime, and pushing it out to $\sim 1\Mpc$ from galaxies, we may expect to see this in a variety of
ways. A class of \HI\ free \OVI\ absorbers may arise, as both the hydrogen and oxygen are ionized in
strong winds. We see no evidence of a large number of \HI-free \OVI\ absorbers, with only two
systems present in our survey. Reasoning that the \HI\ absorption will best trace the local
over-density, we compared the velocity difference between our \OVI\ components and the nearest \HI\
component. A wind origin may leave an imprint on the kinematics of the absorbers, manifest as a
systematic offset between the \HI\ and \OVI\ positions. We see no such effect (see also comparisons
in T08), but this test is likely a weak one. Clearly systematic galaxy redshift surveys around the
QSO lines of sight are required to properly address this.

The location of the missing baryons in the low-redshift is clearly an important issue in modern
cosmology and astronomy. The WHIM is a leading candidate reservoir for this mass with simulations
and suggesting that it contains as much as 50\% of all $z = 0$ baryons. While some have claimed that
the observations support this view, we argue that caution is required. High-resolution spectra that
cover a broad wavelength range are crucial for constraining the ionization, metallicity, and the
temperature of the gas involved, and care must be taken to deduce non-thermal broadening
mechanisms. If the baryons are to be unambiguously located in the WHIM, further observations are
crucial.

\acknowledgements

This research has made use of the NASA/IPAC Extragalactic Database (NED) which is operated by the
Jet Propulsion Laboratory, California Institute of Technology, under contract with the National
Aeronautics and Space Administration.  C.~T. and H.-W.~C.\ acknowledge support from NASA grant
NNG06GC36G.  H.-W.~C. acknowledges partial support from an NSF grant AST-0607510.

\facility{HST(STIS)}

\bibliography{MASTER}
\bibliographystyle{apj}

\clearpage


\begin{deluxetable}{p{1in}rrrcr}
\tabletypesize{\scriptsize}
\tablecaption{\label{tab: journal_observations}Summary of the STIS Echelle Spectra}
\tablewidth{0pt}
\tablehead{
  \colhead{QSO} & 
  \colhead{$z_{\rm QSO}$} & 
  \colhead{$z_{\rm min}$} & 
  \colhead{$z_{\rm max}$\tablenotemark{a}} & 
  \colhead{$t_{\rm exp}$} & 
  \colhead{PID}  \\
  \colhead{(1)} & 
  \colhead{(2)} & 
  \colhead{(3)} & 
  \colhead{(4)} & 
  \colhead{(5)} & 
  \colhead{(6)}
}
\startdata
3C\,249.1       \dotfill  &  0.3115 &    0.1222 &    0.2885 &   68776 &   9184  \\
3C\,273         \dotfill  &  0.1580 &    0.1144 &    0.1417 &   18671 &   8017  \\
3C\,351.0       \dotfill  &  0.3719 &    0.1309 &    0.3483 &   73198 &   8015  \\
HE\,0226$-$4110 \dotfill  &  0.4950 &    0.1154 &    0.4707 &   43772 &   9184  \\
HS\,0624+6907   \dotfill  &  0.3700 &    0.1222 &    0.3464 &   61950 &   9184  \\
H\,1821+643     \dotfill  &  0.2970 &    0.1144 &    0.2741 &   50932 &   8165  \\
PG\,0953+415    \dotfill  &  0.2390 &    0.1144 &    0.2163 &   24478 &   7747  \\
PG\,1116+215    \dotfill  &  0.1765 &    0.1144 &    0.1536 &   39836 &   8165/8097 \\
PG\,1216+069    \dotfill  &  0.3313 &    0.1309 &    0.3078 &   69804 &   9184  \\
PG\,1259+593    \dotfill  &  0.4778 &    0.1144 &    0.4533 &   95760 &   8695  \\
PG\,1444+407    \dotfill  &  0.2673 &    0.1222 &    0.2442 &   48624 &   9184  \\
PHL\,1811       \dotfill  &  0.1917 &    0.1144 &    0.1690 &   33919 &   9418  \\
PKS\,0312$-$77  \dotfill  &  0.2230 &    0.1241 &    0.1999 &   37908 &   8651  \\
PKS\,0405$-$12  \dotfill  &  0.5726 &    0.1241 &    0.5478 &   27208 &   7576  \\
PKS\,1302$-$102 \dotfill  &  0.2784 &    0.1183 &    0.2558 &   22119 &   8306  \\
Ton\,28         \dotfill  &  0.3297 &    0.1231 &    0.3059 &   48401 &   9184  \\
\enddata
\tablecomments{This summary of STIS echelle lines of sight is taken from \pI. It is repeated
  here for completeness.}
\tablenotetext{a}{The maximum redshift is defined for O\,VI absorbers at velocity separation $>
  5000$ \kms\ from the background QSO, but the line search is conducted through the emission
  redshift of the QSO. In this paper we do not consider system inside this 5000\kms limit.}
\end{deluxetable}

\input{tab2}

\input{tab3.tex}

\input{tab4}

\clearpage

\input{tab5}

\input{tab6}
   \clearpage
                                 
\clearpage

\input{tab7}

\input{tab8}

\input{tab9}

\input{tab10}

\clearpage

\input{tab11}

\input{tab12}

\input{tab13}
    \clearpage
                      

\input{tab14}

\clearpage

\input{tab15}

\input{tab16}

\input{tab17}

\begin{deluxetable}{p{1in}rrr}
\tabletypesize{\scriptsize}
\tablecaption{\label{tab: unconfirmed_systems}Systems not confirmed by our search}
\tablewidth{0pt}
\tablehead{
  \colhead{QSO} & 
  \colhead{$z_{abs}$} & 
  \colhead{\NOVI} &
  \colhead{\NHI} \\
  \colhead{(1)} & 
  \colhead{(2)} & 
  \colhead{(3)} & 
  \colhead{(4)}
}
\startdata
3C\,249.1        \dotfill  &   0.23641  & 13.99 & 13.23                    \\
3C\,351.0        \dotfill  &   0.21811  & 13.96 & 13.50                    \\
3C\,351.0        \dotfill  &   0.22111  & 14.25 & $>$14.56                 \\
H\,1821+643      \dotfill  &   0.12143  & 13.94 & 14.31                    \\
H\,1821+643      \dotfill  &   0.21331  & 13.57 & $>$14.29                 \\           
HE\,0226$-$4110  \dotfill  &   0.42670  & 14.34 & \nodata\tablenotemark{a} \\
PG\,1216+069     \dotfill  &   0.26768  & 13.30 & $>$13.86                 \\
PG\,1259+593     \dotfill  &   0.31972  & 13.92 & 14.04                    \\
PG\,1444+407     \dotfill  &   0.22032  & 13.95 & 13.63                    \\
PHL\,1811        \dotfill  &   0.13240  & 14.39 & $>$14.30                 \\
PKS\,0312$-$77   \dotfill  &   0.15890  & 13.94 & 13.90                    \\
PKS\,0312$-$77   \dotfill  &   0.19827  & 13.84 & $>$14.26                 \\
PKS\,0405$-$12   \dotfill  &   0.36156  & 14.00 & $>$14.14                 \\
PKS\,1302$-$102  \dotfill  &   0.19159  & 13.93 & $>$14.17                 \\
Ton\,28          \dotfill  &   0.13783  & 13.99 & $>$14.37                 \\
Ton\,28          \dotfill  &   0.20524  & 13.69 & 13.20                    \\           
\enddata
\tablecomments{Summary of \scOVI\ candidates which we do not recover. See text for details on
  individual systems.}

\tablenotetext{a}{\citet{tripp-etal-08-OVI} report no \scHI\ with this system. \Lya\ is redshifted out
  of the STIS band, and \Lyb\ is not detected.}

\end{deluxetable}

\begin{figure}
  \plotone{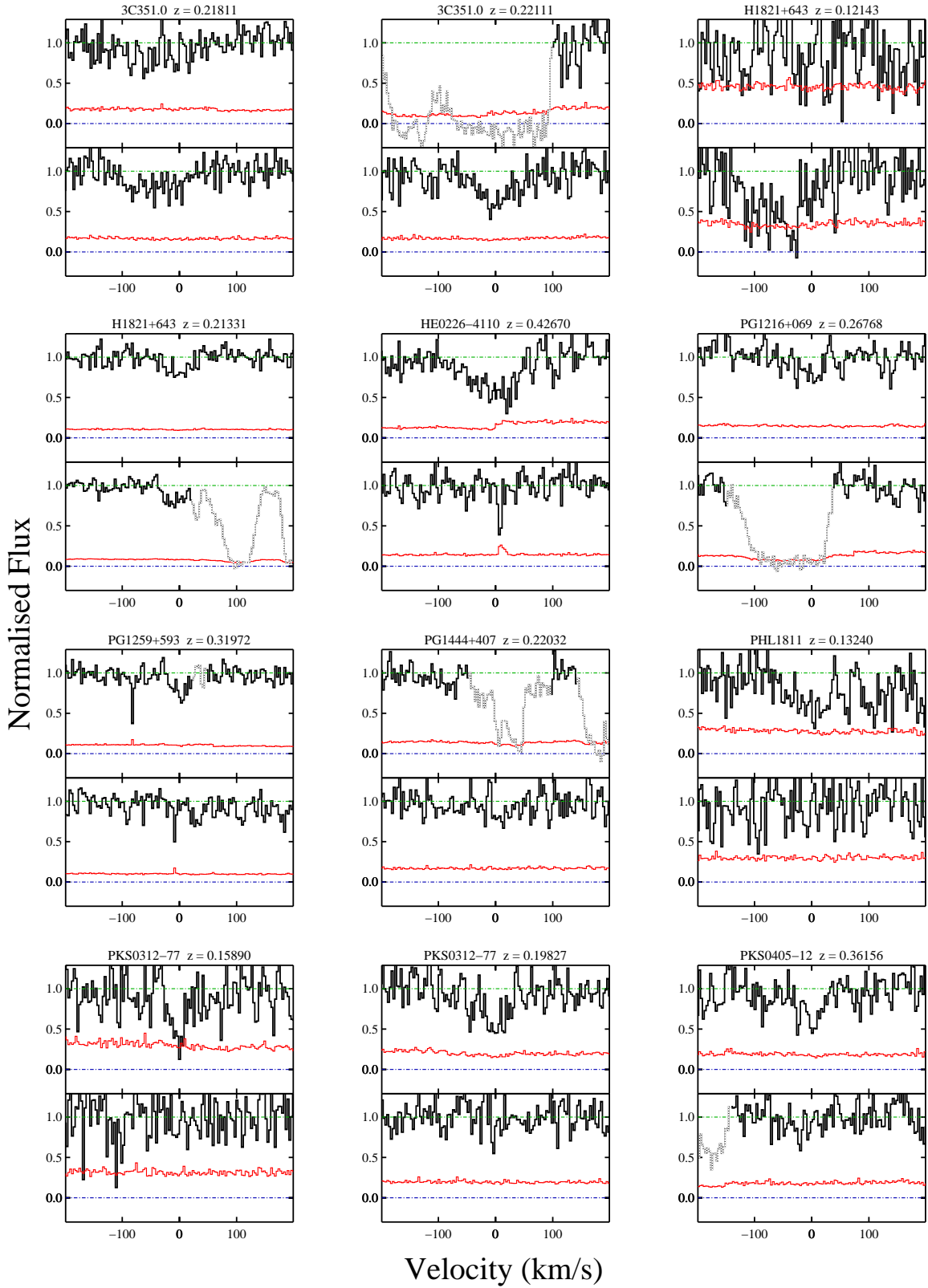}
  \caption{Spectra of the \OVI\,1031 (top) and \OVI\,1037 (bottom) spectral regions for absorbers
    we do not confirm. Blended regions are shown by dotted spectral regions.}
  \label{fig: OVI_rejected_1}
\end{figure}

\begin{figure}
  \includegraphics[angle=0, scale=0.8]{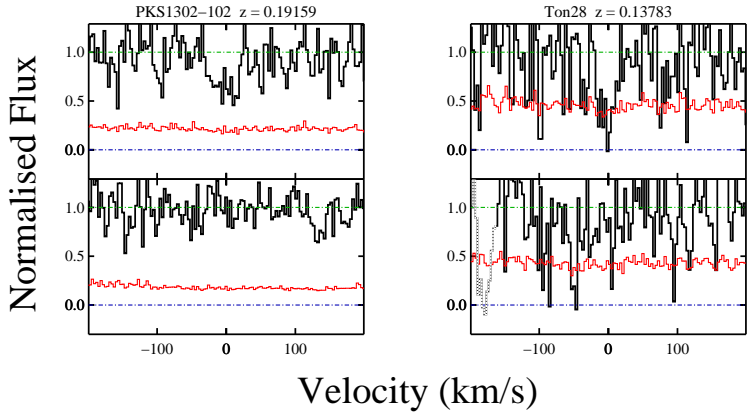}
  \caption{As for Figure~\ref{fig: OVI_rejected_1}, but remaining systems not shown in that figure.}
  \label{fig: OVI_rejected_2}
\end{figure}

\begin{figure}
  \plotone{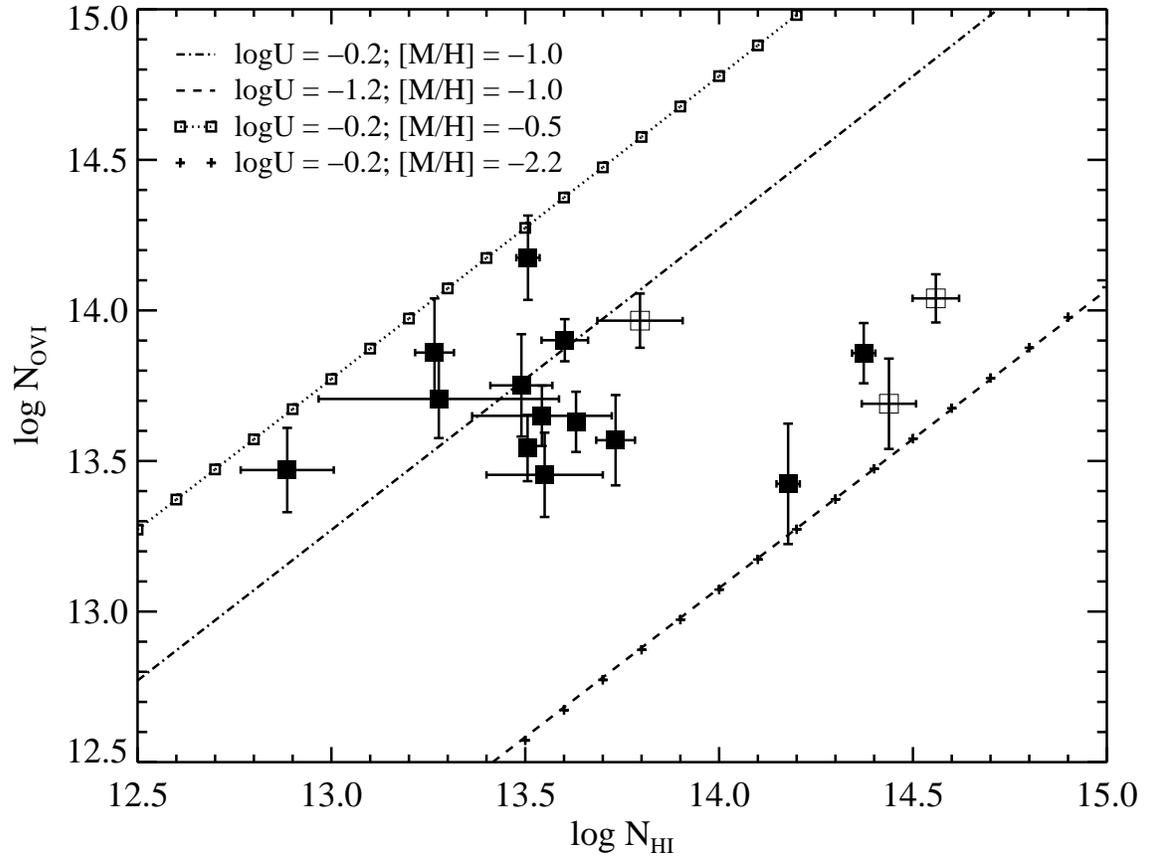}
  \caption{Distribution of \OVI\ and \HI\ column densities for well aligned absorption components.
    Full symbols are regular components, while open circles designate components whose alignment is
    flagged as uncertain.}
  \label{fig: column_density_component_plots}
\end{figure}

\begin{figure}
  \epsscale{1.0} 
  \plotone{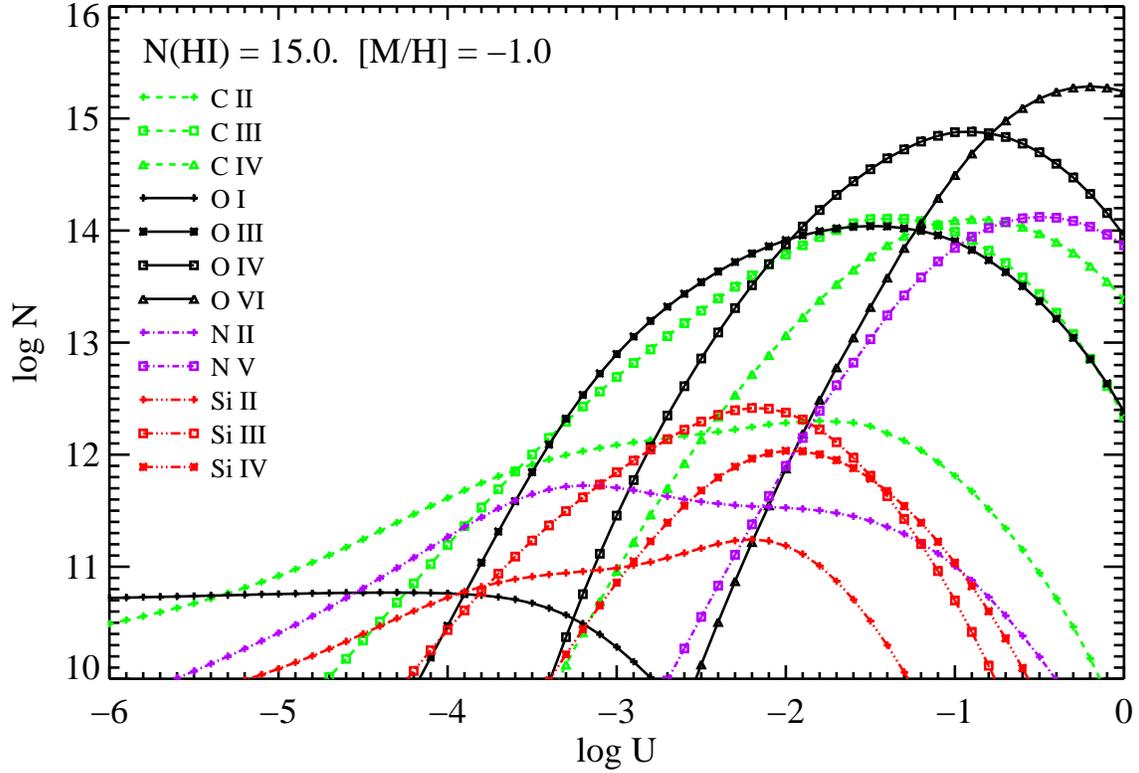}
  \caption{Example of column densities as a function of ionization parameter derived from {\it
      Cloudy} photoionization models. With increasing ionization parameter, species become more
    highly ionized, as expected. The models were computed for \logN = 15.0, and a nominal abundance
    $[{\rm M}/{\rm H}] = -1.0$. \OVI\ dominates the Oxygen ionization at $\logU > -0.8$ and peaks at
    $\logU = -0.2$.}
  \label{fig: photo_model}
\end{figure}

\begin{figure}
  \plotone{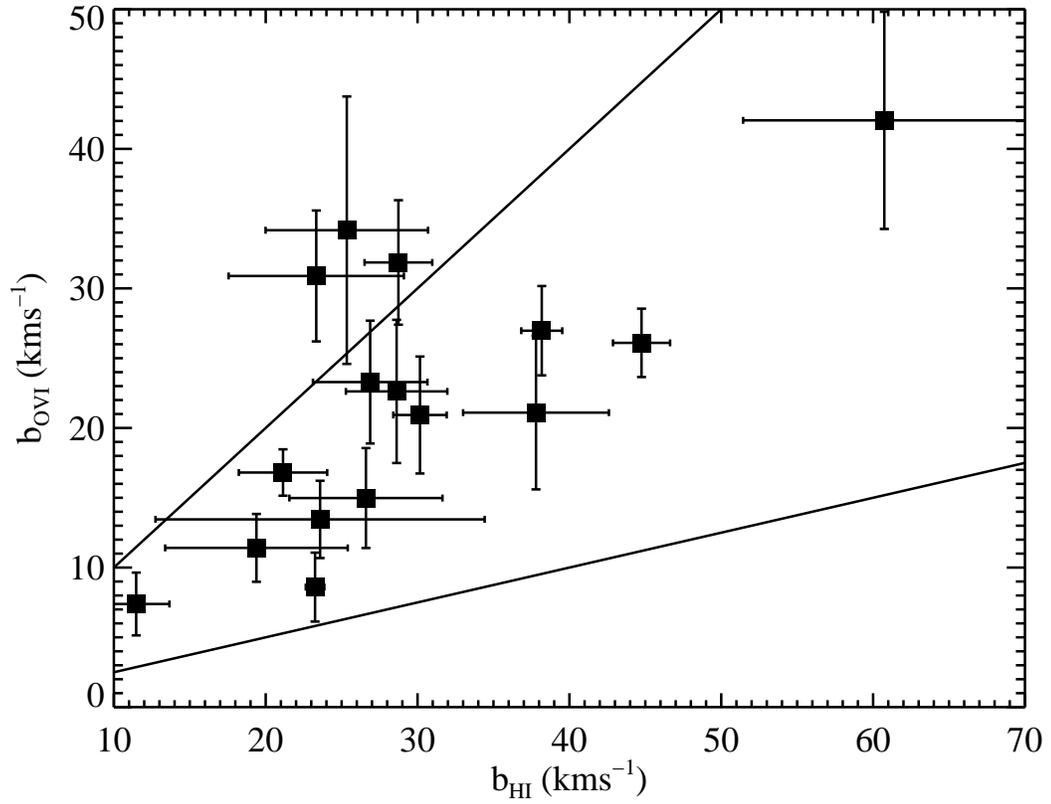}
  \caption{Distribution of Doppler parameters for aligned components. Symbols are the same as for
    Fig~\ref{fig: column_density_component_plots}. The lower line shows the boundary for $\bnt = 0$,
    below which we should not see any components. The upper line shows $\bOVI = \bHI$, where the
    Doppler parameter is dominated by turbulent motions.}
  \label{fig: doppler_parameter_component_plots}
\end{figure}

\input{tab19}

\end{document}

%% file: tab2.tex
\begin{deluxetable}{lrrrrrrrrr}
\tabletypesize{\scriptsize}
\tablecaption{\label{tab: ovi_absorbers}Summary of \OVI\ absorbers}
\tablewidth{0pt}
\tablehead{
\colhead{QSO} &
\colhead{$z_{\rm abs}$} & 
\colhead{$W_r$ (1031)}  &
\colhead{$\sigma_{W_r (1031)}$} &
\colhead{$W_r$ (1037)}  &
\colhead{$\sigma_{W_r (1037)}$}  &
\colhead{$R_{\tabOVI}$} &
\colhead{$\sigma_{R_{\tabOVI}}$} &
\colhead{$N_{\scOVI}$} &
\colhead{$\sigma_{N_{\scOVI}}$} \\
\colhead{(1)} & 
\colhead{(2)} & 
\colhead{(3)} & 
\colhead{(4)} &
\colhead{(5)} & 
\colhead{(6)} &
\colhead{(7)} & 
\colhead{(8)} &
\colhead{(9)} &
\colhead{(10)} 
 }
\startdata
     3C\,249.1   &  0.24676 &   72.1 &    9.0 &   31.0 &    9.0 &   2.33 &   0.74 &   13.9 &    0.1 \\
       3C\,273   &  0.12003 &   22.4 &    5.6 &   21.5 &    5.6 &   1.04 &   0.38 &   13.5 &    0.1 \\
     3C\,351.0   &  0.31659 &  232.5 &   13.8 &  139.8 &   13.8 &   1.66 &   0.19 &   14.4 &    0.1 \\
   H\,1821+643   &  0.22498 &  162.2 &    6.7 &  108.5 &    6.7 &   1.49 &   0.11 &   14.3 &    0.0 \\
   H\,1821+643   &  0.22638 &   29.8 &    3.9 &   19.7 &    3.9 &   1.51 &   0.36 &   13.5 &    0.1 \\
   H\,1821+643   &  0.24532 &   53.9 &    4.9 &   36.0 &    4.9 &   1.50 &   0.25 &   13.7 &    0.1 \\
   H\,1821+643   &  0.26656 &   44.2 &    4.0 &   23.7 &    4.0 &   1.86 &   0.36 &   13.6 &    0.1 \\
 HE\,0226$-$4110 &  0.20702 &  165.3 &    8.9 &  106.5 &    8.9 &   1.55 &   0.15 &   14.4 &    0.1 \\
 HE\,0226$-$4110 &  0.32639 &   43.0 &    8.0 &   15.8 &    8.0 &   2.72 &   1.47 &   13.6 &    0.2 \\
 HE\,0226$-$4110 &  0.34034 &   62.1 &    6.6 &   41.5 &    6.6 &   1.50 &   0.29 &   13.9 &    0.1 \\
 HE\,0226$-$4110 &  0.35529 &   45.5 &    7.8 &   11.5 &    7.8 &   3.96 &   2.77 &   13.6 &    0.2 \\
 HS\,0624+6907   &  0.31796 &   43.7 &    6.0 &   26.9 &    6.0 &   1.62 &   0.43 &   13.7 &    0.1 \\
 HS\,0624+6907   &  0.33984 &   27.1 &    8.3 &   22.5 &    8.3 &   1.20 &   0.58 &   13.4 &    0.3 \\
  PG\,0953+415   &  0.14232 &  121.2 &   22.3 &   89.1 &   22.3 &   1.36 &   0.42 &   14.2 &    0.1 \\
  PG\,1116+215   &  0.13846 &   75.7 &   16.0 &   38.0 &   16.0 &   1.99 &   0.94 &   13.9 &    0.1 \\
  PG\,1216+069   &  0.28232 &   26.4 &    5.6 &   16.9 &    5.6 &   1.56 &   0.61 &   13.4 &    0.2 \\
  PG\,1259+593   &  0.21950 &   98.9 &    7.6 &   21.1 &    7.6 &   4.68 &   1.73 &   13.9 &    0.1 \\
  PG\,1259+593   &  0.25981 &   77.0 &    8.3 &   34.1 &    8.3 &   2.26 &   0.60 &   13.9 &    0.1 \\
     PHL\,1811   &  0.15786 &   63.5 &   13.2 &   39.2 &   13.2 &   1.62 &   0.64 &   13.9 &    0.2 \\
  PKS\,0312$-$77 &  0.20275 &  655.3 &   20.1 &  336.5 &   20.1 &   1.95 &   0.13 &   15.0 &    0.2 \\
  PKS\,0405$-$12 &  0.16697 &  360.7 &   41.4 &  236.7 &   41.4 &   1.52 &   0.32 &   14.7 &    0.1 \\
  PKS\,0405$-$12 &  0.18291 &  103.7 &   14.7 &   77.5 &   14.7 &   1.34 &   0.32 &   14.2 &    0.2 \\
  PKS\,0405$-$12 &  0.36333 &   30.4 &    4.8 &   12.4 &    4.8 &   2.45 &   1.02 &   13.5 &    0.1 \\
  PKS\,0405$-$12 &  0.49506 &  212.6 &   15.6 &  151.0 &   15.6 &   1.41 &   0.18 &   14.5 &    0.1 \\
 PKS\,1302$-$102 &  0.22565 &   70.9 &    9.3 &   64.4 &    9.3 &   1.10 &   0.21 &   13.9 &    0.1 \\
 PKS\,1302$-$102 &  0.22744 &   30.3 &    5.5 &   14.4 &    5.5 &   2.10 &   0.89 &   13.5 &    0.1 \\
       Ton\,28   &  0.27340 &   25.6 &    7.0 &   18.1 &    7.0 &   1.41 &   0.67 &   13.4 &    0.2 \\
\enddata 
\tablecomments{Redshift of absorbers is generally the position of the strongest
  component. Equivalent widths are rest-frame \mA. Column densities are total column 
  density for absorber, taken from component fitting. See individual systems for details.}
\end{deluxetable}

%% file: tab3.tex
\begin{deluxetable}{lrrrrrrrc}
\tabletypesize{\tiny}
\tablecaption{\label{tab: 3C249.1_transitions}\OVI\ absorber measurements along the line of sight towards 3C\,249.1}
\tablewidth{0pt}
\tablecolumns{9}
\tablehead{
\colhead{ion} & 
\colhead{$z_{\mbox{comp}}$} & 
\colhead{$v$} &
\colhead{$\sigma_v$} &
\colhead{$b$} & 
\colhead{$\sigma_b$} &
\colhead{$\logN$} & 
\colhead{$\sigma_{\logN}$} &
\colhead{Flag} \\
\colhead{} &
\colhead{} &
\colhead{$\kms$} &
\colhead{$\kms$} &
\colhead{$\kms$} &
\colhead{$\kms$} &
\colhead{log($\cm$)} &
\colhead{log($\cm$)} &
\colhead{}
}
\startdata

\cutinhead{$z = 0.24676$}
    \tabOVI &  0.24676 &    0 &    0 &   27.0 &    3.2 &   13.9 &       0.1 &   \nodata  \\
     \tabHI &  0.24676 &    0 &    1 &   38.2 &    1.4 &   14.4 &       0.1 &   \nodata  \\
  \tabSiIII &  0.24681 &   11 &    3 &    9.9 &    5.6 &   12.1 &       0.2 &         Z  \\

\enddata
\end{deluxetable}

%% file: tab4.tex
\begin{deluxetable}{lrrrrrrrc}
\tabletypesize{\tiny}
\tablecaption{\label{tab: 3C273_transitions}\OVI\ absorber measurements along the line of sight towards 3C\,273}
\tablewidth{0pt}
\tablecolumns{9}
\tablehead{
\colhead{ion} & 
\colhead{$z_{\mbox{comp}}$} & 
\colhead{$v$} &
\colhead{$\sigma_v$} &
\colhead{$b$} & 
\colhead{$\sigma_b$} &
\colhead{$\logN$} & 
\colhead{$\sigma_{\logN}$} &
\colhead{Flag} \\
\colhead{} &
\colhead{} &
\colhead{$\kms$} &
\colhead{$\kms$} &
\colhead{$\kms$} &
\colhead{$\kms$} &
\colhead{log($\cm$)} &
\colhead{log($\cm$)} &
\colhead{}
}
\startdata

\cutinhead{$z = 0.12003$}
    \tabOVI &  0.12003 &    0 &    1 &    8.6 &    2.5 &   13.5 &       0.1 &   \nodata  \\
     \tabHI &  0.12004 &    3 &    0 &   23.2 &    0.6 &   13.5 &       0.1 &   \nodata  \\

\enddata
\end{deluxetable}

%% file: tab5.tex
\begin{deluxetable}{lrrrrrrrc}
\tabletypesize{\tiny}
\tablecaption{\label{tab: 3C351.0_transitions}\OVI\ absorber measurements along the line of sight towards 3C\,351.0}
\tablewidth{0pt}
\tablecolumns{9}
\tablehead{
\colhead{ion} & 
\colhead{$z_{\mbox{comp}}$} & 
\colhead{$v$} &
\colhead{$\sigma_v$} &
\colhead{$b$} & 
\colhead{$\sigma_b$} &
\colhead{$\logN$} & 
\colhead{$\sigma_{\logN}$} &
\colhead{Flag} \\
\colhead{} &
\colhead{} &
\colhead{$\kms$} &
\colhead{$\kms$} &
\colhead{$\kms$} &
\colhead{$\kms$} &
\colhead{log($\cm$)} &
\colhead{log($\cm$)} &
\colhead{}
}
\startdata

\cutinhead{$z = 0.31659$}
    \tabOVI &  0.31635 &  -53 &    4 &   21.1 &    5.5 &   13.7 &       0.2 &   \nodata  \\
    \tabOVI &  0.31657 &   -3 &    2 &   23.3 &    4.4 &   14.0 &       0.1 &   \nodata  \\
    \tabOVI &  0.31686 &   62 &    3 &   30.9 &    4.7 &   14.0 &       0.1 &   \nodata  \\
     \tabHI &  0.31633 &  -58 &    5 &   37.8 &    4.8 &   14.4 &       0.1 &   \nodata  \\
     \tabHI &  0.31660 &    2 &    2 &   26.9 &    3.8 &   14.6 &       0.1 &   \nodata  \\
     \tabHI &  0.31686 &   62 &    5 &   23.3 &    5.8 &   13.8 &       0.1 &   \nodata  \\

\enddata
\end{deluxetable}

%% file: tab6.tex
\begin{deluxetable}{lrrrrrrrc}
\tabletypesize{\tiny}
\tablecaption{\label{tab: H1821+643_transitions}\OVI\ absorber measurements along the line of sight towards H\,1821+643}
\tablewidth{0pt}
\tablecolumns{9}
\tablehead{
\colhead{ion} & 
\colhead{$z_{\mbox{comp}}$} & 
\colhead{$v$} &
\colhead{$\sigma_v$} &
\colhead{$b$} & 
\colhead{$\sigma_b$} &
\colhead{$\logN$} & 
\colhead{$\sigma_{\logN}$} &
\colhead{Flag} \\
\colhead{} &
\colhead{} &
\colhead{$\kms$} &
\colhead{$\kms$} &
\colhead{$\kms$} &
\colhead{$\kms$} &
\colhead{log($\cm$)} &
\colhead{log($\cm$)} &
\colhead{}
}
\startdata

\cutinhead{$z = 0.22496$}
    \tabOVI &  0.22496 &    0 &    1 &   45.3 &    2.0 &   14.3 &       0.1 &   \nodata  \\
    \tabOVI &  0.22521 &   62 &    1 &    9.6 &    2.7 &   13.2 &       0.1 &   \nodata  \\
     \tabHI &  0.22453 & -104 &    4 &   34.5 &    3.2 &   14.0 &       0.1 &   \nodata  \\
     \tabHI &  0.22479 &  -41 &    1 &   24.9 &    1.8 &   15.2 &   \nodata &         L  \\
     \tabHI &  0.22502 &   15 &    1 &   21.6 &    0.8 &   15.2 &   \nodata &         L  \\
     \tabHI &  0.22526 &   74 &    4 &   15.1 &    5.1 &   12.8 &       0.1 &   \nodata  \\
   \tabCIII &  0.22493 &   -6 &    1 &   15.4 &    3.0 &   13.7 &   \nodata &         L  \\
   \tabCIII &  0.22506 &   23 &    0 &    5.2 &    2.5 &   13.5 &   \nodata &         L  \\
   \tabCIII &  0.22517 &   50 &    4 &   20.8 &   12.0 &   13.0 &       0.2 &         Z  \\
   \tabCIII &  0.22528 &   78 &    2 &    4.5 &    5.0 &   12.3 &       0.3 &         Z  \\
   \tabCIII &  0.22479 &  -41 &    1 &   13.1 &    2.0 &   13.4 &       0.1 &   \nodata  \\
   \tabCIII &  0.22466 &  -73 &    2 &    5.8 &    4.5 &   12.2 &       0.2 &         Z  \\
   \tabCIII &  0.22538 &  102 &    3 &    9.7 &    6.0 &   12.3 &       0.2 &         Z  \\
  \tabSiIII &  0.22496 &    0 &    0 &    9.1 &    0.8 &   12.5 &       0.1 &   \nodata  \\
  \tabSiIII &  0.22507 &   26 &    0 &    4.5 &    0.7 &   12.4 &       0.1 &   \nodata  \\
  \tabSiIII &  0.22478 &  -43 &    1 &    6.8 &    2.1 &   11.9 &       0.1 &   \nodata  \\
   \tabSiIV &  0.22496 &    0 &    3 &    5.2 &    5.8 &   12.0 &       0.3 &         Z  \\
   \tabSiIV &  0.22508 &   30 &    2 &    9.9 &    4.2 &   12.4 &       0.1 &         Z  \\
\cutinhead{$z = 0.22638$}
    \tabOVI &  0.22638 &    0 &    1 &   16.5 &    1.8 &   13.5 &       0.1 &   \nodata  \\
     \tabHI &  0.22616 &  -53 &    2 &   53.1 &    3.4 &   13.5 &       0.1 &   \nodata  \\
\cutinhead{$z = 0.24532$}
    \tabOVI &  0.24525 &  -17 &    4 &   17.4 &    4.6 &   13.4 &       0.2 &   \nodata  \\
    \tabOVI &  0.24536 &   10 &    2 &   15.1 &    3.5 &   13.4 &       0.2 &   \nodata  \\
     \tabHI &  0.24530 &   -5 &    3 &   43.2 &    5.2 &   13.2 &       0.1 &   \nodata  \\
     \tabNV &  0.24536 &   10 &    0 &   19.4 &    5.2 &   13.0 &       0.1 &         Z  \\
\cutinhead{$z = 0.26656$}
    \tabOVI &  0.26656 &    0 &    1 &   26.1 &    2.5 &   13.6 &       0.1 &   \nodata  \\
     \tabHI &  0.26658 &    4 &    1 &   44.8 &    1.9 &   13.6 &       0.1 &   \nodata  \\
   \tabCIII &  0.26656 &    1 &    0 &   30.1 &    0.0 &   12.2 &   \nodata &         U  \\

\enddata
\end{deluxetable}

%% file: tab7.tex
\begin{deluxetable}{lrrrrrrrc}
\tabletypesize{\tiny}
\tablecaption{\label{tab: HE0226-4110_transitions}\OVI\ absorber measurements along the line of sight towards HE\,0226$-$4110}
\tablewidth{0pt}
\tablecolumns{9}
\tablehead{
\colhead{ion} & 
\colhead{$z_{\mbox{comp}}$} & 
\colhead{$v$} &
\colhead{$\sigma_v$} &
\colhead{$b$} & 
\colhead{$\sigma_b$} &
\colhead{$\logN$} & 
\colhead{$\sigma_{\logN}$} &
\colhead{Flag} \\
\colhead{} &
\colhead{} &
\colhead{$\kms$} &
\colhead{$\kms$} &
\colhead{$\kms$} &
\colhead{$\kms$} &
\colhead{log($\cm$)} &
\colhead{log($\cm$)} &
\colhead{}
}
\startdata

\cutinhead{$z = 0.20702$}
    \tabOVI &  0.20702 &    0 &    1 &   33.0 &    1.8 &   14.4 &       0.1 &   \nodata  \\
     \tabHI &  0.20695 &  -18 &    2 &   23.7 &    2.7 &   15.0 &   \nodata &         L  \\
     \tabHI &  0.20706 &   10 &   15 &   35.9 &    6.7 &   14.7 &   \nodata &         L  \\
   \tabCIII &  0.20697 &  -12 &    1 &   28.0 &    2.8 &   13.9 &   \nodata &         L  \\
     \tabNV &  0.20712 &   25 &    2 &   11.3 &    3.4 &   13.2 &       0.1 &         Z  \\
     \tabNV &  0.20698 &  -10 &    2 &    9.6 &    3.5 &   13.1 &       0.1 &         Z  \\
  \tabSiIII &  0.20699 &   -6 &    2 &    7.7 &    4.8 &   12.1 &       0.2 &   \nodata  \\
  \tabSiIII &  0.20692 &  -24 &    1 &    6.0 &    2.0 &   12.3 &       0.1 &   \nodata  \\
\cutinhead{$z = 0.32639$}
    \tabOVI &  0.32639 &    0 &    3 &   24.8 &    4.8 &   13.6 &       0.2 &   \nodata  \\
     \tabHI &  0.32639 &    0 &    0 &   27.9 &    0.0 &   12.5 &   \nodata &         U  \\
\cutinhead{$z = 0.34034$}
    \tabOVI &  0.34035 &    2 &    1 &   16.8 &    1.7 &   13.9 &       0.1 &   \nodata  \\
     \tabHI &  0.34033 &   -2 &    2 &   21.1 &    2.9 &   13.6 &       0.1 &         Z  \\
     \tabHI &  0.34054 &   45 &    2 &    5.4 &    4.3 &   12.7 &       0.2 &   \nodata  \\
   \tabCIII &  0.34033 &   -2 &    1 &    6.7 &    2.3 &   12.5 &       0.1 &   \nodata  \\
\cutinhead{$z = 0.35525$}
    \tabOVI &  0.35525 &    0 &    5 &   22.6 &    5.1 &   13.6 &       0.2 &   \nodata  \\
     \tabHI &  0.35524 &   -3 &    2 &   28.6 &    3.3 &   13.7 &       0.1 &   \nodata  \\

\enddata
\end{deluxetable}

%% file: tab8.tex
\begin{deluxetable}{lrrrrrrrc}
\tabletypesize{\tiny}
\tablecaption{\label{tab: HS0624+6907_transitions}\OVI\ absorber measurements along the line of sight towards HS\,0624+6907}
\tablewidth{0pt}
\tablecolumns{9}
\tablehead{
\colhead{ion} & 
\colhead{$z_{\mbox{comp}}$} & 
\colhead{$v$} &
\colhead{$\sigma_v$} &
\colhead{$b$} & 
\colhead{$\sigma_b$} &
\colhead{$\logN$} & 
\colhead{$\sigma_{\logN}$} &
\colhead{Flag} \\
\colhead{} &
\colhead{} &
\colhead{$\kms$} &
\colhead{$\kms$} &
\colhead{$\kms$} &
\colhead{$\kms$} &
\colhead{log($\cm$)} &
\colhead{log($\cm$)} &
\colhead{}
}
\startdata

\cutinhead{$z = 0.31796$}
    \tabOVI &  0.31796 &    0 &    2 &   24.0 &    3.1 &   13.7 &       0.1 &   \nodata  \\
     \tabHI &  0.31789 &  -15 &    4 &   35.9 &    6.5 &   13.3 &       0.1 &   \nodata  \\
\cutinhead{$z = 0.33984$}
    \tabOVI &  0.33984 &    0 &    7 &   34.8 &   11.6 &   13.4 &       0.3 &         Z  \\
     \tabHI &  0.33978 &  -14 &    0 &   40.9 &    0.9 &   14.5 &       0.1 &   \nodata  \\

\enddata
\end{deluxetable}

%% file: tab9.tex
\begin{deluxetable}{lrrrrrrrc}
\tabletypesize{\tiny}
\tablecaption{\label{tab: PG0953+415_transitions}\OVI\ absorber measurements along the line of sight towards PG\,0953+415}
\tablewidth{0pt}
\tablecolumns{9}
\tablehead{
\colhead{ion} & 
\colhead{$z_{\mbox{comp}}$} & 
\colhead{$v$} &
\colhead{$\sigma_v$} &
\colhead{$b$} & 
\colhead{$\sigma_b$} &
\colhead{$\logN$} & 
\colhead{$\sigma_{\logN}$} &
\colhead{Flag} \\
\colhead{} &
\colhead{} &
\colhead{$\kms$} &
\colhead{$\kms$} &
\colhead{$\kms$} &
\colhead{$\kms$} &
\colhead{log($\cm$)} &
\colhead{log($\cm$)} &
\colhead{}
}
\startdata

\cutinhead{$z = 0.14232$}
    \tabOVI &  0.14232 &   -1 &    3 &   31.9 &    4.5 &   14.2 &       0.1 &   \nodata  \\
     \tabHI &  0.14232 &    1 &    2 &   28.7 &    2.2 &   13.5 &       0.1 &   \nodata  \\
    \tabCII &  0.14233 &    2 &    3 &    8.3 &    4.6 &   12.8 &       0.2 &         Z  \\

\enddata
\end{deluxetable}

%% file: tab10.tex
\begin{deluxetable}{lrrrrrrrc}
\tabletypesize{\tiny}
\tablecaption{\label{tab: PG1116+215_transitions}\OVI\ absorber measurements along the line of sight towards PG\,1116+215}
\tablewidth{0pt}
\tablecolumns{9}
\tablehead{
\colhead{ion} & 
\colhead{$z_{\mbox{comp}}$} & 
\colhead{$v$} &
\colhead{$\sigma_v$} &
\colhead{$b$} & 
\colhead{$\sigma_b$} &
\colhead{$\logN$} & 
\colhead{$\sigma_{\logN}$} &
\colhead{Flag} \\
\colhead{} &
\colhead{} &
\colhead{$\kms$} &
\colhead{$\kms$} &
\colhead{$\kms$} &
\colhead{$\kms$} &
\colhead{log($\cm$)} &
\colhead{log($\cm$)} &
\colhead{}
}
\startdata

\cutinhead{$z = 0.13847$}
    \tabOVI &  0.13847 &    0 &    6 &   40.5 &    8.7 &   13.9 &       0.1 &   \nodata  \\
     \tabHI &  0.13850 &    8 &    0 &   30.0 &    0.6 &   15.0 &   \nodata &         L  \\
    \tabCII &  0.13847 &    0 &    0 &   11.1 &    0.9 &   13.9 &       0.1 &   \nodata  \\
    \tabNII &  0.13848 &    2 &    1 &   10.7 &    1.4 &   13.7 &       0.1 &   \nodata  \\
     \tabNV &  0.13848 &    2 &    5 &   31.3 &    7.6 &   13.0 &       0.1 &   \nodata  \\
   \tabSiII &  0.13846 &   -1 &    0 &    6.6 &    0.3 &   12.8 &       0.1 &   \nodata  \\
  \tabSiIII &  0.13848 &    2 &    0 &    8.6 &    0.5 &   13.0 &       0.1 &   \nodata  \\
   \tabSiIV &  0.13845 &   -4 &    1 &    9.5 &    2.2 &   12.7 &       0.1 &   \nodata  \\

\enddata
\end{deluxetable}

%% file: tab11.tex
\begin{deluxetable}{lrrrrrrrc}
\tabletypesize{\tiny}
\tablecaption{\label{tab: PG1216+069_transitions}\OVI\ absorber measurements along the line of sight towards PG\,1216+069}
\tablewidth{0pt}
\tablecolumns{9}
\tablehead{
\colhead{ion} & 
\colhead{$z_{\mbox{comp}}$} & 
\colhead{$v$} &
\colhead{$\sigma_v$} &
\colhead{$b$} & 
\colhead{$\sigma_b$} &
\colhead{$\logN$} & 
\colhead{$\sigma_{\logN}$} &
\colhead{Flag} \\
\colhead{} &
\colhead{} &
\colhead{$\kms$} &
\colhead{$\kms$} &
\colhead{$\kms$} &
\colhead{$\kms$} &
\colhead{log($\cm$)} &
\colhead{log($\cm$)} &
\colhead{}
}
\startdata

\cutinhead{$z = 0.28232$}
    \tabOVI &  0.28232 &    0 &    1 &   12.6 &    2.6 &   13.4 &       0.2 &   \nodata  \\
     \tabHI &  0.28216 &  -38 &    1 &   48.3 &    1.0 &   15.2 &   \nodata &         L  \\
     \tabHI &  0.28229 &   -8 &    0 &   20.2 &    0.6 &   16.5 &   \nodata &         L  \\
   \tabCIII &  0.28228 &   -9 &    0 &   10.1 &    1.6 &   14.2 &   \nodata &         L  \\
  \tabSiIII &  0.28229 &   -6 &    0 &   11.7 &    1.3 &   12.9 &       0.1 &   \nodata  \\

\enddata
\end{deluxetable}

%% file: tab12.tex
\begin{deluxetable}{lrrrrrrrc}
\tabletypesize{\tiny}
\tablecaption{\label{tab: PG1259+593_transitions}\OVI\ absorber measurements along the line of sight towards PG\,1259+593}
\tablewidth{0pt}
\tablecolumns{9}
\tablehead{
\colhead{ion} & 
\colhead{$z_{\mbox{comp}}$} & 
\colhead{$v$} &
\colhead{$\sigma_v$} &
\colhead{$b$} & 
\colhead{$\sigma_b$} &
\colhead{$\logN$} & 
\colhead{$\sigma_{\logN}$} &
\colhead{Flag} \\
\colhead{} &
\colhead{} &
\colhead{$\kms$} &
\colhead{$\kms$} &
\colhead{$\kms$} &
\colhead{$\kms$} &
\colhead{log($\cm$)} &
\colhead{log($\cm$)} &
\colhead{}
}
\startdata

\cutinhead{$z = 0.21950$}
    \tabOVI &  0.21934 &  -39 &    2 &   18.9 &    3.2 &   13.7 &       0.1 &         Z  \\
    \tabOVI &  0.21950 &    0 &    1 &   13.6 &    2.5 &   13.6 &       0.1 &   \nodata  \\
     \tabHI &  0.21932 &  -44 &   31 &   46.0 &   13.6 &   14.0 &       0.5 &   \nodata  \\
     \tabHI &  0.21948 &   -5 &    1 &   27.5 &    1.6 &   15.1 &   \nodata &         L  \\
   \tabCIII &  0.21949 &   -2 &    0 &   10.8 &    2.0 &   13.7 &   \nodata &         L  \\
  \tabSiIII &  0.21950 &    0 &    1 &    8.7 &    2.6 &   12.1 &       0.1 &   \nodata  \\
\cutinhead{$z = 0.25981$}
    \tabOVI &  0.25962 &  -44 &    3 &   15.0 &    3.6 &   13.5 &       0.1 &   \nodata  \\
    \tabOVI &  0.25982 &    4 &    6 &   34.2 &    9.6 &   13.6 &       0.1 &   \nodata  \\
     \tabHI &  0.25963 &  -43 &    7 &   26.6 &    5.0 &   13.6 &       0.2 &   \nodata  \\
     \tabHI &  0.25982 &    1 &    6 &   25.3 &    5.3 &   13.5 &       0.2 &   \nodata  \\

\enddata
\end{deluxetable}

%% file: tab13.tex
\begin{deluxetable}{lrrrrrrrc}
\tabletypesize{\tiny}
\tablecaption{\label{tab: PHL1811_transitions}\OVI\ absorber measurements along the line of sight towards PHL\,1811}
\tablewidth{0pt}
\tablecolumns{9}
\tablehead{
\colhead{ion} & 
\colhead{$z_{\mbox{comp}}$} & 
\colhead{$v$} &
\colhead{$\sigma_v$} &
\colhead{$b$} & 
\colhead{$\sigma_b$} &
\colhead{$\logN$} & 
\colhead{$\sigma_{\logN}$} &
\colhead{Flag} \\
\colhead{} &
\colhead{} &
\colhead{$\kms$} &
\colhead{$\kms$} &
\colhead{$\kms$} &
\colhead{$\kms$} &
\colhead{log($\cm$)} &
\colhead{log($\cm$)} &
\colhead{}
}
\startdata

\cutinhead{$z = 0.15786$}
    \tabOVI &  0.15786 &    0 &    5 &   42.0 &    7.8 &   13.9 &       0.2 &   \nodata  \\
     \tabHI &  0.15785 &   -3 &    5 &   49.9 &    7.1 &   13.3 &       0.1 &   \nodata  \\

\enddata
\end{deluxetable}

%% file: tab14.tex
\begin{deluxetable}{lrrrrrrrc}
\tabletypesize{\tiny}
\tablecaption{\label{tab: PKS0312-77_transitions}\OVI\ absorber measurements along the line of sight towards PKS\,0312$-$77}
\tablewidth{0pt}
\tablecolumns{9}
\tablehead{
\colhead{ion} & 
\colhead{$z_{\mbox{comp}}$} & 
\colhead{$v$} &
\colhead{$\sigma_v$} &
\colhead{$b$} & 
\colhead{$\sigma_b$} &
\colhead{$\logN$} & 
\colhead{$\sigma_{\logN}$} &
\colhead{Flag} \\
\colhead{} &
\colhead{} &
\colhead{$\kms$} &
\colhead{$\kms$} &
\colhead{$\kms$} &
\colhead{$\kms$} &
\colhead{log($\cm$)} &
\colhead{log($\cm$)} &
\colhead{}
}
\startdata

\cutinhead{$z = 0.20275$}
    \tabOVI &  0.20177 & -243 &    8 &   80.8 &   13.1 &   14.1 &       0.1 &   \nodata  \\
    \tabOVI &  0.20276 &    1 &    1 &   68.4 &    2.3 &   14.9 &       0.2 &   \nodata  \\
     \tabHI &  0.20178 & -241 &    1 &   14.6 &    1.2 &   14.7 &       0.1 &   \nodata  \\
     \tabHI &  0.20192 & -205 &    1 &   11.8 &    2.7 &   14.7 &       0.1 &   \nodata  \\
     \tabHI &  0.20210 & -161 &    1 &   20.4 &    2.7 &   14.5 &       0.1 &   \nodata  \\
     \tabHI &  0.20276 &    1 &    0 &   38.2 &    0.4 &   18.4 &   \nodata &         U  \\
    \tabCII &  0.20275 &    0 &    0 &    5.4 &    2.7 &   14.4 &   \nodata &         L  \\
    \tabCII &  0.20288 &   32 &    0 &   17.3 &    5.6 &   13.9 &       0.1 &   \nodata  \\
    \tabCII &  0.20297 &   53 &    0 &   10.4 &    1.6 &   14.4 &   \nodata &         L  \\
    \tabCII &  0.20261 &  -35 &    0 &   19.8 &    5.3 &   14.5 &   \nodata &         L  \\
    \tabCII &  0.20254 &  -53 &    0 &   17.3 &    2.0 &   14.7 &   \nodata &         L  \\
    \tabCII &  0.20181 & -235 &    1 &    7.2 &    1.9 &   13.5 &       0.1 &   \nodata  \\
   \tabCIII &  0.20279 &    8 &    3 &    9.8 &    6.1 &   13.6 &       0.4 &   \nodata  \\
   \tabCIII &  0.20299 &   60 &    2 &   15.5 &    5.9 &   14.3 &   \nodata &         L  \\
   \tabCIII &  0.20255 &  -50 &    2 &   17.2 &    7.2 &   15.6 &   \nodata &         L  \\
   \tabCIII &  0.20319 &  108 &    1 &    4.1 &    7.2 &   14.4 &   \nodata &         L  \\
   \tabCIII &  0.20327 &  130 &    2 &    4.8 &    5.5 &   13.0 &   \nodata &         L  \\
   \tabCIII &  0.20334 &  147 &    2 &    6.1 &    5.2 &   13.1 &       0.3 &   \nodata  \\
   \tabCIII &  0.20212 & -157 &    1 &    4.9 &    6.7 &   14.3 &   \nodata &         L  \\
   \tabCIII &  0.20186 & -221 &    2 &    8.1 &    1.6 &   16.8 &   \nodata &         L  \\
    \tabNII &  0.20275 &    0 &    0 &    6.8 &    5.5 &   13.6 &       0.3 &   \nodata  \\
    \tabNII &  0.20288 &   32 &    0 &   14.4 &   16.6 &   13.5 &       0.4 &   \nodata  \\
    \tabNII &  0.20297 &   53 &    0 &    9.0 &    3.0 &   14.2 &   \nodata &         L  \\
    \tabNII &  0.20261 &  -35 &    0 &   19.4 &   10.4 &   14.1 &   \nodata &         L  \\
    \tabNII &  0.20254 &  -53 &    0 &   15.9 &    4.2 &   14.5 &   \nodata &         L  \\
    \tabNII &  0.20211 & -160 &    0 &   35.6 &    6.7 &   15.2 &   \nodata &         L  \\
    \tabNII &  0.20193 & -205 &    0 &   19.1 &    7.6 &   14.3 &   \nodata &         L  \\
     \tabNV &  0.20272 &   -7 &    9 &   13.8 &   18.8 &   13.1 &       0.5 &   \nodata  \\
     \tabNV &  0.20291 &   41 &    8 &   27.6 &   13.6 &   13.5 &       0.2 &   \nodata  \\
     \tabNV &  0.20262 &  -31 &    8 &    7.9 &   12.1 &   12.9 &       0.7 &   \nodata  \\
     \tabNV &  0.20254 &  -52 &    3 &    4.3 &    5.9 &   12.9 &       0.3 &   \nodata  \\
     \tabNV &  0.20180 & -236 &    2 &    7.2 &    3.7 &   13.2 &       0.2 &   \nodata  \\
    \tabSII &  0.20266 &  -21 &    2 &    9.0 &    3.1 &   14.1 &       0.1 &   \nodata  \\
    \tabSII &  0.20254 &  -53 &    2 &   10.7 &    3.4 &   14.1 &       0.1 &   \nodata  \\
   \tabSiII &  0.20275 &   -1 &    0 &   10.5 &    1.8 &   13.1 &       0.1 &   \nodata  \\
   \tabSiII &  0.20291 &   39 &    9 &   15.3 &    7.4 &   12.8 &       0.3 &   \nodata  \\
   \tabSiII &  0.20297 &   55 &    0 &    6.3 &    1.2 &   13.6 &       0.2 &   \nodata  \\
   \tabSiII &  0.20261 &  -33 &    5 &   10.1 &    3.7 &   13.8 &       0.3 &   \nodata  \\
   \tabSiII &  0.20254 &  -52 &    2 &   14.2 &    1.2 &   14.2 &       0.2 &   \nodata  \\
  \tabSiIII &  0.20273 &   -5 &    0 &   30.4 &    2.3 &   15.7 &   \nodata &         L  \\
  \tabSiIII &  0.20194 & -201 &    1 &    7.9 &    1.8 &   12.5 &       0.1 &   \nodata  \\
  \tabSiIII &  0.20181 & -235 &    0 &   11.5 &    1.6 &   12.8 &       0.1 &   \nodata  \\
   \tabSiIV &  0.20274 &   -1 &   22 &   56.6 &   14.5 &   13.7 &       0.2 &   \nodata  \\
   \tabSiIV &  0.20295 &   49 &    2 &   11.1 &    3.9 &   13.5 &       0.2 &   \nodata  \\
   \tabSiIV &  0.20257 &  -44 &    2 &   17.0 &    5.1 &   14.6 &   \nodata &         L  \\
   \tabSiIV &  0.20194 & -201 &    0 &   15.2 &    4.7 &   13.2 &       0.1 &   \nodata  \\
   \tabSiIV &  0.20181 & -235 &    0 &    2.6 &    2.3 &   13.1 &       0.5 &   \nodata  \\

\enddata
\end{deluxetable}

%% file: tab15.tex
\begin{deluxetable}{lrrrrrrrc}
\tabletypesize{\tiny}
\tablecaption{\label{tab: PKS0405-12_transitions}\OVI\ absorber measurements along the line of sight towards PKS\,0405$-$12}
\tablewidth{0pt}
\tablecolumns{9}
\tablehead{
\colhead{ion} & 
\colhead{$z_{\mbox{comp}}$} & 
\colhead{$v$} &
\colhead{$\sigma_v$} &
\colhead{$b$} & 
\colhead{$\sigma_b$} &
\colhead{$\logN$} & 
\colhead{$\sigma_{\logN}$} &
\colhead{Flag} \\
\colhead{} &
\colhead{} &
\colhead{$\kms$} &
\colhead{$\kms$} &
\colhead{$\kms$} &
\colhead{$\kms$} &
\colhead{log($\cm$)} &
\colhead{log($\cm$)} &
\colhead{}
}
\startdata

\cutinhead{$z = 0.16703$}
    \tabOVI &  0.16660 & -110 &    1 &    7.4 &    2.2 &   13.8 &       0.2 &   \nodata  \\
    \tabOVI &  0.16703 &    0 &    3 &   58.7 &    5.1 &   14.7 &       0.1 &   \nodata  \\
     \tabHI &  0.16662 & -105 &    1 &   11.5 &    2.2 &   13.5 &       0.1 &   \nodata  \\
     \tabHI &  0.16674 &  -75 &    3 &   12.8 &    6.8 &   13.1 &       0.2 &   \nodata  \\
     \tabHI &  0.16712 &   23 &    1 &   36.8 &    2.2 &   15.7 &   \nodata &         L  \\
    \tabCII &  0.16713 &   26 &    0 &   12.2 &    1.3 &   14.3 &   \nodata &         L  \\
    \tabCII &  0.16699 &  -11 &    1 &    7.4 &    1.4 &   13.7 &       0.1 &   \nodata  \\
    \tabNII &  0.16713 &   26 &    1 &   11.2 &    1.3 &   14.2 &   \nodata &         L  \\
     \tabOI &  0.16712 &   22 &    1 &    2.5 &    2.2 &   13.8 &       0.3 &         Z  \\
   \tabSiII &  0.16713 &   25 &    0 &    9.4 &    1.0 &   13.3 &       0.1 &   \nodata  \\
   \tabSiII &  0.16697 &  -15 &    1 &    6.9 &    2.1 &   12.5 &       0.1 &   \nodata  \\
  \tabSiIII &  0.16700 &   -7 &    1 &   13.1 &    2.2 &   12.7 &       0.1 &   \nodata  \\
  \tabSiIII &  0.16716 &   32 &    1 &   16.6 &    1.8 &   13.3 &       0.1 &   \nodata  \\
   \tabSiIV &  0.16712 &   22 &    2 &   30.5 &    3.5 &   13.4 &       0.1 &   \nodata  \\
\cutinhead{$z = 0.18291$}
    \tabOVI &  0.18256 &  -87 &    4 &   21.5 &    6.5 &   13.7 &       0.2 &   \nodata  \\
    \tabOVI &  0.18291 &    0 &    2 &   21.4 &    3.8 &   14.0 &       0.2 &   \nodata  \\
     \tabHI &  0.18258 &  -82 &    4 &   35.3 &    3.1 &   14.5 &   \nodata &         L  \\
     \tabHI &  0.18288 &   -8 &    5 &   30.9 &    4.6 &   14.2 &   \nodata &         L  \\
\cutinhead{$z = 0.36333$}
    \tabOVI &  0.36334 &    2 &    0 &    8.8 &    1.8 &   13.5 &       0.1 &   \nodata  \\
     \tabHI &  0.36334 &    2 &    0 &   22.8 &    4.1 &   13.6 &       0.1 &         Z  \\
   \tabCIII &  0.36334 &    2 &    0 &    6.9 &    3.8 &   12.4 &       0.2 &         Z  \\
\cutinhead{$z = 0.49514$}
    \tabOVI &  0.49514 &    0 &    1 &   43.8 &    1.9 &   14.5 &       0.1 &   \nodata  \\
     \tabHI &  0.49515 &    2 &   10 &   75.0 &    0.0 &   14.3 &       0.1 &         Z  \\
   \tabCIII &  0.49507 &  -13 &    1 &   10.1 &    1.7 &   13.1 &       0.1 &   \nodata  \\
   \tabOIII &  0.49499 &  -29 &    8 &   29.9 &   11.1 &   13.7 &       0.1 &         Z  \\
   \tabOIII &  0.49558 &   88 &    3 &    9.2 &    4.2 &   13.5 &       0.2 &         Z  \\
    \tabOIV &  0.49509 &  -10 &    6 &   33.0 &    0.0 &   14.5 &       0.1 &         Z  \\
    \tabOIV &  0.49556 &   85 &    5 &    6.9 &    5.7 &   13.6 &       0.2 &         Z  \\

\enddata
\end{deluxetable}

%% file: tab16.tex
\begin{deluxetable}{lrrrrrrrc}
\tabletypesize{\tiny}
\tablecaption{\label{tab: PKS1302-102_transitions}\OVI\ absorber measurements along the line of sight towards PKS\,1302$-$102}
\tablewidth{0pt}
\tablecolumns{9}
\tablehead{
\colhead{ion} & 
\colhead{$z_{\mbox{comp}}$} & 
\colhead{$v$} &
\colhead{$\sigma_v$} &
\colhead{$b$} & 
\colhead{$\sigma_b$} &
\colhead{$\logN$} & 
\colhead{$\sigma_{\logN}$} &
\colhead{Flag} \\
\colhead{} &
\colhead{} &
\colhead{$\kms$} &
\colhead{$\kms$} &
\colhead{$\kms$} &
\colhead{$\kms$} &
\colhead{log($\cm$)} &
\colhead{log($\cm$)} &
\colhead{}
}
\startdata

\cutinhead{$z = 0.22565$}
    \tabOVI &  0.22553 &  -28 &    5 &   19.4 &    6.9 &   13.5 &       0.3 &   \nodata  \\
    \tabOVI &  0.22565 &    0 &    2 &   13.4 &    2.8 &   13.7 &       0.1 &   \nodata  \\
     \tabHI &  0.22553 &  -28 &    0 &   51.6 &    3.7 &   13.9 &       0.1 &   \nodata  \\
     \tabHI &  0.22565 &   -1 &    6 &   23.6 &   10.8 &   13.3 &       0.3 &   \nodata  \\
\cutinhead{$z = 0.22744$}
    \tabOVI &  0.22744 &    0 &    1 &   11.4 &    2.4 &   13.5 &       0.1 &   \nodata  \\
     \tabHI &  0.22743 &   -1 &    4 &   19.4 &    6.0 &   12.9 &       0.1 &   \nodata  \\

\enddata
\end{deluxetable}

%% file: tab17.tex
\begin{deluxetable}{lrrrrrrrc}
\tabletypesize{\tiny}
\tablecaption{\label{tab: Ton28_transitions}\OVI\ absorber measurements along the line of sight towards Ton\,28}
\tablewidth{0pt}
\tablecolumns{9}
\tablehead{
\colhead{ion} & 
\colhead{$z_{\mbox{comp}}$} & 
\colhead{$v$} &
\colhead{$\sigma_v$} &
\colhead{$b$} & 
\colhead{$\sigma_b$} &
\colhead{$\logN$} & 
\colhead{$\sigma_{\logN}$} &
\colhead{Flag} \\
\colhead{} &
\colhead{} &
\colhead{$\kms$} &
\colhead{$\kms$} &
\colhead{$\kms$} &
\colhead{$\kms$} &
\colhead{log($\cm$)} &
\colhead{log($\cm$)} &
\colhead{}
}
\startdata

\cutinhead{$z = 0.27340$}
    \tabOVI &  0.27340 &    0 &    3 &   20.9 &    4.2 &   13.4 &       0.2 &   \nodata  \\
     \tabHI &  0.27337 &   -8 &    1 &   30.2 &    1.8 &   14.2 &       0.1 &   \nodata  \\

\enddata
\end{deluxetable}

%% file: tab19.tex
\begin{deluxetable}{lrrrrrrrrrr}
\tabletypesize{\scriptsize}
\tablecaption{\label{tab: ovi_components_temperature}Derived temperatures of \OVI/\HI\ components}
\tablewidth{0pt}
\tablecolumns{11}
\tablehead{
\colhead{LOS} & 
\colhead{$z_{\mbox{abs}}$} & 
\colhead{$v$} & 
\colhead{$b_{\tabHI}$} &
\colhead{$b_{\tabOVI}$} &
\colhead{$\mbox{log}\,T$} & 
\colhead{$\sigma_{\mbox{log}\,T}$} &
\colhead{$b_{nt}$} &
\colhead{$b_{\tabHI, therm}$} &
\colhead{$b_{\tabOVI, therm}$} & 
\colhead{$\mbox{log}\,T_{nt = 0}$} \\
}
\startdata

     3C\,249.1 & 0.24676 &     0 &   38.2 &   27.0 &    4.7 &    0.1 &   26.0 &   27.9 &    7.0 &    5.8  \\
       3C\,273 & 0.12003 &     0 &   23.2 &    8.6 &    4.5 &    0.1 &    6.6 &   22.3 &    5.6 &    4.8  \\
     3C\,351.0 & 0.31659 &    62 &   23.3 &   30.9 &   \nodata &   \nodata &   \nodata &   \nodata &   \nodata &    6.0  \\
     3C\,351.0 & 0.31659 &    -3 &   26.9 &   23.3 &    4.1 &    0.4 &   23.0 &   13.9 &    3.5 &    5.7  \\
     3C\,351.0 & 0.31659 &   -53 &   37.8 &   21.1 &    4.8 &    0.2 &   19.5 &   32.4 &    8.1 &    5.6  \\
   H\,1821+643 & 0.26656 &     0 &   44.8 &   26.1 &    4.9 &    0.1 &   24.4 &   37.5 &    9.4 &    5.8  \\
HE\,0226$-$411 & 0.34034 &     2 &   21.1 &   16.8 &    4.0 &    0.3 &   16.5 &   13.2 &    3.3 &    5.4  \\
HE\,0226$-$411 & 0.35525 &     0 &   28.6 &   22.6 &    4.3 &    0.3 &   22.2 &   18.1 &    4.5 &    5.7  \\
  PG\,0953+415 & 0.14232 &    -1 &   28.7 &   31.9 &   \nodata &   \nodata &   \nodata &   \nodata &   \nodata &    6.0  \\
  PG\,1259+593 & 0.25981 &   -44 &   26.6 &   15.0 &    4.5 &    0.2 &   13.9 &   22.7 &    5.7 &    5.3  \\
  PG\,1259+593 & 0.25981 &     4 &   25.3 &   34.2 &   \nodata &   \nodata &   \nodata &   \nodata &   \nodata &    6.1  \\
     PHL\,1811 & 0.15786 &     0 &   49.9 &   42.0 &    4.7 &    0.4 &   41.5 &   27.8 &    6.9 &    6.2  \\
PKS\,0405$-$12 & 0.16703 &  -110 &   11.5 &    7.4 &    3.7 &    0.2 &    7.0 &    9.1 &    2.3 &    4.7  \\
PKS\,1302$-$10 & 0.22565 &     0 &   23.6 &   13.4 &    4.4 &    0.4 &   12.5 &   20.0 &    5.0 &    5.2  \\
PKS\,1302$-$10 & 0.22744 &     0 &   19.4 &   11.4 &    4.2 &    0.3 &   10.7 &   16.2 &    4.1 &    5.1  \\
       Ton\,28 & 0.27340 &     0 &   30.2 &   20.9 &    4.5 &    0.2 &   20.2 &   22.4 &    5.6 &    5.6  \\
\enddata
\tablecomments{Results of temperature and non-thermal broadening calculation for well-matched 
  \scOVI\ and \scHI\ components. The line-of-sight, system and component are listed, along with the
  measured doppler parameters. The temperature, \logT, and non-thermal broadening, \bnt, are
  given. We also list the thermal component of the observed doppler parameter. The final column
  gives the temperature we would have derived is we assumed $\bnt = 0$.  In three cases these
  parameters cannot be detemined since $\bOVI\ > \bHI$.}

\end{deluxetable}